%% file: main.tex
\documentclass[conference]{IEEEtran}

\usepackage[noadjust]{cite}

\usepackage[utf8]{inputenc}
\usepackage{amssymb}
\usepackage{multirow}
\usepackage{enumitem}
\usepackage{xcolor}
\usepackage{comment}
\usepackage{graphicx}
\usepackage{hyperref}

\newtheorem{problem}{Problem}

\begin{document}
\title{\huge{Benchmarking Filtering Techniques for Entity Resolution}}

\author{\IEEEauthorblockN{George Papadakis}
\IEEEauthorblockA{\textit{National and Kapodistrian University of Athens, Greece} \\
gpapadis@di.uoa.gr}
\and
\IEEEauthorblockN{Marco Fisichella}
\IEEEauthorblockA{\textit{L3S Research Center, Germany} \\
mfisichella@l3s.de}
\and
\IEEEauthorblockN{Franziska Schoger}
\IEEEauthorblockA{\textit{L3S Research Center, Germany} \\
schoger@l3s.de}
\and
\IEEEauthorblockN{George Mandilaras}
\IEEEauthorblockA{\textit{National and Kapodistrian University of Athens, Greece} \\
gmandi@di.uoa.gr}
\and
\IEEEauthorblockN{Nikolaus Augsten}
\IEEEauthorblockA{\textit{University of Salzburg, Austria} \\
nikolaus.augsten@plus.ac.at}
\and
\IEEEauthorblockN{Wolfgang Nejdl}
\IEEEauthorblockA{\textit{L3S Research Center, Germany} \\
nejdl@l3s.de}
}

\maketitle

\begin{abstract}
Entity Resolution is the task of identifying pairs of entity profiles that represent the same real-world object. To avoid checking a quadratic number of entity pairs, various filtering techniques have been proposed that fall into two main categories: (i) \textit{blocking workflows} group together entity profiles with identical or similar signatures, and (ii) \textit{nearest-neighbor methods} convert all entity profiles into vectors and identify the closest ones to every query entity. Unfortunately, the main techniques from these two categories have rarely been compared in the literature and, thus, their relative performance is unknown. We perform the first systematic experimental study that investigates the relative performance of the main representatives per category over numerous established datasets. Comparing techniques from different categories turns out to be a non-trivial task due to the various configuration parameters that are hard to fine-tune, but have a significant impact on performance. We consider a plethora of parameter configurations, optimizing each technique with respect to recall and precision targets. Both schema-agnostic and schema-based settings are evaluated. The experimental results provide novel insights into the effectiveness, the time efficiency and the scalability of the considered techniques.
\end{abstract}

\input{introduction}
\input{relatedWork}
\input{preliminaries}
\input{filtering}
\input{blockingMethods}
\input{nnMethods}
\input{qualitativeAnalysis}
\input{experimentalAnalysis}

\vspace{-6pt}
\section{Conclusions}
\label{sec:conclusions}
\vspace{-6pt}
Our experimental results lead to the following conclusions:

\textbf{1) Fine-tuning vs default parameters.} For all types of methods, optimizing the internal parameters with respect to a performance goal  significantly raises the performance of filtering. This problem is poorly addressed in the literature \cite{DBLP:journals/csur/PapadakisSTP20}, and the few proposed tuning methods require the involvement of experts \cite{DBLP:conf/edbt/LiKCDSPKDR18,DBLP:journals/tlsdkcs/MaskatPE16}. \textit{More emphasis should be placed on a-priori fine-tuning the 
filtering methods through an automatic, data-driven approach that requires no labelled set}. 

\textbf{2) Schema-based vs schema-agnostic settings.} The former significantly improve the time efficiency at the cost of unstable effectiveness, while the latter offer robust effectiveness, as they inherently address heterogeneous schemata as well as misplaced and missing values that are common in ER  \cite{DBLP:journals/pvldb/Thirumuruganathan21,DBLP:conf/sigmod/MudgalLRDPKDAR18}. Even when the schema-based settings exhibit high recall, their maximum precision outperforms the schema-agnostic settings in just three cases ($D_1$--$D_3$). The schema-agnostic settings also exceed the target recall even in combination with default configurations (the baseline blocking workflows), unlike the schema-based settings, where all baseline methods fall short of the target recall at least once. For these reasons, \textit{the schema-agnostic settings are preferable over the schema-based ones}.

\textbf{3) Similarity vs cardinality thresholds.} Poor performance is typically achieved by all similarity-based NN methods (cf. Table \ref{tb:functionality}). 
The LSH variants achieve high recall only by producing an excessively large number of candidates: \textsf{MH-}, \textsf{CP-} and \textsf{HP-LSH} reduce the candidate pairs of the brute-force approach by 48\%, 91\% and 89\%, respectively, on average, across all datasets in Table \ref{tb:ccerDatasets}. This might seem high (a whole order of magnitude for \textsf{CP-} and \textsf{HP-LSH}), but is consistently inferior to the cardinality-based NN methods. This applies even to $\varepsilon$-\textsf{Join}, which is the best similarity-based approach, reducing the candidate pairs of the brute-force approach by 99\% (i.e., multiple orders of magnitude): it underperforms \textsf{kNN-Join} in 9 out of 16 cases. Most importantly, the number of candidates produced by similarity-based methods depends \underline{quadratically} on the total size of the input. For the cardinality-based methods, it depends \underline{linearly} on the size of the query dataset, which is usually the smallest one, i.e., $|C|=k\cdot min{(|E_1|, |E_2|)}$; in almost all cases, $k\ll100$ for all cardinality-based methods, especially \textsf{kNN-Join}. Therefore, \textit{cardinality thresholds are preferable over similarity thresholds.}

\textbf{4) Syntactic vs semantic representations.}
The blocking workflows and the sparse NN methods assume that the pairs of duplicates share textual content; the rarer this content is, the more likely are two entities to be matching. In contrast, most dense NN methods assume that the duplicate entities share syntactically different, but semantically similar content that can be captured by pre-trained character-level embeddings. The latter assumption is true for the matching step of ER \cite{DBLP:journals/pvldb/Thirumuruganathan21}, but our experiments advocate that filtering violates this assumption: semantic-based representations outperform the syntactic ones only in two cases ($D_{b4}$ and $D_{b9}$). Comparing \textsf{kNN-Join} with cardinality-based dense NN methods, we observe that the former consistently uses a lower threshold (see Tables \ref{tb:joinConfiguration} and \ref{tb:nnConfiguration}). This means that the semantic representations introduce more false positives than the syntactic ones, due to the out-of-vocabulary, domain-specific terms in ER datasets -- see Section B in the Appendix for an in depth analysis. Hence, the \textit{syntactic representations are preferable over semantic ones}.

\textbf{5) Most effective} filtering method. The only method that combines a cardinality threshold with a syntactic representation is \textsf{kNN-Join}. Although the Standard Blocking workflow (SBW) performs better in the schema-agnostic settings, \textsf{kNN-Join} offers two qualitative advantages: (i) Unlike \textsf{SBW}, the number of candidates is linear in the input size. (ii)  \textsf{kNN-Join} is easy to configure as shown by the high performance of the \textsf{DkNN-Join} baseline: even though its recall fluctuates in $[0.8, 0.9]$ for three datasets, it outperforms \textsf{PBW}, the default configuration of \textsf{SBW}, in almost all other cases.

\textbf{6) Most scalable filtering method.} Among the considered techniques, only the blocking workflows, \textsf{FAISS} and \textsf{SCANN} scale to all synthetic, Dirty ER datasets within a reasonable time ($<$4 hrs). As the number of input entities increases from 10$^4$ ($D_{10K}$) to 2$\cdot$10$^6$ ($D_{2M}$), the run-times of all techniques scales superlinearly ($>$200 times), but subquadratically ($<$40,000 times). For the blocking workflows, the increase actually ranges from $\sim$8,000 times (\textsf{EQBW}) to $\sim$20,000 times (\textsf{SBW}). This increase is just $\sim$1,600 times for \textsf{SCANN} and $\sim$700 times for \textsf{FAISS}. As a result, \textsf{FAISS} is by far the fastest and most scalable filtering technique when processing large datasets, due to its approximate indexing scheme, leaving \textsf{SCANN} in the second place.

In the future, we will enrich the \textit{Continuous Benchmark of Filtering methods for ER} with new datasets and will update the rankings per dataset with new filtering methods. We will also explore filtering techniques that consider not only textual information but also geographic, numeric etc.

\bibliographystyle{IEEEtran}
\bibliography{references}

\pagebreak
\input{appendix}

\end{document}

%% file: introduction.tex
\section{Introduction}
\label{sec:intro}

Entity Resolution (ER) is a well-studied problem that aims to identify so-called \textit{duplicates} or \textit{matches}, i.e., different entity profiles that describe the same real-world object \cite{DBLP:journals/pvldb/GetoorM12}. ER constitutes a crucial task in a number of data integration tasks, which range from Link Discovery for interlinking the sources of the Linked Open Data Cloud to data analytics, query answering and object-oriented searching~\cite{DBLP:series/synthesis/2015Dong,DBLP:books/daglib/0030287,DBLP:journals/tkde/ElmagarmidIV07}.

Due to its quadratic time complexity, ER typically scales to large data through a Filtering-Verification framework \cite{DBLP:books/daglib/0030287,DBLP:journals/pvldb/GetoorM12}. The first step (\emph{filtering}) constitutes a coarse-grained, rapid phase that restricts the computational cost to the most promising matches, a.k.a. \textit{candidate pairs}. This is followed by the \emph{verification} step,  which examines every candidate pair to decide whether it is a duplicate.

Verification is called \emph{matching} in the ER literature \cite{DBLP:books/daglib/0030287,DBLP:series/synthesis/2015Dong,DBLP:series/synthesis/2015Christophides}. Numerous matching methods have been proposed; most of them rely on similarity functions that compare the textual values of entity profiles \cite{DBLP:series/synthesis/2015Dong,DBLP:series/synthesis/2015Christophides,DBLP:series/synthesis/2021Papadakis,DBLP:books/daglib/0030287}. Early attempts were rule-based, comparing similarity values with thresholds, but more recent techniques rely on learning, i.e., they usually model matching as a binary classification task (\texttt{match}, \texttt{non-match}) \cite{DBLP:series/synthesis/2021Papadakis}. Using a labelled training dataset, supervised, active, and deep learning techniques are adapted to ER \cite{DBLP:journals/tkdd/BarlaugG21}. In some cases, a clustering step is subsequently applied on the resulting similarity scores to refine the output~\cite{DBLP:journals/pvldb/HassanzadehCML09}.

In this work, we are interested in filtering methods, which significantly reduce the 
search space of ER.
The performance of a filtering technique is assessed along  three dimensions: recall, precision and time efficiency. A good filter stands~out~by:

\begin{enumerate}
    \item \emph{High recall.} The candidate pairs should involve many duplicates to reduce the number of false negatives.
    \item \emph{High precision.} The candidate pairs should involve few false positives (i.e., non-matching pairs) to significantly reduce the search space.
    \item \emph{Low run-time.} The overhead added by the filtering step to the overall run-time of ER should be low.
\end{enumerate}

Ideally, filtering should also be directly applicable to the input data.
For this reason, we exclusively consider methods that require no labelled training instances, which are often not available or expensive to produce \cite{DBLP:journals/pvldb/Thirumuruganathan21}. We also focus on techniques for textual entity profiles. These methods are organized into two types~\cite{DBLP:journals/csur/PapadakisSTP20}:
\vspace{-2pt}
\begin{enumerate}
 \item  \textit{Blocking workflows} extract one or more signatures from every entity profile and form clusters of profiles with identical or similar signatures \cite{DBLP:journals/pvldb/0001SGP16,DBLP:journals/tkde/Christen12,DBLP:books/daglib/0030287}. 


 \item \textit{Nearest neighbor (NN) methods} index part or all of the input data and treat every entity profile as a query, to which they return the most similar candidates. They rely on sparse \cite{DBLP:journals/pvldb/MannAB16,DBLP:journals/pvldb/JiangLFL14} or dense numeric vectors \cite{DBLP:journals/is/AumullerBF20}. 
\end{enumerate}
\vspace{-2pt}
Although the two types of filtering techniques follow very different approaches, they all receive the same input (the entity profiles) and produce the same output (candidate~pairs). 

To the best of our knowledge, no previous work systematically examines the relative performance across the two different kinds of methods. The main blocking methods are empirically evaluated in \cite{DBLP:journals/tkde/Christen12,DBLP:journals/pvldb/0001APK15,DBLP:journals/pvldb/0001SGP16}. The studies on sparse \cite{DBLP:journals/pvldb/FierABLF18,DBLP:journals/pvldb/MannAB16,DBLP:journals/pvldb/JiangLFL14} and dense vector-based NN methods~\cite{DBLP:journals/is/AumullerBF20} evaluate run-time and approximation quality, but do not evaluate the performance on ER tasks (cf. Section~\ref{sec:relatedWork}). Even in the few cases where blocking is used as baseline for an NN method (e.g., Standard/Token Blocking is compared to DeepBlocker in \cite{DBLP:journals/pvldb/Thirumuruganathan21}), the comparison is not fair: the blocking method is treated as an independent approach, instead of applying it as part of a blocking workflow, which is common in the literature~\cite{DBLP:journals/pvldb/0001SGP16}. 

Comparing techniques from different categories turns out to be a challenging task, due to their fundamentally different functionality and the diversity of configuration parameters that significantly affect their performance. Yet, there is \emph{no} systematic fine-tuning approach that is generic enough to apply to all filtering methods -- e.g., the step-by-step configuration optimization in \cite{DBLP:journals/pvldb/0001SGP16} applies to blocking pipelines, but not to the single-stage NN~methods.

In this work, we perform the first thorough and systematic experimental study that covers both types of filtering methods. To ensure a fair comparison, every approach is represented by its best performance per dataset, as it is determined after an exhaustive fine-tuning with respect to a common performance target that considers several thousand different parameter configurations. This approach is applied to 5 blocking workflows 
and 8 NN methods over 10 real-world datasets. For each filter type, we also consider two baseline methods with default parameters. We explore both \textit{schema-agnostic} and \textit{schema-based settings}: the latter focus exclusively on the values of the most informative attribute, while the former take into account all information within an entity, essentially treating it as a long textual value -- as a result, they are inherently applicable to datasets with heterogeneous schemata.

Overall, we make the following contributions:

\begin{itemize}[leftmargin=*]
    \item We perform the first thorough experimental analysis on filtering techniques from different categories. We evaluate 14 state-of-the-art filters and 4 baselines on 10 real-world datasets in both schema-based and schema-agnostic settings.
    \item Our configuration optimization process enables the meaningful comparison of blocking workflows and NN methods in the context of serial processing (single-core execution).  
    \item We present a qualitative analysis of the filtering techniques based on
    their \emph{scope} and \emph{internal functionality}.
    \item We perform a thorough scalability analysis that involves seven synthetic datasets of increasing size.
    \item Our work provides new insights into the relative performance and scalability of the considered techniques. We show that the blocking workflows and the cardinality-based sparse NN methods consistently excel in performance. 
    \item Two of the tested NN methods, SCANN and kNN-Join, are applied to ER for the first time. SCANN is one of~the~most scalable techniques, and kNN-Join one of the best~performing ones, while sticking out by its intuitive fine-tuning.
    \item All code and data used in this work are publicly available through a new, open initiative that is called \textsf{Continuous Benchmark of Filtering methods for ER}: \url{https://github.com/gpapadis/ContinuousFilteringBenchmark}.
\end{itemize} 

The main part of the paper is structured as follows: Section \ref{sec:relatedWork} discusses the related works, while Section \ref{sec:prelim} provides background knowledge on filtering and defines formally the configuration optimization task. We elaborate on the filtering methods in Section \ref{sec:filteringMethods} and present their qualitative and quantitative analyses in Sections \ref{sec:qualitativeAnalysis} and \ref{sec:quantitativeAnalysis}, respectively. Section \ref{sec:conclusions} concludes with the main findings of our experimental analysis along with directions for future research.

%% file: relatedWork.tex
\section{Related Work}
\label{sec:relatedWork}

There has been a plethora of works examining the relative performance of blocking methods. The earliest systematic studies were presented in \cite{DBLP:journals/tkde/Christen12,DBLP:books/daglib/0030287}. They focus exclusively on schema-based settings in combination with several user-defined parameter configurations. They also consider exclusively the first step of blocking workflows: block building.

These studies were extended in \cite{DBLP:journals/pvldb/0001APK15}, which examines the same block building methods and configurations, but applies them to schema-agnostic settings, too. The experimental outcomes suggest that recall raises significantly, when compared to the schema-based settings, while requiring no background knowledge about the given data and the quality of its schema. They also suggest that the sensitivity to parameter configuration is significantly reduced.

Building on these works, the experimental analysis in \cite{DBLP:journals/pvldb/0001SGP16} examines the relative performance of the blocking workflows in schema-agnostic settings. In particular, it considers blocking workflows formed by exactly three steps: block building, block filtering and comparison cleaning. In our work, we extend this analysis by considering the top performing block building methods, all of which cluster together entities that share identical signatures. We combine them with three consecutive, but optional steps: block purging, block filtering and comparison cleaning. These steps give rise to seven different filtering pipelines, out of which only one was examined in \cite{DBLP:journals/pvldb/0001SGP16}.

Our work also differs from \cite{DBLP:journals/pvldb/0001SGP16} in that it has only one dataset in common -- excluding the scalability ones. Most importantly,
\cite{DBLP:journals/pvldb/0001SGP16} optimizes the configuration parameters in a heuristic step-by-step manner: first, the performance of block building is heuristically optimized and then, block filtering is heuristically fine-tuned by receiving as input the output of optimized block building as so on for comparison cleaning. In contrast, we consider a holistic approach to configuration optimization, simultaneously fine-tuning all steps in a blocking workflow. As explained in \cite{DBLP:journals/sigmod/PapadakisTTGPK19,DBLP:journals/is/PapadakisMGSTGB20}, this approach consistently outperforms the step-by-step fine-tuning, because it is not confined to local maxima per workflow step, while it considers a significantly larger set of possible configurations.

Finally, we also go beyond \cite{DBLP:journals/pvldb/0001SGP16,DBLP:journals/pvldb/0001SGP16,DBLP:journals/tkde/Christen12,DBLP:books/daglib/0030287} in two more ways: (i) we systematically fine-tune blocking workflows in the context schema-based settings, and (ii) we compare blocking workflows with sparse and dense NN methods. 

The sparse vector-based NN methods essentially correspond to similarity joins. The relative performance of the main methods is examined in \cite{DBLP:journals/pvldb/FierABLF18,DBLP:journals/pvldb/MannAB16,DBLP:journals/pvldb/JiangLFL14} with respect to run-time. Recall and precision are not considered, because they are identical across all approaches, i.e., all methods retrieve the same pairs of entities that exceed a similarity threshold. These pairs, which are not necessarily matching, have not been evaluated with respect to the recall and precision of ER. As a result, these studies are not useful in assessing the performance of string similarity joins for ER. Note also that none of the sparse NN methods we consider has been examined in prior experimental analyses. The reason is that the local kNN-Join lies out of the focus of \cite{DBLP:journals/pvldb/FierABLF18,DBLP:journals/pvldb/MannAB16,DBLP:journals/pvldb/JiangLFL14}, while the range joins for ER involve very low similarity thresholds, unlike the string similarity joins examined in \cite{DBLP:journals/pvldb/FierABLF18,DBLP:journals/pvldb/MannAB16,DBLP:journals/pvldb/JiangLFL14}.

The dense vector-based NN methods are experimentally compared in \cite{DBLP:journals/is/AumullerBF20}, but the evaluation measures are restricted to throughput, i.e., executed queries per second, and to recall, i.e., the portion of retrieved vectors that are indeed the nearest ones with respect to a specific distance function (e.g., the Euclidean one). This is different from ER recall, as the closest vectors are not necessarily matching. 
Nevertheles, we rely on the experimental results of \cite{DBLP:journals/is/AumullerBF20} in order to select the top-performing dense NN methods: Cross-polytope and Hyperplane LSH, FAISS and SCANN. We also consider MinHash LSH, a popular filtering technique \cite{DBLP:journals/csur/PapadakisSTP20}, and DeepBlocker, the most recent learning-based approach, which consistently outperforms all others \cite{DBLP:journals/pvldb/Thirumuruganathan21}.

%% file: preliminaries.tex
\section{Preliminaries}
\label{sec:prelim}

We define an entity profile $e_i$ as the set of textual name-value pairs, i.e., $e_i = \{\langle n_j, v_j\rangle\}$, that describes a real-world object \cite{DBLP:journals/pvldb/0001APK15,DBLP:journals/pvldb/0001SGP16}. This model covers most established data formats, such as the structured records in relational databases and the semi-structured instance descriptions in RDF data. Two entities, $e_i$ and $e_j$, that pertain to the same real-world object ($e_i \equiv e_j$) are called \textit{duplicates} or \textit{matches}.

ER is distinguished into two main tasks \cite{DBLP:series/synthesis/2015Christophides,DBLP:books/daglib/0030287}: 
\begin{enumerate}
    \item  \textit{Clean-Clean ER} or \textit{Record Linkage}, which receives as input two sets of entity profiles, $\mathcal{E}_1$ and $\mathcal{E}_2$ that are individually duplicate-free, but overlapping, and
    \item \textit{Dirty ER} or \textit{Deduplication}, whose input comprises a single set of entity profiles, $\mathcal{E}$, with duplicates in itself. 
\end{enumerate}
In both cases, the output consists of the detected duplicate profiles. 
In the context of Clean-Clean ER, 
the filtering methods
receive as input $\mathcal{E}_1$ and $\mathcal{E}_2$ and produce a set of candidate pairs $\mathcal{C}$, which are highly likely to be duplicates and should be analytically examined during the verification step. To measure the \textbf{effectiveness} of filtering, the following measures are typically used~\cite{DBLP:series/synthesis/2015Christophides,DBLP:journals/pvldb/0001APK15,DBLP:journals/pvldb/0001SGP16,DBLP:journals/tkde/Christen12,DBLP:conf/icde/ElfekyEV02}:
\begin{enumerate}
 \item \textit{Pair Completeness ($PC$)} expresses \textit{recall}, estimating the portion of the duplicate pairs in $\mathcal{C}$ with respect to those in the groundtruth: $PC(\mathcal{C},\mathcal{E}_1,\mathcal{E}_2) = |\mathcal{D}(\mathcal{C})| / |\mathcal{D}(\mathcal{E}_1 \times \mathcal{E}_2)|$, where $\mathcal{D}(x)$ denotes the set of duplicates in set $x$.
 
  \item \textit{Pairs Quality ($PQ$)} captures \textit{precision}, estimating the portion of comparisons in $\mathcal{C}$ that correspond to real duplicates:
  $PQ(\mathcal{C})= |\mathcal{D}(\mathcal{C})|/|\mathcal{C}|$. 
 \end{enumerate}

All measures result in values in the range $[0, 1]$, with higher values indicating higher effectiveness. Note that there is a trade-off between $PC$ and $PQ$: the larger $C$ is, the higher $PC$ gets at the cost of lower $PQ$, and vice versa for a smaller set of candidates. The goal of filtering is to achieve a good balance between these measures.

In this context, we formalize the following configuration optimization task, which enables the comparison of fundamentally different filtering techniques on an equal basis:

\begin{problem}[Configuration Optimization]
Given two sets of entity profiles, $\mathcal{E}_1$ and $\mathcal{E}_2$, a filter method, and a threshold $\tau$ on recall ($PC$), configuration optimization fine-tunes the parameters of the filtering method such that the resulting set of candidates $\mathcal{C}$ on $\mathcal{E}_1$ and $\mathcal{E}_2$ maximizes $PQ$ for $PC\geq\tau$.
\label{pb:optimizationTask}
\end{problem}

Note that we set a threshold on recall, because ER solutions typically consist of two consecutive steps: Filtering and Matching. The recall of the former step determines the overall recall of ER, since the duplicate pairs that are not included in the resulting set of candidates $C$ cannot be detected by most matching methods. This applies both to Clean-Clean ER, where there is no transitivity, due to the 1-1 matching constraint, and to Dirty ER, where the matching algorithms typically consider local information, e.g., the (deep) learning-based methods that treat matching as a binary classification task \cite{DBLP:journals/pvldb/KondaDCDABLPZNP16,DBLP:conf/sigmod/MudgalLRDPKDAR18}.
As a result, we chose $\mathbf{\tau=0.9}$ as the $PC$ threshold that ensures high recall for the overall ER process.
Preliminary experiments demonstrated that a lower threshold,
e.g., 0.85, didn’t alter the relative performance of the considered techniques. Besides, in more than 40\% of considered cases, 
the filtering techniques address Problem 1 with $PC \geq 91\%$, i.e., much higher than our threshold.

Regarding \textbf{time efficiency}, the \textit{run-time} ($RT$) measures the time between receiving the set(s) of entity profiles as input and producing the set of candidate pairs as output. $RT$ should be minimized to restrict the overhead of filtering on ER.

%% file: filtering.tex
\section{Filtering Methods}
\label{sec:filteringMethods}

\subsection{The two paradigms of filtering}

Blocking methods first associate every input entity with one or (usually) more signatures and then, they cluster together entities with identical or similar signatures into blocks. Every pair of entities that appears in at least one block is considered a matching candidate. The resulting blocks contain two \underline{unnecessary} types of candidates, i.e., pairs whose verification lowers precision, without any benefit for recall:

\begin{enumerate}[leftmargin=*]
    \item The \textit{redundant} candidates are repeated across different blocks, because every entity typically participates in multiple ones and, thus, the blocks are overlapping.
    \item The \textit{superfluous} candidates involve non-matching entities.
\end{enumerate}

\begin{figure}[t]
\centerline{\includegraphics[width=0.47\textwidth]{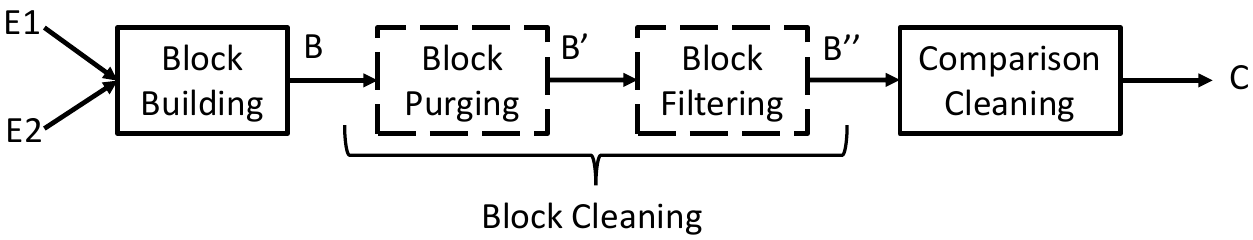}}
\vspace{-8pt}
\caption{The blocking workflow \cite{DBLP:conf/sigmod/GalhotraFSS21}. Dotted contours indicate optional steps.}
\vspace{-15pt}
\label{fig:bworkflow}
\end{figure}

To eliminate the former and reduce the latter, block and comparison cleaning are applied to the initial blocks, restructuring them based on global patterns \cite{DBLP:journals/pvldb/0001SGP16,DBLP:journals/vldb/GalhotraFSS21}. Figure \ref{fig:bworkflow} depicts the complete workflow for blocking \cite{DBLP:journals/csur/PapadakisSTP20,DBLP:conf/sigmod/GalhotraFSS21}. Initially, a set of blocks is created by at least one block building method. The initial block collection is then processed by two coarse-grained block cleaning techniques, Block Purging and Block Filtering. Both produce a new, smaller block collection, $B'$ and $B''$, respectively, but are optional and can be omitted, e.g., in the case of schema-based blocks with low levels of redundant and/or superfluous comparisons. Finally, a comparison cleaning technique is applied, whose fine-grained functionality decides for individual comparisons whether they should be retained or discarded; in this mandatory step, at least the redundant candidates are discarded, but the superfluous ones are also subject to removal. 

A fundamentally different approach is followed by NN methods. Instead of extracting signatures from the input entity profiles, they organize the input set $\mathcal{E}_1$ into an index $I$ (e.g., an inverted index) and use the other dataset as a query set, as shown in Figure \ref{fig:nnWorkflow}. This means that the set of candidate pairs $C$ is formed by probing the index $I$ for every entity profile $e_2 \in \mathcal{E}_2$ and aggregating all query results. To restrict the noise in this process, the textual attribute values are typically cleaned from stop-words and every word is reduced to its base/root form through lemmatization or stemming (e.g., ``blocks'' becomes ``block'') \cite{DBLP:books/daglib/0021593}. The attribute values may also be transformed into (pre-trained) embeddings, i.e., into fixed-length, dense and distributed representations that give rise to semantic similarities \cite{DBLP:journals/computing/WangZJ20}. These optional, preprocessing steps apply to both inputs. After cleaning, we get $\mathcal{E}'_1$ and $\mathcal{E}'_2$, which remain in textual form, but after embedding, we get two sets of dense numeric vectors, $\mathcal{VE}_1$ and $\mathcal{VE}_2$, respectively.

Note that the blocking workflows produce redundant candidates as intermediate results of block building, which are eliminated during comparison cleaning. Contrariwise, NN methods produce no redundant candidates, as every query entity from $\mathcal{E}_2$ is associated with a subset of the indexed entities from~$\mathcal{E}_1$.

%% file: blockingMethods.tex
\subsection{Blocking workflows}
\label{sec:blockingMethods}

\textbf{Block building.} We consider the following state-of-the-art techniques, based on the results of the past experimental studies \cite{DBLP:journals/tkde/Christen12, DBLP:journals/pvldb/0001APK15, DBLP:journals/pvldb/0001SGP16}, 

1) \textit{Standard Blocking.} Given an entity, it tokenizes the considered attribute values on whitespace and uses the resulting tokens as signatures. Hence, every block corresponds to a distinct token, involving all entities that contain it in the selected attribute value(s).

2) \textit{Q-Grams Blocking.} To accommodate character-level errors, it defines as signatures the set of $q$-grams that are extracted from the tokens of Standard Blocking. In other words, every block corresponds to a distinct $q$-gram, encompassing all entities with that $q$-gram in any of the considered values.

3) \textit{Extended Q-Grams Blocking.} Instead of individual $q$-grams, the signatures of this approach are constructed by concatenating at least $L$ $q$-grams, where $L=max(1, \lfloor k \cdot t\rfloor)$, $k$ is the number of $q$-grams extracted from the original key/token and $t\in[0,1)$ is a threshold that reduces the number of combinations as its value increases. Compared to Q-Grams Blocking, the resulting blocks are smaller, but contain candidate pairs that share more content.

4) \textit{Suffix Arrays Blocking.} Another way of accommodating character-level errors in the signatures of Standard Blocking is to consider their suffixes, as long as they comprise a minimum number of characters $l_{min}$. Every block corresponds to a token suffix that is longer than $l_{min}$ and appears in $<b_{max}$ entities.

5) \textit{Extended Suffix Arrays Blocking.} This approach generalizes the previous one by converting the signatures of Standard Blocking in all substrings longer than $l_{min}$ and less frequent than $b_{max}$ entities.

\vspace{2pt}
\noindent
\textit{Example. To illustrate the difference between these blocking methods, consider as an example the attribute value ``\textit{Joe Biden}''. Standard Blocking produces 2 blocking keys: \{Joe, Biden\}. With $q=3$, Q-Grams Blocking produces 4 keys: \{Joe, Bid, ide, den\}. For $T$=0.9, Extended Q-Grams Blocking combines at least two $q$-grams from each token, defining the following 5 blocking keys:  \{Joe, Bid\_ide\_den, Bid\_ide, Bid\_den, ide\_den\}. Using $l_{min}$=3 and a large enough $b_{max}$, Suffix Arrays Blocking yields 4 keys: \{Joe, Biden, iden, den\}, while Extended Suffix Arrays extracts 7 keys: \{Joe, Biden, Bide, iden, Bid, ide, den\}.
}
\vspace{2pt}

Note that we exclude Attribute Clustering Blocking \cite{DBLP:journals/tkde/PapadakisIPNN13}, because it is incompatible with the schema-based settings we are considering in this work. Note also that \textit{we experimented with Sorted Neighborhood \cite{DBLP:journals/pvldb/0001APK15,DBLP:books/daglib/0030287,DBLP:journals/tkde/Christen12}, but do not report its performance, since it consistently underperforms the above methods}. The reason is that this method is incompatible with block and comparison cleaning techniques that could reduce its superfluous comparisons \cite{DBLP:journals/csur/PapadakisSTP20,DBLP:journals/pvldb/0001SGP16}.

\noindent
\textbf{Block cleaning.} We consider two complementary methods:

\begin{figure}[t]
\centering
\includegraphics[width=0.47\textwidth]{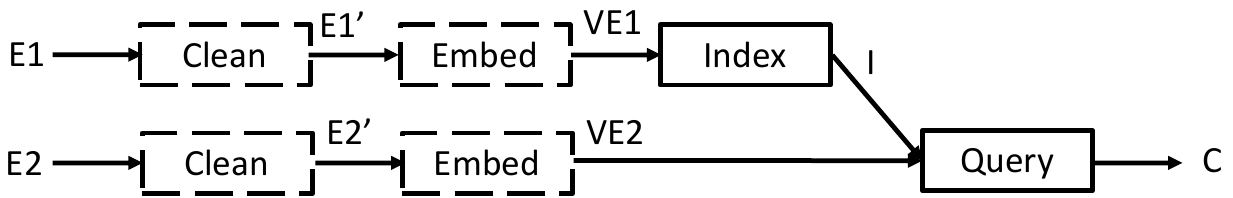}
\vspace{-8pt}
\caption{The workflow of NN methods. Dotted contours indicate optional steps.}
\vspace{-16pt}
\label{fig:nnWorkflow}
\end{figure}

1) \textit{Block Purging \cite{DBLP:journals/tkde/PapadakisIPNN13}.} This parameter-free approach assumes that the larger a block is, the less likely it is to convey matching pairs that share no other block. Such blocks emanate from signatures that are stop-words. Therefore, it removes the largest blocks (e.g., those containing more than half the input entities) in order to significantly increase precision at a negligible (if any) cost in recall.

2) \textit{Block Filtering \cite{DBLP:journals/pvldb/0001APK15}.} It assumes that for a particular entity $e$, its largest blocks are less likely to associate $e$ with its matching entity. For every entity $e$, it orders its blocks in increasing size and retains it in $r$\% of the top (smaller) ones -- $r$ is called filtering ratio. This increases precision to a significant extent for slightly lower recall.

\noindent
\textbf{Comparison cleaning.} We consider two established methods, but only one can be applied in a blocking workflow~\cite{DBLP:journals/pvldb/0001SGP16,DBLP:conf/sigmod/GalhotraFSS21}:

1) \textit{Comparison Propagation \cite{DBLP:journals/tkde/PapadakisIPNN13}.} This parameter-free approach removes all redundant pairs from any block collection without missing any matches, i.e., it increases precision at no cost in recall. It associates every entity with the list of its block ids and retains every candidate pair only in the block with the least~common~id.

2) \textit{Meta-blocking \cite{DBLP:journals/tkde/PapadakisKPN14}.} It targets both redundant and superfluous comparisons using (i) a \textit{weighting scheme}, which associates every candidate pair with a numerical score proportional to the matching likelihood that is extracted from the blocks shared by its constituent entities, and (ii) a \textit{pruning algorithm}, which leverages these scores to decide which candidate pairs will be retained in the restructured block collection that is returned as output.

The rationale behind weighting schemes is that the more and smaller blocks two entities share (i.e., the more and less frequent signatures they share), the more likely they are to be matching. In this context, the following schemes have been proposed \cite{DBLP:journals/tkde/PapadakisKPN14,DBLP:journals/pvldb/SimoniniBJ16,DBLP:journals/csur/PapadakisSTP20}: ARCS promotes pairs that share smaller blocks; CBS counts the blocks the two entities have in common; ECBS extends CBS by discounting the contribution of entities participating in many blocks; JS computes the Jaccard coefficient of the block ids associated with the two entities; EJS extends JS by discounting the contribution of entities participating in many non-redundant pairs; $\chi^2$ estimates to which degree the two entities appear independently in blocks.

These schemes can be combined with the following pruning algorithms \cite{DBLP:journals/tkde/PapadakisKPN14,DBLP:journals/pvldb/SimoniniBJ16}: BLAST retains a pair if its weight exceeds the average maximum weight of its constituent entities; CEP and CNP retain the overall top-K pairs and the top-k pairs per entity, respectively (K and k are automatically configured according to input blocks characteristics); Reciprocal CNP (RCNP) requires that every retained pair is ranked in the top-k positions of both constituent entities; WEP discards all pairs with a weight lower than the overall average one; WNP keeps only pairs with a weight higher than the average one of at least one of their entities; Reciprocal WNP (RWNP) requires a weight higher than the average of both entities.

%% file: nnMethods.tex
\subsection{Sparse vector-based NN methods}
\label{sec:simJoins}

This type includes set-based similarity joins methods, which represent each entity by a set of \emph{tokens} such that the similarity of two entities is derived from their token sets. The similarity between two token sets $A$ and $B$ is computed through one of the following measures, normalized in $[0,1]$~\cite{DBLP:journals/pvldb/MannAB16}:
\begin{enumerate}
    \item Cosine similarity $C(A,B)=|A \cap B|/\sqrt{|A|\cdot |B|}$.
    \item Dice similarity $D(A,B)=2 \cdot |A\cap B|/(|A|+|B|)$.
    \item Jaccard coefficient $J(A,B)=|A\cap B|/|A \cup B|$.
\end{enumerate}
The tokens are extracted from string attributes (the concatenation of all attribute values in the schema-agnostic settings) by considering the character $n$-grams~\cite{DBLP:conf/vldb/GravanoIJKMS01} (as in Q-Grams Blocking) or by splitting the strings on whitespace (as in Standard Blocking). Duplicate tokens within one string are either ignored or de-duplicated by attaching a counter to each token~\cite{DBLP:series/synthesis/2013Augsten} (e.g., $\{a, a, b\}\rightarrow \{a_1, a_2, b_1\}$).

Candidate pairs are formed based on the similarity of two entities according to some \emph{matching principles}~\cite{DBLP:conf/edbt/Augsten18}. We combine two well-known principles with all the aforementioned similarity measures and tokenization schemes~\cite{DBLP:journals/vldb/SilvaALPA13}:

1) \emph{Range join ($\varepsilon$-Join) \cite{DBLP:journals/vldb/SilvaALPA13}.}  It pairs all entities that have a similarity no smaller than a user-defined threshold $\varepsilon$. Numerous efficient algorithms for $\varepsilon$-Join between two collections of token sets have been proposed~\cite{DBLP:conf/www/BayardoMS07,DBLP:conf/icde/ChaudhuriGK06,DBLP:journals/pvldb/BourosGM12,DBLP:journals/pvldb/DengLWF15,DBLP:conf/sigmod/DengT018,DBLP:conf/sigmod/ZhuDNM19,DBLP:journals/pvldb/MannAB16,DBLP:journals/tods/XiaoWLYW11}. All of these techniques produce the exact same set of candidates, but most of them are crafted for high similarity thresholds (above 0.5), which is not the case in ER, as shown in Table \ref{tb:joinConfiguration}. For this reason, we employ \emph{ScanCount}~\cite{DBLP:conf/icde/LiLL08}, which is suitable for low similarity thresholds. In essence, it builds an inverted list on all tokens in the entity collection $\mathcal{E}_1$ and for the lookup of a query entity/token set $e_i\in \mathcal{E}_2$, it performs merge-counts on the posting lists of all tokens in $e_i$. Then, it returns all pairs that exceed the similarity threshold $\varepsilon$.

2) \emph{k-nearest-neighbor join (kNN-Join) \cite{DBLP:journals/vldb/SilvaALPA13}.} Given two collections, $\mathcal{E}_1$ and $\mathcal{E}_2$, it pairs each entity in $e_i\in\mathcal{E}_2$ with the $k$ most similar elements in $\mathcal{E}_1$ that have distinct similarity values, i.e., $e_i$ may be paired with more than $k$ entities if some of them are equidistant from $e_i$. The kNN-Join is not commutative, i.e., the order of the join partners matters. An efficient technique that leverages an inverted list on tokens that are partitioned into size stripes is the \emph{Cone} algorithm~\cite{DBLP:conf/sigmod/KocherA19}, which is crafted for label sets in the context of top-$k$ subtree similarity queries. To increase the limited scope of the original algorithm, we adapted it to leverage ScanCount.

Note that the top-$k$ set similarity joins~\cite{DBLP:conf/icde/XiaoWLS09,DBLP:conf/icde/YangZLZ0J20} compute the $k$ entity pairs between $\mathcal{E}_1$ and $\mathcal{E}_2$ with the highest similarities among all possible pairs. This means that they perform a \underline{global} join that returns the $k$ top-weighted pairs. This is equivalent to $\varepsilon$-Join, if the $k^{th}$ has a similarity equal to $\varepsilon$. Instead, the kNN-Join performs a \underline{local} join that returns at least $k$ pairs per~element~$e_i \in \mathcal{E}_2$. 


\subsection{Dense vector-based NN methods}

\textbf{LSH.} Locality Sensitive Hashing~\cite{lsh98, DBLP:conf/dexa/FisichellaCDN14} constitutes an established solution to the approximate nearest neighbor problem in high-dimensional spaces. 
Its goal is to find entities/vectors that are within $c \cdot R$ distance from a query vector, where $c>1$ is a real number that represents a user-specified approximation ratio, while $R$ is the maximum distance of any nearest neighbor vector from the query. LSH is commonly used as a filtering technique for ER \cite{DBLP:conf/edbt/KarapiperisVVC16, DBLP:conf/edbt/KimL10, DBLP:conf/wsdm/ZhangWSDFP20, DBLP:journals/pvldb/EbraheemTJOT18} because of its sub-linear query performance, which is coupled with a fast and small index maintenance, and its mathematical guarantee on the query accuracy. We consider three popular versions:

1) \textit{MinHash LSH (MH-LSH)} \cite{DBLP:conf/sequences/Broder97,leskovec2020mining}. Given two token sets, this approach approximates their Jaccard coefficient by representing each set as a minhash, i.e., a sequence of hash values that are derived from the minimum values of random permutations. The minhashes are decomposed into a series of \underline{bands} consisting of an equal number of \underline{rows}. This decomposition has a direct impact on performance:  if there are few bands with many rows, there will be collisions between pairs of objects with a very high Jaccard similarity; in contrast, when there are many bands with few rows, collisions occur between pairs of objects with very low similarity. The selected number of bands ($\#bands$) and rows ($\#rows$) approximates a step function, i.e., a high-pass filter, which indicates the probability that two objects share the same hash value:
$(1/\# bands)^{(1/\# rows)}$.

2) \textit{Hyperplane LSH (HP-LSH)}~\cite{Charikar02similarityestimation}.
The vectors are assumed to lie on a unit hypersphere divided by a random hyperplane at the center, formed by a randomly sampled normal vector $r$. This creates two equal parts of the hypersphere with $+1$ on the one side and $-1$ on the other. A vector $v$ is hashed into $h(v) = sgn ( r \cdot v )$. For two vectors $v_1$ and $v_2$ with an angle $\alpha$ between them, the probability of collision is $Pr[h(v_1) = h(v_2)] = 1 - \frac{\alpha}{\pi}$

3) \textit{Cross-Polytope LSH (CP-LSH)}~\cite{multiprobelsh}. It is a generalization of HP-LSH.
At their core, both HP- and CP-LSH are random spatial partitions of a d-dimensional unit sphere centered at the origin. The two hash families differ in how granular these partitions are. The cross-polytope is also known as an l1-unit ball, where all vectors on the surface of the cross-polytope have the l1-norm. In CP-LSH, the hash value is the vertex of the cross-polytope closest to the (randomly) rotated vector. Thus, a cross-polytope hash function partitions the unit sphere according to the Voronoi cells of the vertices of a randomly rotated cross-polytope. In the 1-dimensional case, the cross-polytope hash becomes the hyperplane LSH family.

\input{tables/scope}

\textbf{kNN-Search.} We consider three popular frameworks:

1) \textit{FAISS~\cite{faiss}.} This framework provides methods for kNN searches. Given two sets of (embedding) vectors, it associates every entry $q$ from the query set with the $k$ entries from the indexed set that have the smallest distance to $q$. Two \underline{approximate} methods are provided: (i) a hierarchical, navigable small world graph method, and (ii) a cell probing method with Voroni cells, possibly in combination with product quantization. We experimented with both of them, but they do not outperform the Flat index with respect to Problem \ref{pb:optimizationTask}. The Flat index is also recommended by \cite{faiss}. For these reasons, we exclude the approximate methods in the following. Note that FAISS also supports range, i.e., similarity, search, but our experiments showed that it consistently underperforms kNN search. 

2) \textit{SCANN \cite{DBLP:conf/icml/GuoSLGSCK20}.} This is another versatile framework with very high throughput. Two are the main similarity measures it supports: dot product and Euclidean distance. It also supports two types of scoring: brute-force, which performs exact computations, and asymmetric hashing, which performs approximate computations, trading higher efficiency for slightly lower accuracy. In all cases, SCANN leverages partitioning, splitting the indexed dataset into disjoint sets during training so that every query is answered by applying scoring to the most relevant partitions.

3) \textit{DeepBlocker \cite{DBLP:journals/pvldb/Thirumuruganathan21}.} It is the most recent method based on deep learning, consistently outperforming all others, e.g., AutoBlock \cite{DBLP:conf/wsdm/ZhangWSDFP20} and DeepER \cite{DBLP:journals/pvldb/EbraheemTJOT18}. It converts attribute values into embedding vectors using fastText and performs indexing and querying with FAISS. Its novelty lies in the tuple embedding module, which converts the set of embeddings associated with an individual entity into a representative vector. Several different modules are supported, with the Autoencoder constituting the most effective one under the schema-based settings. In the schema-agnostic settings, the Autoencoder ranks second, lying in close distance of the top-performing Hybrid module, which couples Autoencoder with cross-tuple training.

Note that FAISS and SCANN also use 300-dimensional fastText embeddings. In fact, they are equivalent to the simple average tuple embedding module of DeepBlocker.

%% file: tables/scope.tex
\begin{table}[t]\centering
\footnotesize
 \setlength{\tabcolsep}{5pt}
    \caption{The scope per type of filtering methods.}
    \vspace{-8pt}
	\begin{tabular}{ | c  c | c | c | c |}
		\hline
		\multicolumn{2}{|c|}{\textbf{Scope}} & \textbf{Blocking} &
		\multicolumn{1}{c|}{\textbf{Sparse NN}} & 
		\multicolumn{1}{c|}{\textbf{Dense NN}} \\
		\hline
        \hline
        Syntactic & Schema-based & \checkmark & \checkmark & \checkmark \\
        Representation &  Schema-agnostic & \checkmark & \checkmark & \checkmark \\
        \hline
        \hline
        Semantic & Schema-based & - & - & \checkmark \\
        Representation &  Schema-agnostic & - & - & \checkmark \\
        \hline
	\end{tabular}
	\label{tb:scope}
	\vspace{-12pt}
\end{table}

%% file: qualitativeAnalysis.tex
\section{Qualitative Analysis}
\label{sec:qualitativeAnalysis}

\textbf{Taxonomies.} To facilitate the use and understanding of filtering methods, we organize them into two novel taxonomies.

\smallskip

\emph{Scope.} The first taxonomy pertains to scope, i.e., the entity representation that lies at the core of the filtering method:

\begin{enumerate}[leftmargin=*]
    \item The \textit{syntactic or symbolic representations} consider the actual text in an entity profile, leveraging the co-occurrences of tokens or character n-grams.
    \item The \textit{semantic representations} consider the embedding vectors that encapsulate a textual value, leveraging word-, character- or transformer-based models. We exclusively consider the unsupervised, pre-trained embeddings of fastText  \cite{DBLP:journals/tacl/BojanowskiGJM17} that have been experimentally verified to effectively address the out-of-vocabulary cases in ER tasks, due to domain-specific terminology \cite{DBLP:journals/pvldb/Thirumuruganathan21,Mudgal2018sigmod,DBLP:conf/ijcai/FuHHS20,DBLP:conf/acl/YaoLDLY0LZD20,DBLP:conf/www/ZhangNWST20}. 
\end{enumerate}

These types are combined with schema-based and schema-agnostic settings, yielding the four fields of scope in Table~\ref{tb:scope}. 

The distinctions introduced by this taxonomy are crucial for two reasons: (i) Syntactic representations have the advantage of producing intelligible and interpretable models. That is, it is straightforward to justify a candidate pair, unlike the semantic representations, whose interpretation is obscure to non-experts. (ii) Semantic representations involve a considerable overhead for transforming the textual values into embeddings, even when using pre-trained models. They also require external resources, which are typically loaded in main memory, increasing space complexity. Instead, the methods using syntactic representations are directly applicable to the input data.

We observe that dense NN methods have the broadest scope, being compatible with all four combinations. The syntactic representations are covered by MinHash LSH; its dimensions stem from character k-grams, which are called \textit{k-shingles} and are weighted according to term frequency \cite{leskovec2020mining}. All other dense NN methods employ semantic representations in the form of fixed-size numeric vectors that are derived from fastText.


The blocking and the sparse NN methods cover only the 
syntactic similarities, as they operate directly on the input data.

\input{tables/functionality}

\emph{Internal functionality.} The second taxonomy pertains to the internal functionality of filtering methods. Blocking techniques have been categorized into \textit{lazy} and \textit{proactive} (see \cite{DBLP:journals/pvldb/0001SGP16} for more details).
For NN methods, we define the taxonomy in Table \ref{tb:functionality}, which comprises two dimensions: 
\begin{enumerate}[leftmargin=*]
    \item The \textit{type of operation}, which can be \underline{deterministic}, lacking any randomness, or \underline{stochastic}, relying on randomness. 
    \item The \textit{type of threshold}, which can be \underline{similarity-} or \underline{cardinality-}based. The former specifies the minimum similarity of candidate pairs, while the latter determines the maximum number of candidates per query entity.
\end{enumerate}

The distinctions introduced by this taxonomy are important for two reasons: (i)  The stochastic methods yield slightly different results in each run, unlike the deterministic ones, which yield a stable performance. This is crucial in the context of Problem \ref{pb:optimizationTask}, which sets a specific limit on a particular evaluation measure. For this reason, we set the performance of stochastic methods as the average one after 10 repetitions. (ii) The configuration of cardinality-based methods is straightforward and can be performed a-priori, because it merely depends on the number of input entities. In contrast, the similarity-based methods depend on data characteristics -- the distribution of similarities, in particular.


\input{tables/blockingConfiguration}

\textbf{Configuration space.}
As explained in Section \ref{sec:intro}, a major aspect of filtering techniques is the fine-tuning of their configuration parameters, which has a decisive impact on their performance. For this reason, we combine every method with a wide range of values for each parameter through grid search. The domains we considered per parameter and method are reported in Tables \ref{tb:bwConfigurationDomain}, \ref{tb:joinConfigurationDomain} and \ref{tb:nnConfigurationDomain}.

Starting with Table \ref{tb:bwConfigurationDomain}, the common parameters of the lazy blocking workflows include the presence or absence of Block Purging and the ratio used by Block Filtering. For the latter, we examined at most 40 values in $[0,1]$, with 1 indicating the absence of Block Filtering. Given that these two steps determine the upper bound of recall for the subsequent steps, we terminate their grid search as soon as the resulting $PC$ drops below the target one (0.9) -- in these cases, the number of tested configurations is lower than the maximum possible one. For comparison cleaning, all methods are coupled with the parameter-free Comparison Propagation (CP) or one of the 42 Meta-blocking configurations, which stem from the six weighting schemes and the seven pruning algorithms. 

The Standard Blocking workflow involves only the common parameters, yielding the fewest configurations. The rest of the blocking workflows use the same settings as in \cite{DBLP:journals/pvldb/0001SGP16}.
Note that the proactive ones, which are based on Suffix Arrays Blocking, are not combined with any block cleaning method.

In Table \ref{tb:joinConfigurationDomain}, we notice that the common parameters of set-based similarity joins include the absence or presence of cleaning (i.e., stop-word removal and stemming), the similarity measure as well as the representation model. For the last two parameters, we consider all options discussed in Section \ref{sec:simJoins}, i.e., three similarity measures in combination with 10 models: whitespace tokenization (T1G) or character $n$-grams (CnG), with $n \in \{2, 3, 4, 5\}$; for each model, we consider both the set and the multiset of its tokens, with the latter denoted by appending M at the end of its name (e.g., T1GM). 

Additionally, $\varepsilon$-Join is combined with up to 100 similarity thresholds. We start with the largest one and terminate the grid search as soon $PC$ drops below target recall. kNN-Join is coupled with at most 100 cardinality thresholds, starting from the smallest one and terminating the grid search as soon $PC$ exceeds the target recall. Another crucial parameter for kNN-Join is $RVS$, which is true ($\checkmark$) if $\mathcal{E}_2$ should be indexed and $\mathcal{E}_1$ should used as the query set, instead of the opposite. In theory, kNN-Join involves double as many configurations as $\varepsilon$-Join, but in practice its cardinality threshold does not exceed 26 (see Table \ref{tb:joinConfiguration}), thus reducing significantly the maximum number of its configurations.


\input{tables/joinConfiguration}
\input{tables/vectorConfiguration}
\input{tables/ccerDatasets}

The parameters of dense NN methods are listed in Table \ref{tb:nnConfigurationDomain}. The common parameter is the absence or presence of cleaning. In MinHash LSH, the number of bands and rows are powers of two such that their product is also a power of two, i.e., 2$^n$ with $n\in\{7, 8, 9\}$. For $k$-shingles, we considered four common values for $k$, i.e., $[2,5]$. For Hyperplane and CrossPolytope LSH, we configure two parameters: (i) the number of hash tables ($\#tables$), i.e., cross-polytopes and hyperplanes, respectively, and (ii) the number of hash functions ($\#hashes$). We tested values within the ranges reported in Table \ref{tb:nnConfigurationDomain}, because further ones increased the query time to a considerable extend for a marginal increase in precision.
The number of probes for multi-probe was automatically set to achieve the target recall using the approach in \cite{falconn}. A parameter applying only to CrossPolytope LSH is the last $cp$ dimension, which is chosen between 1 and the smallest power of two larger than the dimension of the embeddings vector (here 512) \cite{DBLP:conf/nips/AndoniILRS15}.

Among the cardinality-based dense NN methods, there are two more common parameters, which are the same as in kNN-Join (Table \ref{tb:joinConfigurationDomain}): $RVS$ and the cardinality-threshold, $K$. For the latter, we consider all values in $[1, 100]$ with a step of 1, as in kNN-Join.  Given, though, that this is not sufficient in some datasets, we additionally consider all values in $[105, 1000]$ with a step of 5 and all values in $[1010, 5000]$ with a step of 10. In each case, the grid search starts from the lowest value and terminates as soon as $PC$ reaches the target recall.

FAISS does not use any other parameter apart from the common ones. Our experiments also demonstrated that it should use the Flat index, while the embedding vectors should always be normalized and combined with the Euclidean distance.

SCANN adds to the common parameters the type of index -- asymmetric-hashing (AH) or brute-force (BF) -- and the similarity measure -- dot product or Euclidean distance. There is no clear winner among these options  (cf. Table \ref{tb:nnConfiguration}).

Finally, DeepBlocker adds to the common parameters the tuple embedding model. We experimented with both top-performing modules, namely AutoEncoder and Hybrid. In most cases, though, the latter raised out-of-memory exceptions, while being a whole order of magnitude slower than the former, as documented in \cite{DBLP:journals/pvldb/Thirumuruganathan21}. For this reason, we exclusively consider AutoEncoder in the following.

\textbf{Take-away message.} All filtering techniques involve three or more configuration parameters that require fine-tuning, a non-trivial task, given that it typically involves several thousands of different settings. Some parameters are common among the techniques of the same category and, thus, experience with one approach can be useful in fine-tuning another one of the same type. Other parameters are intuitive, i.e., easily configured, such as the number of candidates per entity, which is the main parameter of cardinality-based NN methods. For this reason, these methods offer the highest usability, especially when involving a deterministic functionality. These are kNN-Join, FAISS and SCANN. Among them, only kNN-Join operates on syntactic representations, which allows for taking interpretable decisions, just like the blocking workflows. DeepBlocker uses a cardinality threshold, too, but its tuple embedding model employs neural networks with random initialization that are trained on automatically generated random synthetic data; this renders it a stochastic, and thus less robust approach. Among the similarity-based methods, only $\varepsilon$-Join is deterministic, while the three LSH methods are stochastic by definition: MinHash LSH involves random permutations of the input token sets, whereas Hyperplane and CrossPolytope LSH constitute random spatial partitions of a d-dimensional unit sphere. Hence, despite their smaller configuration space, $\varepsilon$-Joins offer higher usability.

\input{tables/derDatasets}

%% file: tables/functionality.tex
\begin{table}[t]\centering
\footnotesize
    \caption{Functionality per NN method.}
    \vspace{-6pt}
	\begin{tabular}{ | c | c | c |}
		\hline
		\textbf{Operation} & \textbf{Similarity Threshold} &  \textbf{Cardinality Threshold} \\
		\hline
		\hline
        Deterministic & $\varepsilon$-Join & kNN-Join, FAISS, SCANN \\
        \hline
        \hline
        Stochastic & MH-, HP-, CP-LSH & DeepBlocker \\
        \hline
	\end{tabular}
	\label{tb:functionality}
	\vspace{-18pt}
\end{table}

%% file: tables/blockingConfiguration.tex
\begin{table}[t]\centering
\footnotesize
\setlength{\tabcolsep}{1pt}
    \caption{The configuration space per blocking workflow.}
    \vspace{-6pt}
	\begin{tabular}{ | c | l | c |}
		\cline{2-3}
		\multicolumn{1}{c|}{} &
		\multicolumn{1}{c|}{\textbf{Parameter}}&
		\multicolumn{1}{c|}{\textbf{Domain}}\\
		\hline
        \hline
        \multirow{5}{*}{Common} & Block Purging ($BP$) & \{ -, \checkmark \} \\
        & Block Filtering ratio ($BFr$) & [0.025, 1.00] with a step of 0.025 \\
        & Weighting Scheme ($WS$) & \{ARCS, CBS, ECBS, JS, EJS, $\chi^2$\} \\
        \cline{2-3}
        & \multirow{2}{*}{Pruning Algorithm ($PA$)} & CP or \{BLAST, CEP, CNP,  \\
        & & RCNP, RWNP, WEP, WNP\} \\
        \hline
        \hline
        Standard & Block Building & parameter-free \\
        \cline{2-3}
        Blocking & Maximum Configurations & 3,440 \\
        \hline
        \hline
        Q-Grams  & $q$ & $[2, 6]$ with a step of 1 \\
        \cline{2-3}
        Blocking & Maximum Configurations & 17,200 \\
        \hline
        \hline
        Extended & $q$ & $[2, 6]$ with a step of 1 \\
        Q-Grams  & $t$ & $[0.8, 1.0)$ with a step of 0.05 \\
        \cline{2-3}
        Blocking & Maximum Configurations & 68,800 \\
        \hline
        \hline
        (Ex.) Suffix & $l_{min}$ & $[2,6]$ with a step of 1\\
        Arrays & $b_{max}$ & $[2,100]$ with a step of 1\\
        \cline{2-3}
        Blocking & Maximum Configurations & 21,285 \\
        \hline
	\end{tabular}
	\label{tb:bwConfigurationDomain}
	\vspace{-16pt}
\end{table}

%% file: tables/joinConfiguration.tex
\begin{table}[t]\centering
\footnotesize
\setlength{\tabcolsep}{1pt}
    \caption{The configuration space per sparse NN method.}
    \vspace{-8pt}
	\begin{tabular}{ | c | l | c |}
		\cline{2-3}
		\multicolumn{1}{c|}{} &
		\multicolumn{1}{c|}{\textbf{Parameter}}&
		\multicolumn{1}{c|}{\textbf{Domain}}\\
		\hline
        \hline
        \multirow{4}{*}{\textsf{Common}} & Cleaning ($CL$) & \{ -, \checkmark \} \\
        & Similarity Measure ($SM$) & \{Cosine, Dice, Jaccard\}\\
        \cline{2-3}
        & Representation & \{T1G, T1GM, C2G, C2GM, C3G, \\
        & Model ($RM$) & C3GM, C4G, C4GM, C5G, C5GM\}  \\
        \hline
        \hline
        \multirow{2}{*}{\textsf{$\varepsilon$-Join}} & Similarity threshold ($t$) & [0.00, 1.00] with a step of 0.01 \\
        \cline{2-3}
        & Maximum Configurations & 6,000 \\
        \hline
        \hline
        \multirow{3}{*}{\textsf{kNN-Join}} &  Candidates per query ($K$) & [1, 100] with a step of 1\\
        & Reverse Datasets ($RVS$) & \{ -, \checkmark \} \\
        \cline{2-3}
        & Maximum Configurations & 12,000 \\
		\hline
	\end{tabular}
	\label{tb:joinConfigurationDomain}
	\vspace{-16pt}
\end{table}

%% file: tables/vectorConfiguration.tex
\begin{table}[t]\centering
\footnotesize
\setlength{\tabcolsep}{1pt}
    \caption{The configuration space per dense NN~method.}
    \vspace{-8pt}
	\begin{tabular}{ | c | l | c |}
		\cline{2-3}
		\multicolumn{1}{c|}{} &
		\multicolumn{1}{c|}{\textbf{Parameter}}&
		\multicolumn{1}{c|}{\textbf{Domain}}\\
		\hline
        \hline
        Common & Cleaning ($CL$) & \{ -, \checkmark \} \\
        \hline
        \hline
        \multirow{4}{*}{\shortstack[c]{MH-LSH}} & $\#bands$ &  \multirow{2}{*}{\shortstack[l]{$\#bands \in P(2), \#rows \in P(2):$\\$\#bands \times \#rows \in \{128, 256, 512\}$}}\\
        & $\#rows$ & \\
        \cline{2-3}
        & $k$ & $[2,5]$ with a step of 1 \\
        \cline{2-3}
        & Configurations & 168 \\
        \hline
        \hline
        \multirow{4}{*}{\shortstack[c]{HP- \& CP- \\LSH}} 
        & $\#tables$ & $2^n: n \in \{0, 9\}$ \\
        & $\#hashes$ & $[1, 20]$ with a step of 1\\
        & last cp dimension & $2^n: n \in \{0, 9\}$ \\
        \cline{2-3}
        & Configurations & 400 (HP), 2,000 (CP) \\
		\hline
		\multicolumn{3}{c}{\textbf{(a) Threshold-based algorithms}} \\
		\hline
		\multirow{2}{*}{Common} & Rev. Datasets ($RVS$) & \{ -, \checkmark \} \\
		& $K$ & $[1, 5000]$ with an increasing step \\
		\hline
		\hline
		FAISS & Max. Configurations & 2,720\\
		\hline
		\hline
		\multirow{3}{*}{SCANN} & index & \{ AH, BF \} \\
		& similarity & \{ DP, LP$^2$ \} \\
		\cline{2-3}
		& Max. Configurations & 10,880\\
		\hline
		\hline
		DeepBlocker & Max. Configurations & 2,720\\
		\hline
		\multicolumn{3}{c}{\textbf{(b) Cardinality-based algorithms}} \\
	\end{tabular}
	\label{tb:nnConfigurationDomain}
	\vspace{-16pt}
\end{table}

%% file: tables/ccerDatasets.tex
\begin{table*}[t]\centering
\setlength{\tabcolsep}{1pt}
{\scriptsize
    \caption{Technical characteristics of the real datasets for Clean-Clean ER in increasing computational cost. 
    }
    \vspace{-8pt}
	\begin{tabular}{ | l | r | r | r | r | r | r | r | r | r | r |}
		\cline{2-11}
		\multicolumn{1}{c|}{}&
		\multicolumn{1}{c|}{$\mathbf{D_{c1}}$} &
		\multicolumn{1}{c|}{$\mathbf{D_{c2}}$} &
		\multicolumn{1}{c|}{$\mathbf{D_{c3}}$} &
		\multicolumn{1}{c|}{$\mathbf{D_{c4}}$} &
		\multicolumn{1}{c|}{$\mathbf{D_{c5}}$} &
        \multicolumn{1}{c|}{$\mathbf{D_{c6}}$} &
        \multicolumn{1}{c|}{$\mathbf{D_{c7}}$} &
        \multicolumn{1}{c|}{$\mathbf{D_{c8}}$} &
        \multicolumn{1}{c|}{$\mathbf{D_{c9}}$} &
        \multicolumn{1}{c|}{$\mathbf{D_{c10}}$} \\
		\hline
        \hline
        $E_1$ / $E_2$ & Rest. 1 / Rest. 2 & Abt / Buy & Amazon / GB & DBLP / ACM & IMDb / TMDb & IMDb / TVDB & TMDb / TVDB & Walmart / Amazon &  DBLP / GS & IMDb / DBpedia \\
        $E_1$ / $E_2$ entities & 339 / 2,256 & 1,076 / 1,076 & 1,354 / 3,039 & 2,616 / 2,294 & 5,118 / 6,056 & 5,118 / 7,810 & 6,056 / 7,810  & 2,554 / 22,074 & 2,516 / 61,353 & 27,615 / 23,182 \\
		Duplicates & 89 & 1,076 & 1,104 & 2,224 & 1,968 & 1,072 & 1,095 & 853 & 2,308 & 22,863 \\
		Cartesian Product & 7.65$\cdot10^5$ & 1.16$\cdot10^6$ & 4.11$\cdot10^6$ &  6.00$\cdot10^6$ & 3.10$\cdot10^7$ & 4.00$\cdot10^7$ & 4.73$\cdot10^7$ & 5.64$\cdot10^7$ &  1.54$\cdot10^8$ & 6.40$\cdot10^8$ \\
		\hline
        Best Attribute & Name & Name & Title & Title & Title & Name & Name & Title & Title & Title \\
        \hline
	\end{tabular}
	\label{tb:ccerDatasets}
	\vspace{-15pt}
}
\end{table*}

%% file: tables/derDatasets.tex
\begin{table}[t]\centering
\renewcommand{\tabcolsep}{2.5pt}
\caption{Technical characteristics of the synthetic, Dirty ER datasets.}
\vspace{-5pt}
{\scriptsize
	\begin{tabular}{ | l | r | r | r | r | r | r | r |}
		\cline{2-8}
		\multicolumn{1}{c|}{}&
		\multicolumn{1}{c|}{$\mathbf{D_{10K}}$} &
		\multicolumn{1}{c|}{$\mathbf{D_{50K}}$} &
		\multicolumn{1}{c|}{$\mathbf{D_{100K}}$} &
		\multicolumn{1}{c|}{$\mathbf{D_{200K}}$} &
		\multicolumn{1}{c|}{$\mathbf{D_{300K}}$} &
		\multicolumn{1}{c|}{$\mathbf{D_{1M}}$} &
		\multicolumn{1}{c|}{$\mathbf{D_{2M}}$} \\
		\hline
		\hline
		$|E|$ & 10,000 & 50,000 & 100,000 & 200,000  & 300,000 
		& 1,000,000 & 2,000,000 \\
		$|D|$ & 8,705 & 43,071 & 85,497 & 172,403 & 257,034 
        & 857,538 & 1,716,102 \\
		$||E||$ & 5.00$\cdot$$10^7$ & 1.25$\cdot$$10^9$ &
		5.00$\cdot$$10^9$ & 2.00$\cdot$$10^{10}$ & 4.50$\cdot$$10^{10}$ &
		5.00$\cdot$$10^{11}$ & 2.00$\cdot$$10^{12}$ \\
		\hline
	\end{tabular}
}
\vspace{-12pt}
	\label{tb:derDatasets}
\end{table}

%% file: experimentalAnalysis.tex
\section{Quantitative Analysis}
\label{sec:quantitativeAnalysis}

\input{experiments}

\textbf{Baseline methods.} To highlight the impact of fine-tuning, our analysis includes two baseline methods per type that require no parameter configuration. Instead, they employ default parameters that are common across all datasets. 

In fact, we consider two baseline blocking workflows: 

(i) \textit{Parameter-free BW (PBW)}. It combines three methods with no configuration parameter (see Section \ref{sec:blockingMethods}): Standard Blocking, Block Purging and Comparison Propagation. It constitutes a Standard Blocking workflow with no configuration.

(ii) \textit{Default BW (DBW)}. We experimented with the default configurations specified in \cite{DBLP:journals/pvldb/0001SGP16} for the five blocking workflows discussed in Section \ref{sec:blockingMethods} and opted for the one achieving the best performance, on average, across all settings. This configuration is Q-Grams Blocking with $q=6$ for block building, Block Filtering with ratio=0.5 for block cleaning and $WEP+ECBS$ for comparison cleaning. 

For sparse NN methods, we use the \textit{Default kNN-Join (DkNN)}  as a baseline. The reason is that kNN-Join typically outperforms the other algorithms of this type and is easy to configure, since it constitutes a deterministic, cardinality-based approach. Table \ref{tb:joinConfiguration} shows that its best performance is usually achieved when it is combined with cosine similarity, pre-processing to clean the attribute values and a very low number of nearest neighbors per query entity, $K$. We used the smallest input dataset as the query set, minimizing the candidate pairs, set $K$=$5$ and $C5GM$ as the default representation model, since it achieves the best average~performance.

Among the dense NN methods, we selected DeepBlocker as the baseline approach, given that it typically outperforms all others in terms of effectiveness (cf. Figure \ref{fig:heatmaps}). Table \ref{tb:nnConfiguration} shows that it usually works best when cleaning the attribute values with stemming and stop-word removal, when using the smallest input dataset as the query set and when using a small number of candidates per query. We set $K$=$5$ so that the \textit{Default DeepBlocker (DDB)} as with DkNN.

\textbf{Schema-agnostic Settings.} Due to lack of space, the detailed experimental results with respect to $PC$, $PQ$, run-time and the number of candidates over the datasets in Table \ref{tb:ccerDatasets} are reported in Table \ref{tb:results}.
All fine-tuned methods consistently exceed the target recall, i.e., $PC\ge0.9$, regardless of their type -- only \textsf{DkNN} and \textsf{DDB} violate the desired recall level in a few cases. For this reason,  the relative effectiveness of the considered methods is primarily determined by $PQ$.

Figure \ref{fig:heatmaps}(a) presents the ranking of every filtering technique with respect to precision ($PQ$) per dataset, along with its average ranking position across all datasets. The method achieving the highest $PQ$ is ranked first, the next best second etc. Ties receive the same ranking, which is the highest possible; e.g, if FAISS and SCANN have the highest $PQ$ after the third best method, they are both placed fourth and the next best method is placed in the sixth place. Methods that fail to satisfy the recall threshold are placed at the last ranking position. 

Among the blocking workflows, the Standard Blocking one (\textsf{SBW}) has the highest average ranking position (2.2), because it outperforms all others in the eight largest datasets. Its parameter-free counterpart, \textsf{PBW}, exhibits the lowest average ranking (11.9),
thus highlighting the benefits of fine-tuning. The same conclusion is drawn from the comparison between the Q-Grams Blocking workflow (\textsf{QBW}) and its default configuration, \textsf{DBW}, whose average ranking positions are 4.7 and 10.8, respectively. The Extended Q-Grams Blocking workflow (\textsf{EQBW}) follows \textsf{QBW} in close distance, taking the 5$^{th}$ position, on average.
These three methods are outperformed by Suffix Arrays Blocking workflow (\textsf{SABW}), 
which has the second highest average ranking (4.2), 
while being consistently faster.
Finally, the Extended Suffix Arrays Blocking workflow (\textsf{ESABW}) ranks as the last fine-tuned workflow, 
even though it achieves the overall best $PQ$~for~$D_{a2}$. 

These patterns suggest that \textit{attribute value tokens offer the best granularity for blocking signatures in the schema-agnostic settings}. Even though some candidates might be missed by typographical errors, they typically share multiple other tokens, due to the schema-agnostic settings. Using substrings of tokens (i.e., q-grams and suffix arrays) as signatures increases significantly the number of candidate pairs, without any benefit in recall. These pairs are significantly reduced by the block and comparison cleaning, but to a lesser extent than those of \textsf{SBW}, yielding lower precision. The only exceptions are the two smallest datasets, where the maximum block size limit of (\textsf{E})\textsf{SABW} raises precision to the overall highest level. 

Among the sparse NN methods, we observe that kNN-Join (\textsf{kNNJ}) outperforms $\epsilon$-Join in eight datasets. In half of these cases, \textsf{kNNJ} actually achieves the best precision among all considered methods. On average, \textsf{kNNJ} ranks much higher than $\epsilon$-\textsf{Join} (3.2 vs 5.7, respectively),
which suggests that \textit{the cardinality thresholds are significantly more effective in reducing the search space of ER than the similarity ones}. The reason is that the latter apply a global condition, unlike the former, which operate locally, selecting the best candidates per query entity. Comparing \textsf{kNNJ} with its baseline method, \textsf{DkNN}, the former consistently outperforms the latter, as expected,
verifying the benefits of parameter fine-tuning.

\begin{figure*}[t]
\centering
\includegraphics[width=0.32\textwidth]{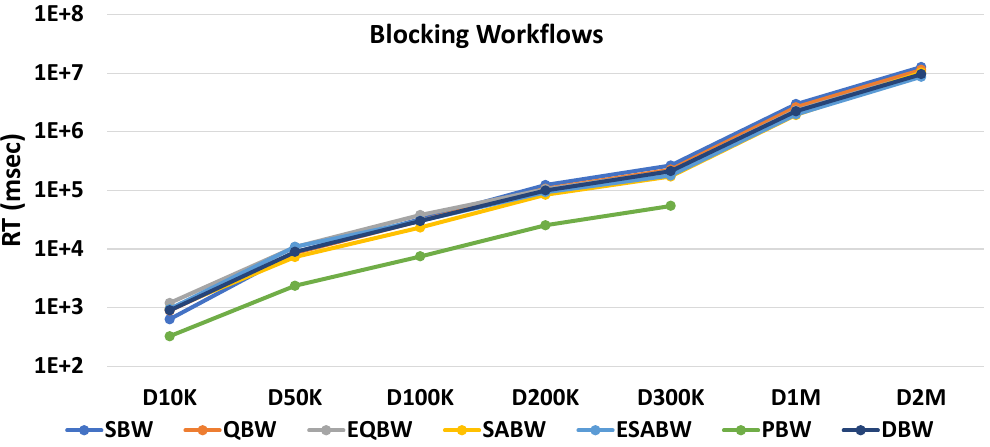}
\includegraphics[width=0.32\textwidth]{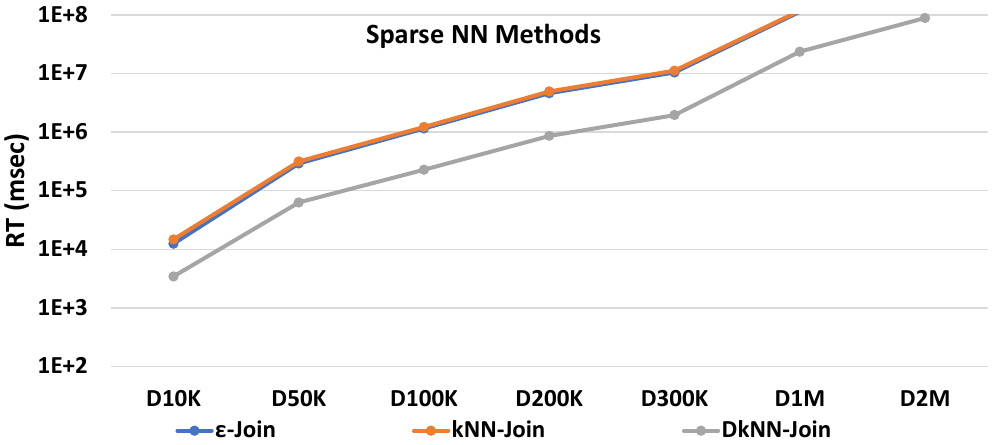}
\includegraphics[width=0.32\textwidth]{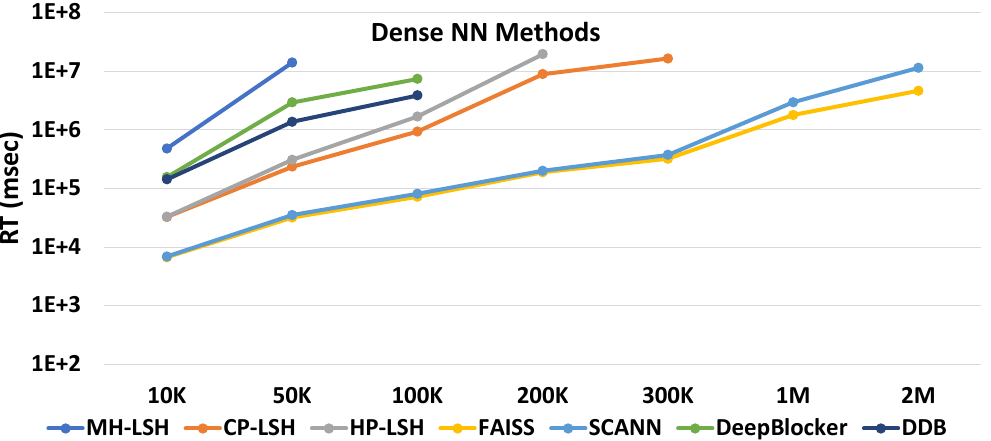}
\vspace{-8pt}
\caption{Scalability analysis with respect to run-time (in milliseconds) over all datasets in Table \ref{tb:derDatasets}. Note the logarithmic scale on the vertical axis.}
\vspace{-16pt}
\label{fig:scalability}
\end{figure*}

Regarding the dense NN methods, we observe that the similarity-based ones consistently achieve the lowest by far precision among all fine-tuned techniques. \textsf{CP-LSH} typically outperforms \textsf{MH-} and \textsf{HP-LSH}, but underperforms all baseline methods (i.e., \textsf{PBW}, \textsf{DBW} and \textsf{DkNN}) in  most of the cases, especially over the largest datasets. The reason is that \textit{the similarity-based methods achieve high recall only by producing an excessively large number of candidate pairs} (\textsf{MH-LSH} actually runs out of memory when processing $D_{a10}$). Their precision raises to high levels only for $PC\ll0.9$. This applies to both sparse syntactic and dense embedding vectors.

Significantly better performance is achieved by the cardinality-based NN methods.
\textsf{FAISS} and \textsf{SCANN} exhibit practically identical performance across all datasets, because they perform an exhaustive search of the nearest neighbors. They differ only in $D_{a3}$, $D_{a8}$ and $D_{a9}$, where \textsf{SCANN} outperforms \textsf{FAISS}, despite using approximate scoring (AH). The two algorithms outperform all NN methods in four datasets, with \textsf{DeepBlocker} being the top performer in the remaining six. As a result, \textsf{DeepBlocker} exhibits the highest average ranking position among all methods of this type (9.0), which
means that \textit{the learning-based tuple embedding module raises significantly the precision of NN methods}. However, \textsf{DeepBlocker} does not scale to $D_{a10}$ with the available memory resources, due to the extremely large set of candidate pairs. 
The same applies to its default configuration, \textsf{DDB}. Note that 
\textsf{DDB} fails to achieve the target recall in four datasets; for the remaining five, its low average ranking position
verifies the need for parameter fine-tuning.

Comparing the top performing methods from each category in terms of precision, we notice that \textsf{SBW} takes a clear lead (2.2), followed in close distance by \textsf{kNNJ} (3.2), leaving \textsf{DeepBlocker} in the last place (9.0). \textsf{SBW} achieves the maximum $PQ$ in four datasets, \textsf{kNNJ} in three and \textsf{DeepBlocker} in none of them. 
Note, though, that \textit{\textsf{kNNJ} constitutes a more robust approach that is easier to configure and apply in practice}. Its default configuration, \textsf{DkNN}, exhibits the highest average ranking position, together with \textsf{DBW} (10.9 and 10.8, respectively),
outperforming the other two baseline methods to large extent: \textsf{PBW} ranks 11.9 and \textsf{DDB} 13.9.

\textbf{Schema-based settings.} Similar to the schema-agnostic settings,
all fine-tuned filtering methods achieve the target $PC$. The baseline methods fail in two datasets, except for \textsf{DkNN}, which fails just once. Along with the four datasets with insufficient coverage, this means that \textit{without fine-tuning, the schema-based settings fall short of recall in half~the~cases}. 

Regarding precision, \textsf{SBW} and \textsf{QBW} outperform all blocking workflows in two datasets each, but the latter achieves the highest average ranking position (4.3), 
leaving \textsf{SBW} in the second place with 4.7.
\textsf{EQBW} ranks third (5.2) and \textsf{SABW} fourth (6.3), even though each method is the top performer in one dataset. \textsf{ESABW} again exhibits the lowest average ranking position among all fine-tuned workflows.

Among the sparse NN methods, there is a balance between $\epsilon$-\textsf{Join} and \textsf{kNNJ}, as each method achieves the top precision in half the datasets. Yet, \textsf{kNNJ} lies very close to $\epsilon$-\textsf{Join} in the cases where the latter is the top performer, but not vice versa: 
as a result, its average ranking position (4.5) is much higher than that of $\epsilon$-\textsf{Join} (5.5).
\textsf{DkNN} exhibits a robust, high performance that remains very close to \textsf{kNNJ} in all datasets, except $D_{b8}$, where it fails to reach the target recall. Note that \textsf{DkNN} outperforms $\epsilon$-\textsf{Join} over $D_{b2}$ and $D_{b3}$. These settings verify that \textit{the cardinality thresholds are superior to the similarity ones, regardless of the schema settings}. 


Regarding the dense NN methods, the similarity-based ones, i.e., the LSH variants, consistently underperform the cardinality-based ones. \textsf{MH-LSH} actually does not scale to $D_{b9}$, due to very large set of candidates it produces. \textsf{FAISS} and \textsf{SCANN} exhibit practically identical performance, outperforming \textsf{DeepBlocker} in four datasets. As a result, they achieve a slightly higher average ranking position (7.2 vs 7.7).
The baseline method \textsf{DDB} exhibits low precision, merely outperforming \textsf{MH-LSH}.

Among the top performing fine-tuned methods per category, \textit{\textsf{QBW} and \textsf{kNNJ} exhibit the best and most robust performance}. The latter actually achieves the overall best $PQ$ in two datasets.
Both methods outperform 
\textsf{FAISS} and \textsf{SCANN} to a significant extent, judging from their average ranking positions (4.3 and 4.5 vs 7.2). Note that
\textsf{DkNN} outperforms all other baselines in most cases, achieving the highest by far average ranking position (10.8 vs 13.7 and 13.8).

\textbf{Scalability Analysis.} We now examine the relative time efficiency of all filtering techniques as the size of the input data increases, using the seven synthetic datasets in Table \ref{tb:derDatasets}. Note that the different programming languages do not allow for comparing them on an equal basis. For this reason, we follow the approach of ANN Benchmark \cite{DBLP:journals/is/AumullerBF20}, which compares implementations rather than algorithms. The reason is that even slight changes (e.g., a different data structure) in the implementation of the same algorithm in the same language might lead to significantly different run-times.

Figure \ref{fig:scalability} reports the experimental results. Note that the scale of the vertical axis is logarithmic, with the maximum value (10$^8$ msec) corresponding to 27.8 hrs. Every filtering technique was fine-tuned on the smallest dataset ($D_{10K}$) with respect to Problem 1 and the same configuration was applied to all seven datasets. We exclusively considered schema-agnostic settings, due to their robustness with respect to recall. The exact configurations are reported in Tables \ref{tb:bwConfigurations}-\ref{tb:nnConfiguration}.

Starting with the blocking workflows on the left, we observe that \textsf{PBW} is the fastest one, due to its simple comparison cleaning, which merely applies Comparison Propagation to eliminate the redundant candidate pairs. As a result, it does not scale to the two largest datasets, $D_{1M}$ and $D_{2M}$, due to the very large number of candidate pairs it generates. All other workflows are coupled with a Meta-blocking approach that assigns a weight to every candidate pair and prunes the lowest-weighted ones in an effort to reduce the superfluous pairs, too. As a result, they trade higher run-times for higher scalability. The differences between most blocking workflows are minor: \textsf{ESABW} is the fastest and \textsf{SBW} the slowest one, requiring 2.4 and 3.5 hrs, respectively, over $D_{2M}$.

Among the sparse NN methods, only \textsf{DkNN} scales to $D_{2M}$, because its large $q$-grams ($q=5$) generate few candidates per entity. In contrast, $\epsilon$-\textsf{Join} and \textsf{kNNJ} rely on character bigrams, yielding a time-consuming functionality, due to the excessive number of candidates: they require more than 30 hrs for $D_{1M}$.

Among the dense NN methods, \textsf{MH-}, \textsf{HP-} and \textsf{CP-LSH} scale up to $D_{50K}$, $D_{200K}$ and $D_{300K}$, respectively, because their large number of candidates does not fit into the available main memory. Similarly, \textsf{DeepBlocker} and \textsf{DDB} scale up to $D_{100K}$, because of their quadratic time complexity (i.e., they requires $\sim$300GB of RAM to create a 200K$\times$200K float array for $D_{200K}$). \textsf{FAISS} and \textsf{SCANN} exhibit much higher scalability, due to the approximate indexes they employ (IVF and AH, respectively). The former actually processes $D_{2M}$ within 1.3 hrs, being the fastest approach by far, while the latter exhibits run-times similar to the blocking workflows, requiring 3.2 hrs for $D_{2M}$.

%% file: experiments.tex
\textbf{Datasets.} We use two sets of datasets. The first one involves 10 real-world datasets for Clean-Clean ER that are popular in the literature \cite{DBLP:journals/pvldb/Thirumuruganathan21,DBLP:conf/sigmod/MudgalLRDPKDAR18,DBLP:journals/pvldb/KopckeTR10,DBLP:journals/pvldb/0001SGP16}. Their technical characteristics are reported in Table \ref{tb:ccerDatasets}. $D_{1}$, which was first used in OAEI 2010 \cite{DBLP:conf/semweb/EuzenatFMPSSSSSS10}, contains restaurant descriptions. $D_{2}$ encompasses duplicate products from the online retailers Abt.com and Buy.com \cite{DBLP:journals/pvldb/KopckeTR10}. $D_{3}$ matches product descriptions from Amazon.com and the Google Base data API (GB) \cite{DBLP:journals/pvldb/KopckeTR10}. $D_{4}$ entails bibliographic data from DBLP and ACM \cite{DBLP:journals/pvldb/KopckeTR10}. $D_{5}$, $D_{6}$ and $D_{7}$ involve descriptions of television shows from TheTVDB.com (TVDB) and of movies from IMDb and themoviedb.org (TMDb) \cite{DBLP:conf/esws/ObraczkaSR21}. $D_{8}$ matches product descriptions from Walmart and Amazon \cite{DBLP:conf/sigmod/MudgalLRDPKDAR18}. $D_{9}$ involves bibliographic data from publications in DBLP and Google Scholar (GS) \cite{DBLP:journals/pvldb/KopckeTR10}. $D_{10}$ interlinks movie descriptions from IMDb and DBpedia \cite{DBLP:journals/is/PapadakisMGSTGB20} -- note that it includes a different snapshot of IMDb than $D_5$~and~$D_6$.

\begin{figure*}[t]
\centering
\includegraphics[width=0.34\textwidth]{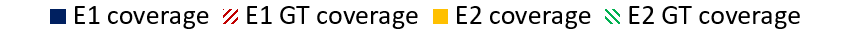}
\includegraphics[width=0.32\textwidth]{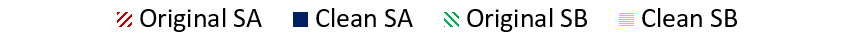}
\includegraphics[width=0.32\textwidth]{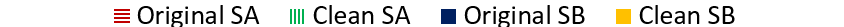}
\includegraphics[width=0.32\textwidth]{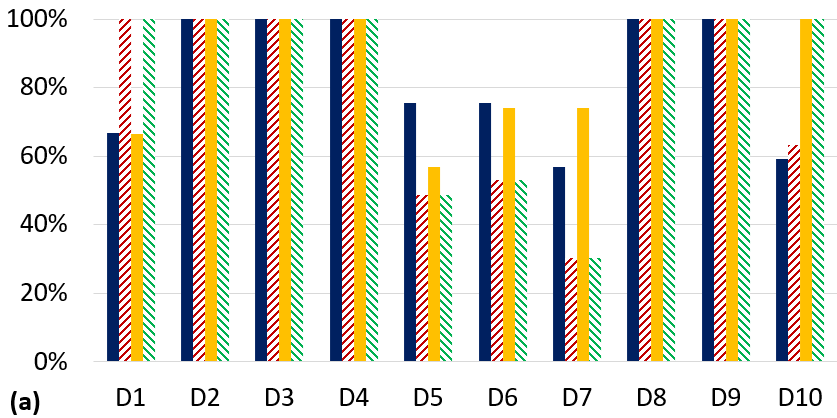}
\includegraphics[width=0.32\textwidth]{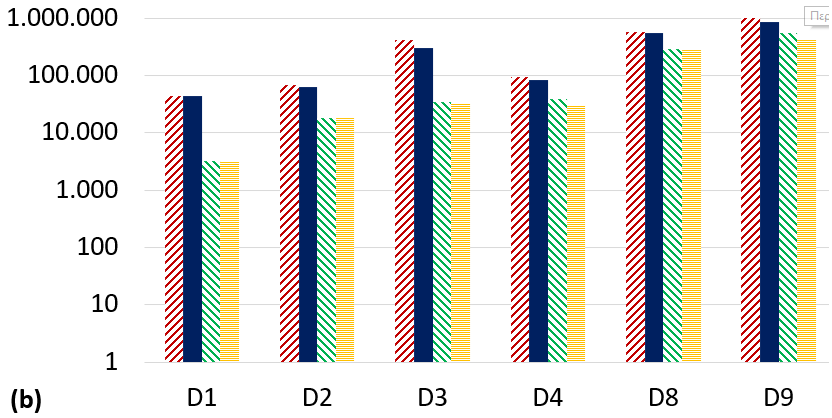}
\includegraphics[width=0.32\textwidth]{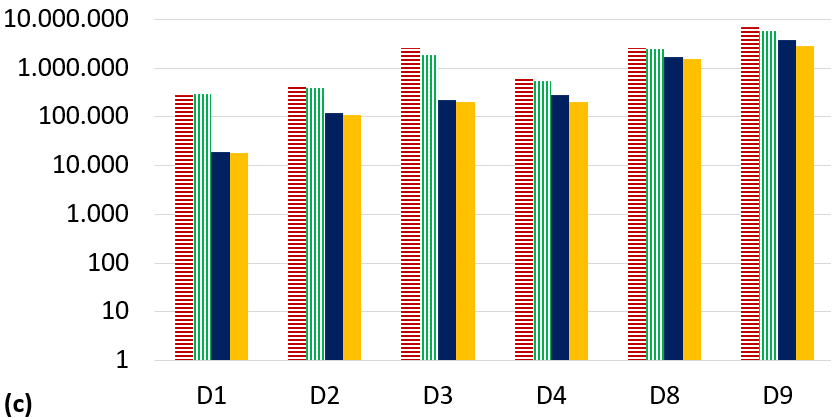}
\vspace{-8pt}
\caption{(a) The coverage of the best attribute per each dataset, (b) the vocabulary size in schema-agnostic and schema-based settings, and (c) the overall character length in the textual content of the datasets for both schema settings.}
\vspace{-16pt}
\label{fig:schemaSettings}
\end{figure*}

The second set of datasets involves seven synthetic ones for Dirty ER of increasing size, from 10 thousand to 2 million entities. They have been widely used in the literature \cite{DBLP:journals/pvldb/0001APK15,DBLP:journals/pvldb/0001SGP16,DBLP:journals/is/KenigG13}, as they are ideal for investigating the scalability of filtering techniques. They have been generated by Febrl \cite{DBLP:conf/kdd/Christen08a} using the guidelines specified in \cite{DBLP:journals/tkde/Christen12}: first, duplicate-free entities describing persons (i.e., their names, addresses etc) were created based on frequency tables of real-world data. Then, duplicates of these entities were randomly generated according to real-world error characteristics and modifications. The resulting datasets contain 40\% duplicate entities with up to 9 duplicates per entity, no more than 3 modifications per attribute, and up to 10 modifications per entity. Table \ref{tb:derDatasets} reports their technical characteristics -- $|D|$ and $||E||$ stand for the number of duplicates and the Cartesian product, resp.

\textbf{Setup.}
All experiments on the Clean-Clean ER datasets were performed on commodity hardware equipped with an Intel i7-4710MQ @ 2.50GHz with 16GB of RAM, running Ubuntu 18.04.3 LTS. The available memory should suffice, given that all datasets occupy few MBs on the disk in their original form. All experiments on the Dirty ER datasets were performed on a server with an Intel Xeon Gold 6238R @ 2.20GHz with 128GB of RAM, running Ubuntu 18.04.6 LTS. For most time measurements, we performed  10 repetitions and report the average value. These measurements do not include the time required to load the input data into main memory.

For the implementation of all methods, we used existing, popular implementations. For blocking workflows and sparse NN methods, we employed JedAI's latest version, 3.2.1 \cite{jedai}. All experiments were run on Java 15. For MH-LSH, we used \texttt{java-LSH}, version 0.12 \cite{minhashlsh}. For HP- and CP-LSH, we used the Python wrapper of FALCONN \cite{DBLP:conf/nips/AndoniILRS15}, version 1.3.1. For FAISS, we used version 1.7.2 of the Python wrapper provided by Facebook Research \cite{faissLIB}. For SCANN, we used version 1.2.5 of the Python implementation provided by Google Research \cite{scannLib}. For DeepBlocker, we used the implementation provided by the authors \cite{dbLib}. FAISS, SCANN and DeepBlocker can exploit GPU optimizations, but all methods were run on a single CPU to ensure a fair comparison. 

\textbf{Schema settings.} In each dataset, we consider both \textit{schema-agnostic} and \textit{schema-based settings}. The former supports heterogeneous datasets, as it takes into account the content of all attributes, regardless of their attribute names, while the latter focuses on the values of the most suitable attribute in terms of coverage and distinctiveness. We define \textit{coverage} of attribute $a$ as the portion of entities that contain a non-empty value for $a$, while \textit{distinctiveness} expresses the portion of different values among these entities (e.g., an attribute like year for publications or movies has very low distinctiveness in contrast to their titles). Based on these criteria, we selected the attributes in Table \ref{tb:ccerDatasets} for the schema-based settings.

The actual coverage of these attributes per dataset is reported in Figure \ref{fig:schemaSettings}(a) along with their \textit{groundtruth coverage}, i.e., the portion of duplicate profiles that have at least one non-empty value for the respective attribute. We observe that for half the datasets ($D_2$-$D_4$, $D_8$, $D_9$), the selected attribute has perfect (groundtruth) coverage. In $D_5$-$D_7$, though, the overall coverage fluctuates between 55\% and 75\%, dropping to 30\%-53\% for duplicates, even though we have selected the most frequent attribute in each case. For these datasets, no filtering technique can satisfy the target recall specified in Section \ref{sec:prelim}. As a result, we exclude the schema-based settings of $D_5$-$D_7$ from our analysis. The same applies to $D_{10}$, even though the inadequate coverage pertains only to one of its constituent datasets. An exception is $D_1$, where the selected attribute covers just 2/3 of all profiles, but all of the duplicate ones. 

Note that the low coverage for distinctive attributes like ``Name'' and ``Title'' does not mean that there are entities missing the corresponding values. Their values are typically misplaced, associated with a different attribute, e.g., due to extraction errors \cite{DBLP:journals/pvldb/Thirumuruganathan21,DBLP:conf/sigmod/MudgalLRDPKDAR18}. \textit{The schema-agnostic settings inherently tackle this form of noise, unlike the schema-based~ones.}

For the datasets that have both settings, it is worth comparing their computational cost in terms of \textit{vocabulary size} (i.e., total number of distinct tokens) and \textit{overall character length} (i.e., total number of characters). These measures appear in Figures \ref{fig:schemaSettings}(b) and (c), resp. In each case, we also consider the values of these measures after \textit{cleaning}, i.e., after removing the stop-words and stemming all tokens, as required by the workflow in Figure \ref{fig:nnWorkflow}. We used \texttt{nltk} \cite{bird2009natural} for this purpose.

We observe that on average, the schema-based settings reduce the vocabulary size and the character length by 66.0\% and 67.7\%, respectively. The reason is that in most cases, the schema-agnostic settings include 3-4 name-value pairs, on average, as indicated in Table \ref{tb:ccerDatasets} (av. profile). The more attributes 
and the more name value (n-v) pairs a dataset includes, the larger is the difference between the two settings. Cleaning further reduces the vocabulary size by 11.9\% and the character length by 13.5\%, on average. Hence,\textit{ the schema-based settings are expected to significantly reduce the run-time of filtering methods, especially when combined with cleaning}. 

\begin{figure}[t]
\centerline{\includegraphics[width=0.5\textwidth]{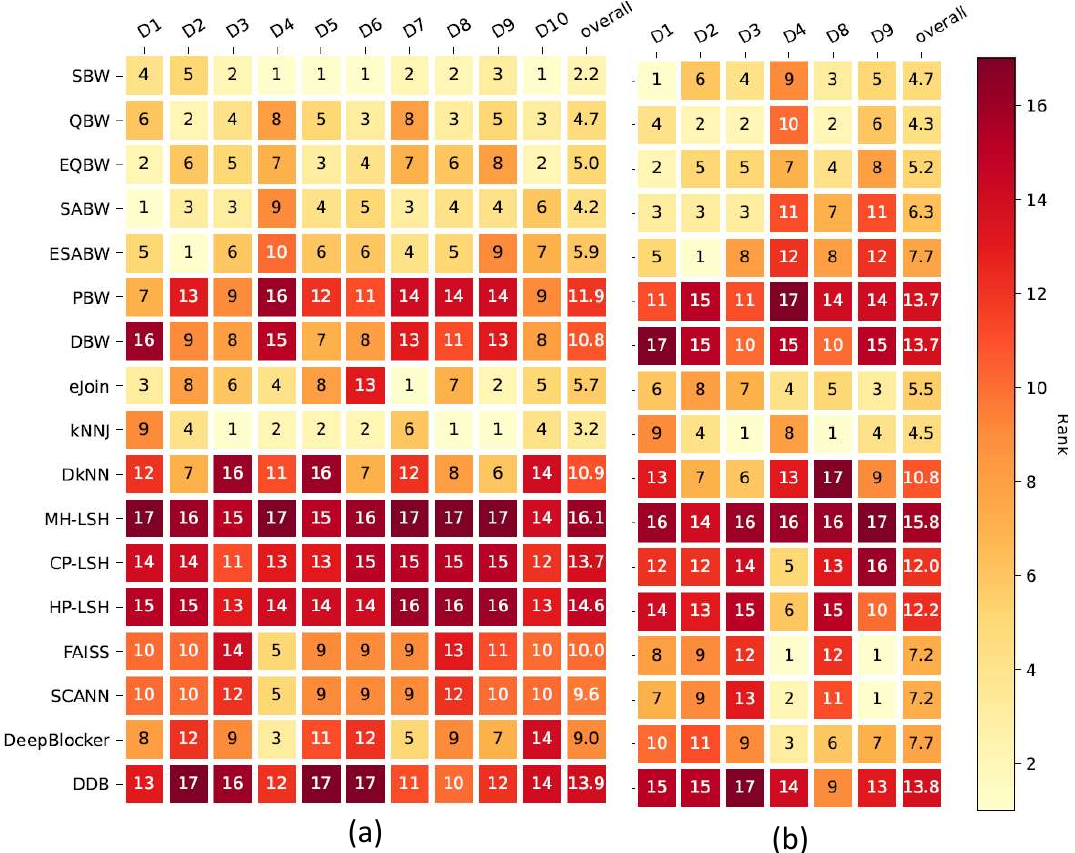}}
\vspace{-8pt}
\caption{The ranking position of each method per dataset in (a) schema-agnostic and (b) schema-based settings. Lower/brighter is better.}
\vspace{-15pt}
\label{fig:heatmaps}
\end{figure}

%% file: appendix.tex
\pagebreak

\section*{Appendix}

\input{tables/results}

\begin{figure*}[t]
\centering
\includegraphics[width=0.95\textwidth]{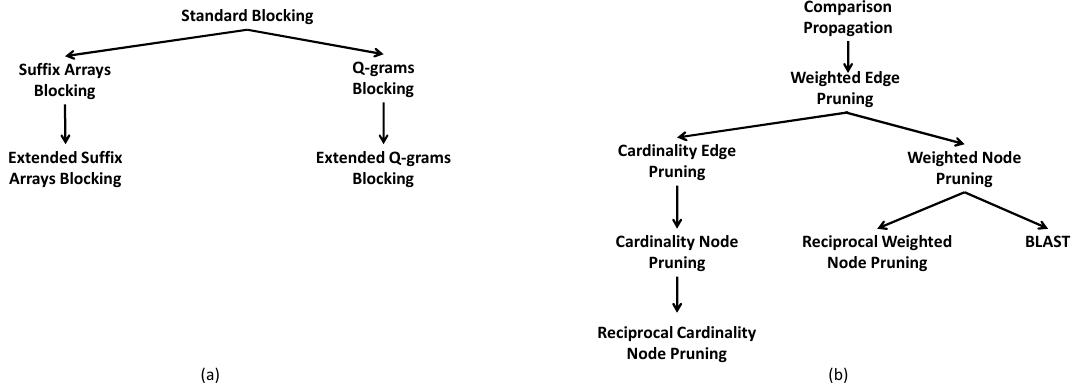}
\vspace{-8pt}
\caption{The genealogy tree of (a) blocking building and (b) comparison cleaning techniques. Every edge points from the original technique to the one that improves it, i.e., it uses transforms the blocking keys into a more noise-tolerant format (in the former case), or alters the threshold or the algorithm that is used for pruning candidate pairs (in the latter case).}
\vspace{-10pt}
\label{fig:bwGenealogy}
\end{figure*}

\begin{figure}[t]
\centering
\includegraphics[width=0.47\textwidth]{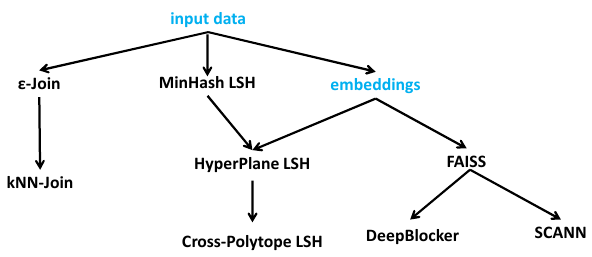}
\vspace{-8pt}
\caption{The genealogy tree of sparse and dense NN methods, on the left and right respectively. Every edge points from an original approach to an extension that alters the type of thresholds or the way the input data is handled.}
\vspace{-10pt}
\label{fig:nnGenealogy}
\end{figure}

\section*{A. Detailed Performance}

Table \ref{tb:results} presents the detailed performance of all filtering techniques over all datasets in Table \ref{tb:ccerDatasets} under schema-agnostic and schema-based settings. Table \ref{tb:results}(a) reports the recall ($PC$), Table \ref{tb:results}(b) the precision ($PQ$), Table \ref{tb:results}(c) the overall run-time ($RT$) and Table \ref{tb:results}(d) the actual number of candidate pairs they generate per dataset. In Tables \ref{tb:results}(b) and (d), the best performance per type of algorithms, dataset and schema settings is underlined, the overall best performance is highlighted in bold, while the cases corresponding to insufficient recall are marked in red.

We notice that the precision of all methods is highly correlated. In fact, the Pearson correlation coefficient between any pair of methods exceeds 0.5. This means that the performance of filtering depends heavily on the dataset characteristics. For example, datasets like $D_{a3}$ yield very low precision for all methods, because their duplicate entities share only generic/noisy content that appears in many non-matching profiles, too (e.g., stop words). In contrast, datasets like $D_{a4}$ entail duplicates with very distinguishing content in common, yielding an almost perfect performance in most cases.

Regarding time efficiency, we observe that the blocking workflows excel in run-time. Most of them require less than a second to process all datasets up to $D_{a8}$, few seconds for $D_{a9}$ and less than a minute for $D_{a10}$. In most cases, the fastest workflow is \textsf{PBW}, due to its simple comparison cleaning, which merely applies Comparison Propagation to eliminate the redundant candidate pairs. All other workflows are always coupled with a Meta-blocking approach that assigns a weight to every candidate pair and prunes the lowest-weighted ones in an effort to reduce the superfluous pairs, too (see the configurations of Table \ref{tb:bwConfigurations}
for more details). \textit{Comparison cleaning actually dominates the run-time of blocking workflows}, with $RT$ being proportional to the number of candidate pairs resulting from block cleaning. For this reason, \textsf{EQBW} and \textsf{ESABW} are typically slower than \textsf{QBW} and \textsf{SABW}, res

Among the sparse NN methods, \textsf{kNNJ} is much faster than $\epsilon$-\textsf{Join} in half the datasets: $D_{a5}$-$D_{a9}$. This is counter-intuitive, given that the former approach involves a more complex functionality, sorting the candidate pairs per query entity. However, as shown in Table \ref{tb:joinConfiguration}, $k$=1 for these datasets (2 in $D_{a8}$), thus minimizing the overhead of sorting. This also explains why \textsf{DkNN}, which uses $k=5$, is slower than both other techniques in $D_{a5}$-$D_{a7}$ (for $D_{a8}$-$D_{a10}$,  \textsf{DkNN} is faster than  \textsf{kNNJ}, because the latter uses shorter $q$-grams that generate more candidate pairs). Overall, \textit{the run-time of sparse NN methods is dominated by the querying time}, which consistently accounts for more than half of $RT$. In contrast, indexing time accounts for less than 10\% in practically all cases, with the rest corresponding to cleaning.
 
Regarding the dense NN-methods, 
\textsf{MH-LSH} is by far the slowest similarity-based one for $D_{a5}$ on, even though it 
does not apply cleaning (i.e., stop-word removal and stemming) in most cases (see Table \ref{tb:nnConfiguration}). This should be attributed to the very large number of candidates it generates during the querying phase, which prevents it from scaling to $D_{a10}$. The lowest run-time corresponds to \textsc{CP-LSH}, because it exhibits the highest $PQ$ among the three methods. Among the cardinality-based NN methods, 
\textsf{FAISS} is consistently the fastest one. It is significantly faster than \textsf{SCANN} in all cases, even though they have an almost identical effectiveness. The reason is that \textsf{FAISS} saves the overhead of data partitioning. Most importantly, both versions of \textit{\textsf{DeepBlocker} are slower than the other NN methods by a whole order of magnitude in most cases.} This is caused by the cost of automatically creating a labelleled dataset and using it for training the tuple embedding module -- the number of candidates per query entity plays a minor role, which can be inferred from the relative run-time of \textsf{DDB} and \textsf{DeepBlocker} and the configurations in Table \ref{tb:nnConfiguration}.
Therefore, we can conclude that \textit{\textsf{DeepBlocker} emphasizes effectiveness at the cost of very low time efficiency}. Note also that the Hybrid tuple embedding module is a whole order of magnitude slower than the Autoencoder~\cite{DBLP:journals/pvldb/Thirumuruganathan21}.

The same patterns apply in the schema-based settings. However, $RT$ has significantly improved in most cases, especially for the largest datasets, due to the lower vocabulary size and character length. As a result, all methods are capable of processing all datasets in less than 1 second (few seconds for $D_{b9}$). \textsf{DeepBlocker} and \textsf{DDB} typically remain a whole order of magnitude slower than all other methods, due to the inelastic computational cost of creating a labelled dataset and using it to train the neural-based tuple embedding module.

\input{tables/blockProcessing}
\input{tables/simJoins}
\input{tables/nnConfiguration}

\begin{figure*}[t]
\centering
\includegraphics[width=0.32\textwidth]{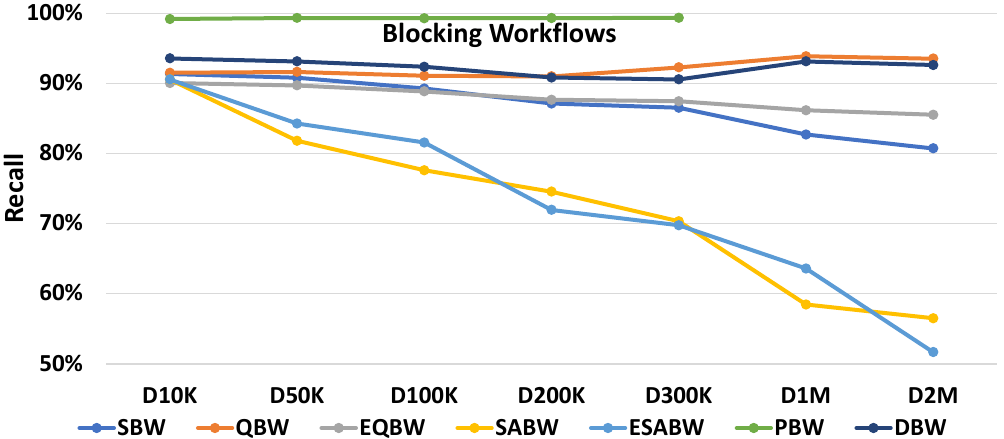}
\includegraphics[width=0.32\textwidth]{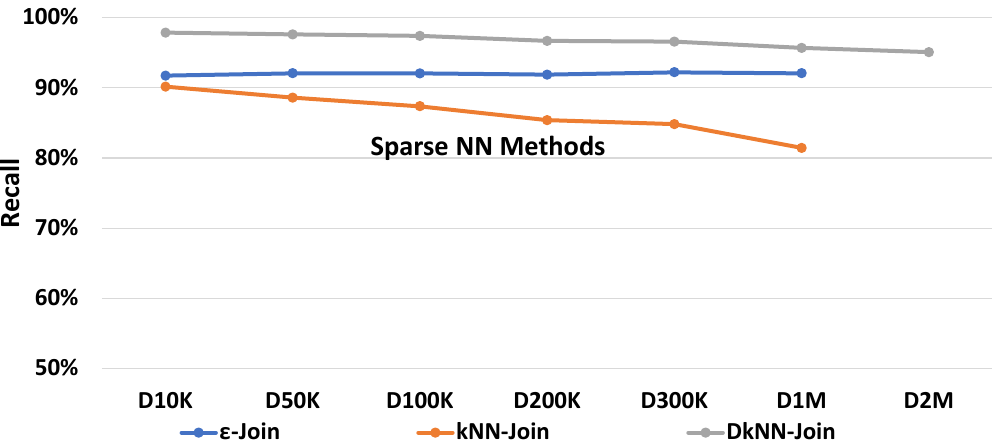}
\includegraphics[width=0.32\textwidth]{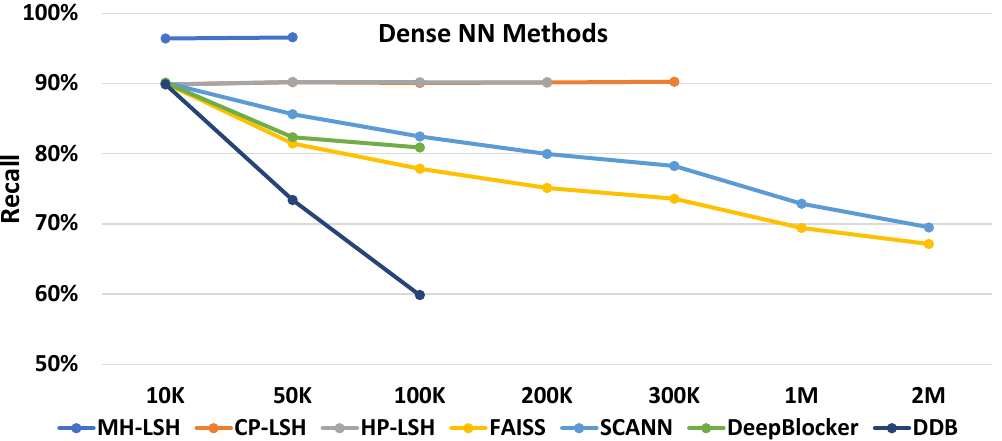}
\vspace{-8pt}
\caption{Scalability analysis with respect to recall ($PC$) over all datasets in Table \ref{tb:derDatasets}.}
\vspace{-10pt}
\label{fig:scalabilityRecall}
\end{figure*}

\begin{figure*}[t]
\centering
\includegraphics[width=0.32\textwidth]{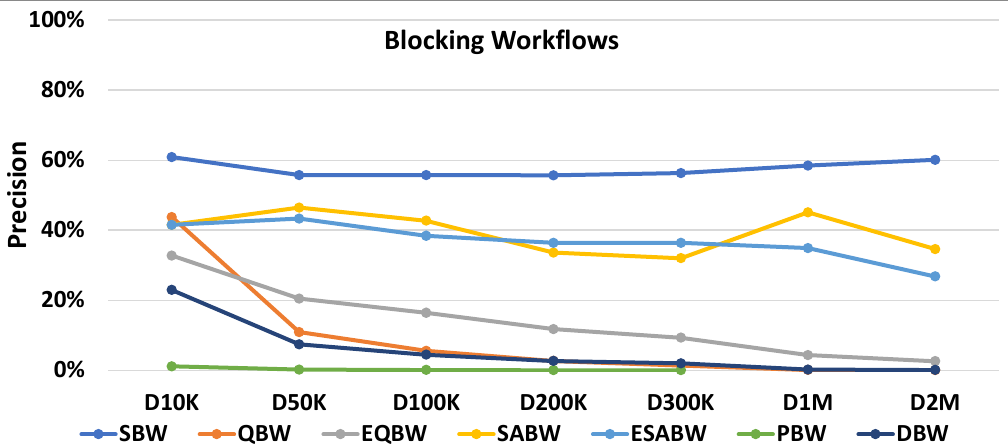}
\includegraphics[width=0.32\textwidth]{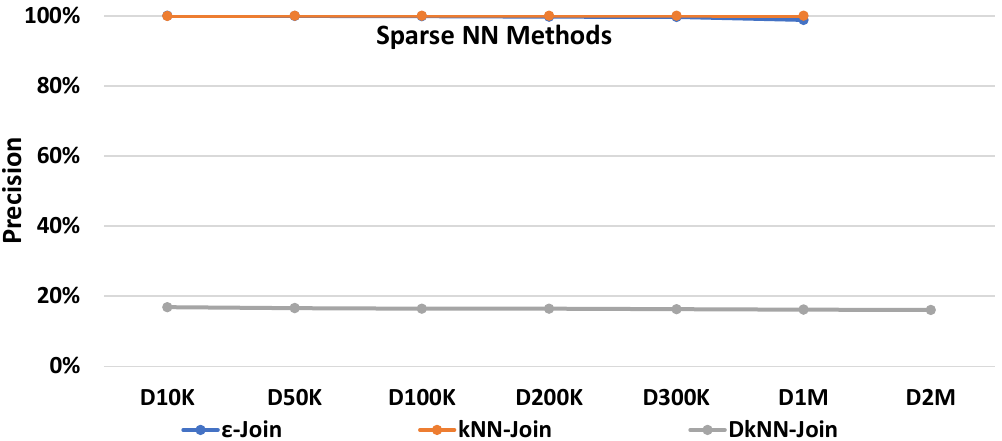}
\includegraphics[width=0.32\textwidth]{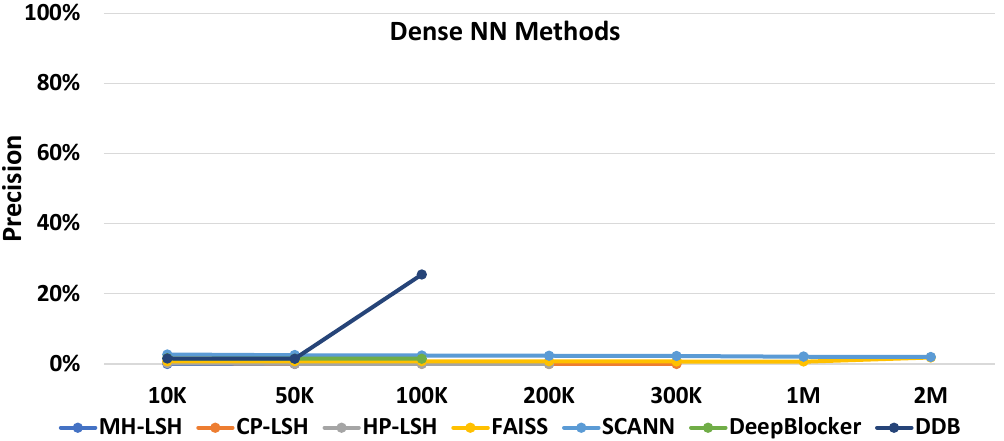}
\vspace{-8pt}
\caption{Scalability analysis with respect to precision ($PQ$) over all datasets in Table \ref{tb:derDatasets}.}
\vspace{-10pt}
\label{fig:scalabilityPrecision}
\end{figure*}

\section*{B. Distances of Duplicates}

In this section, we investigate the relative performance of cardinality-based filtering methods, i.e., \textsf{kNNJ}, which leverages syntactic representations, as well as \textsf{FAISS}, \textsf{SCANN} and \textsf{DeepBlocker}, which rely on semantic representations. All of them use every entity profile from $\mathcal{E}_2$ as a query, whose set of candidates is ranked in decreasing similarity (or increasing distance). For a pair of duplicates $<e_i, e_j>$, if $e_j$ is used as a query, the higher the ranking position of $e_i$ the better is the effectiveness (i.e., higher precision). 

\begin{figure*}[t]
\centering
\includegraphics[width=0.4\textwidth]{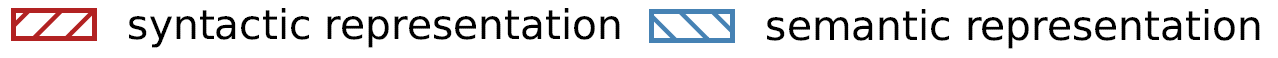}\\
\includegraphics[width=0.19\textwidth]{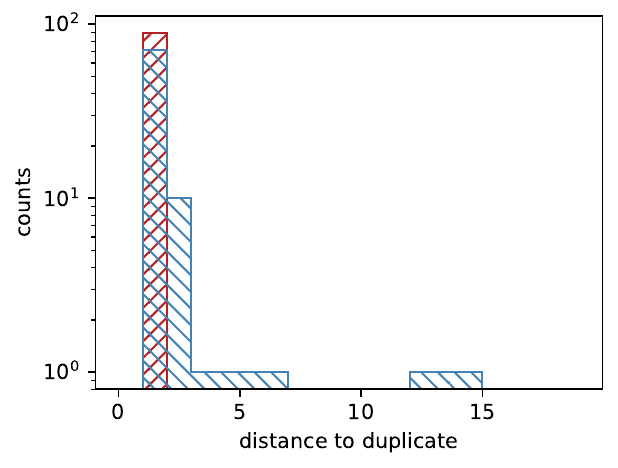}
\includegraphics[width=0.19\textwidth]{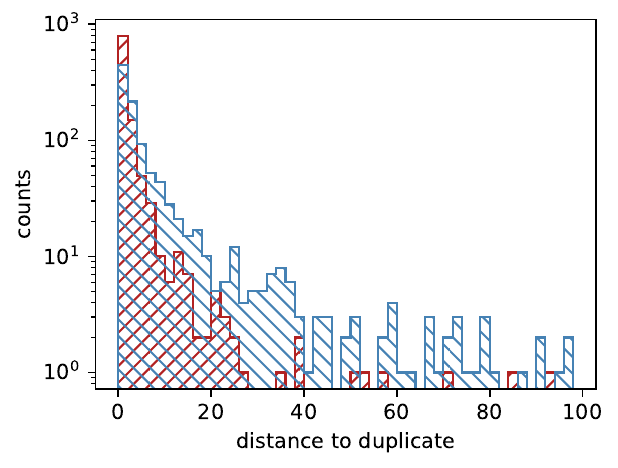}
\includegraphics[width=0.19\textwidth]{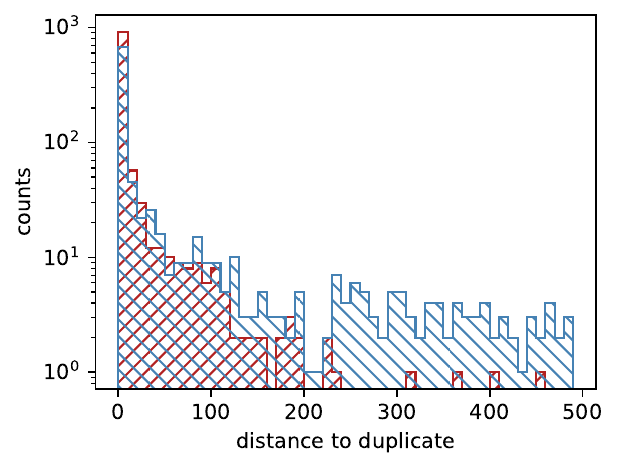}
\includegraphics[width=0.19\textwidth]{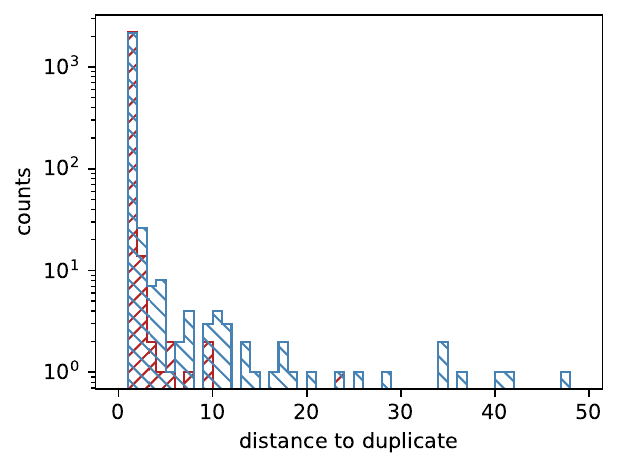}
\includegraphics[width=0.19\textwidth]{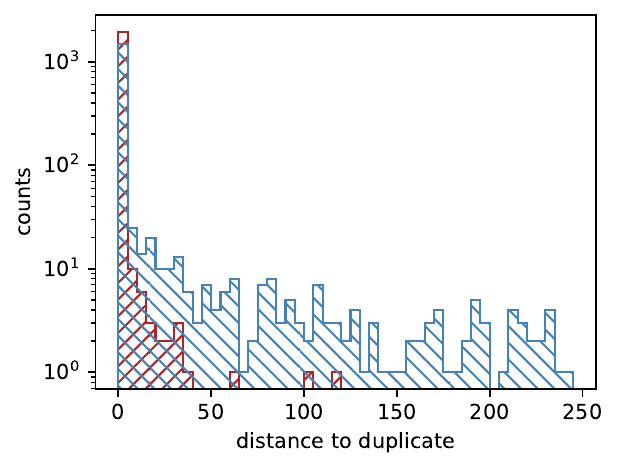}
\includegraphics[width=0.99\textwidth]{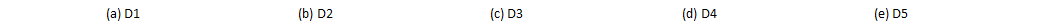}
\includegraphics[width=0.19\textwidth]{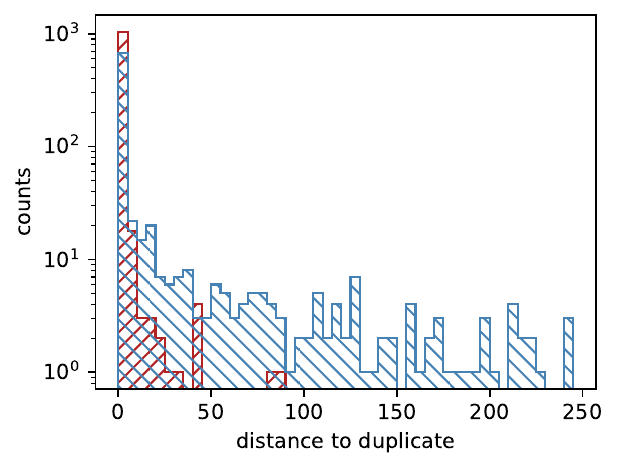}
\includegraphics[width=0.19\textwidth]{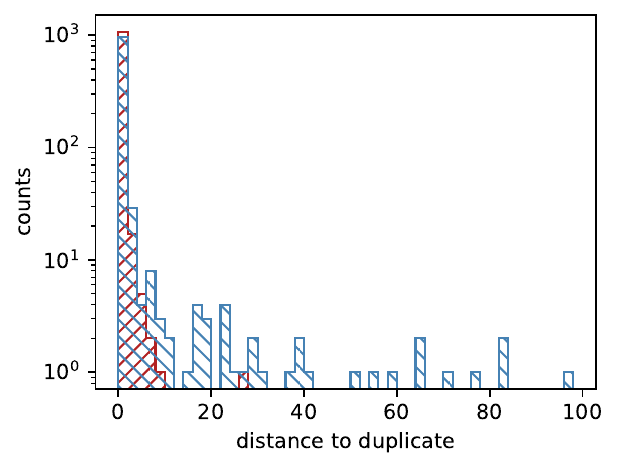}
\includegraphics[width=0.19\textwidth]{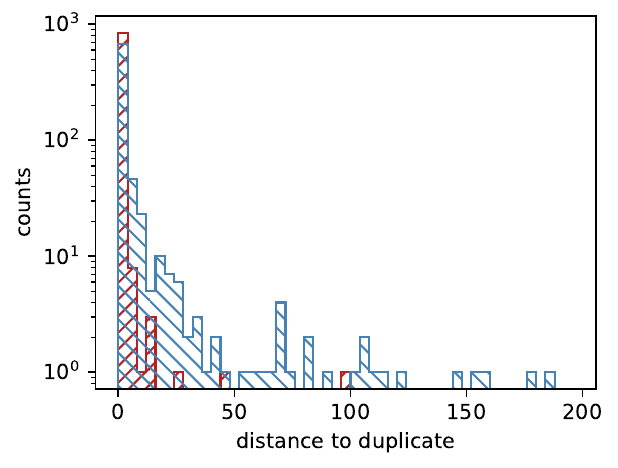}
\includegraphics[width=0.19\textwidth]{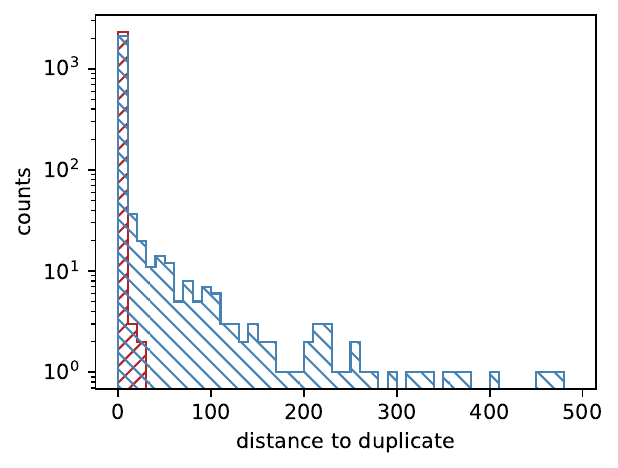}
\includegraphics[width=0.19\textwidth]{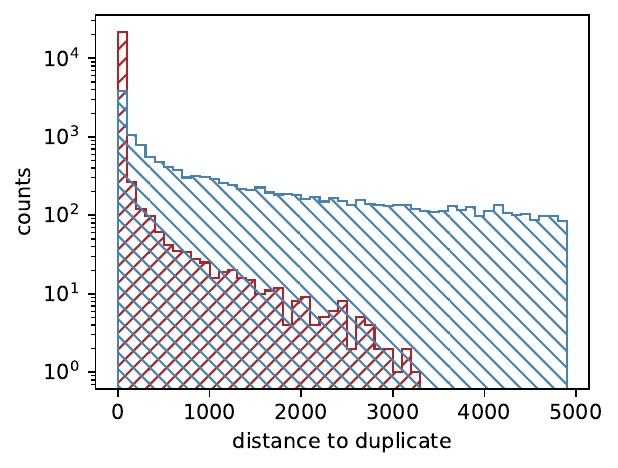}
\includegraphics[width=0.99\textwidth]{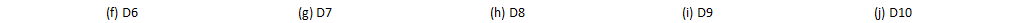}
\vspace{-5pt}
\caption{The distribution of distances between duplicate pairs across all datasets with schema-agnostic settings when indexing $\mathcal{E}_1$ and querying with $\mathcal{E}_2$.}
\vspace{-10pt}
\label{fig:distanceDistributionSA}
\end{figure*}

\begin{figure*}[t]
\centering
\includegraphics[width=0.4\textwidth]{figures/distances/legend.png}\\
\includegraphics[width=0.19\textwidth]{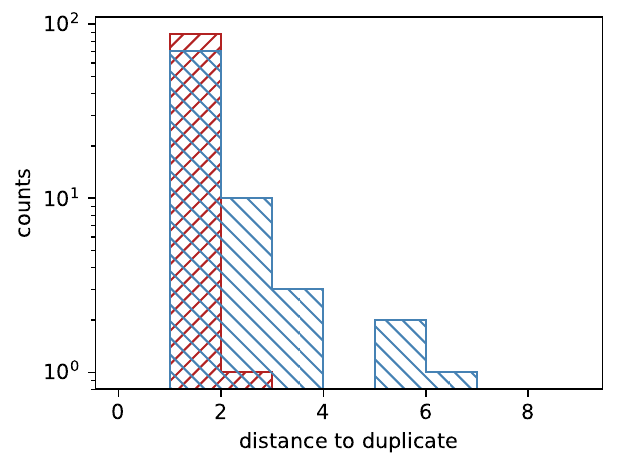}
\includegraphics[width=0.19\textwidth]{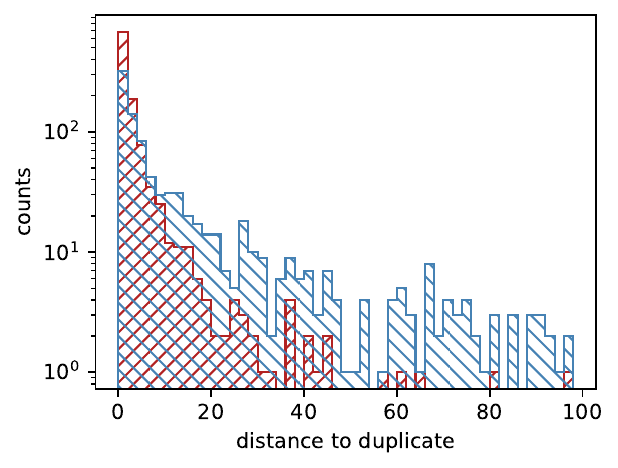}
\includegraphics[width=0.19\textwidth]{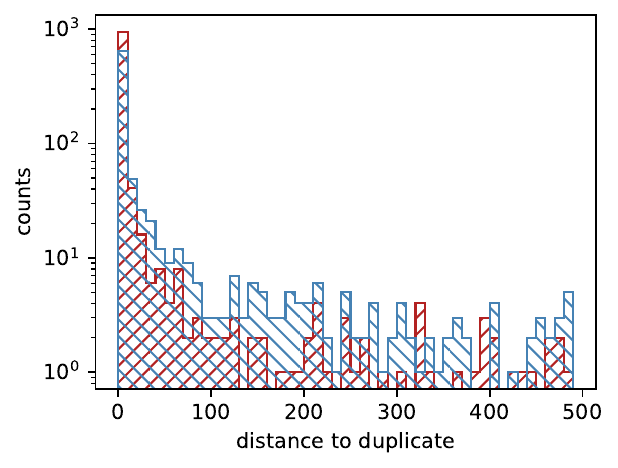}
\includegraphics[width=0.19\textwidth]{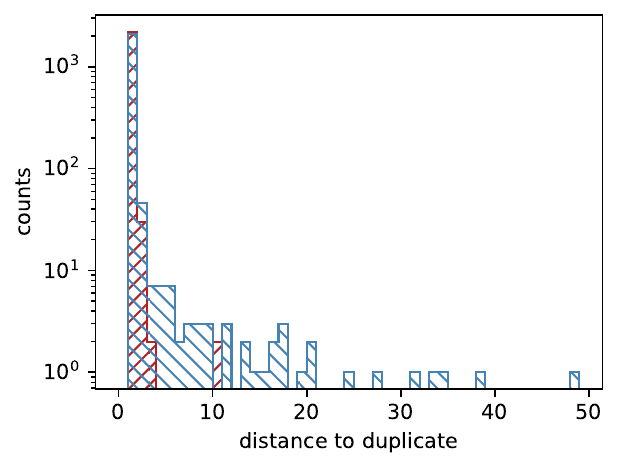}
\includegraphics[width=0.19\textwidth]{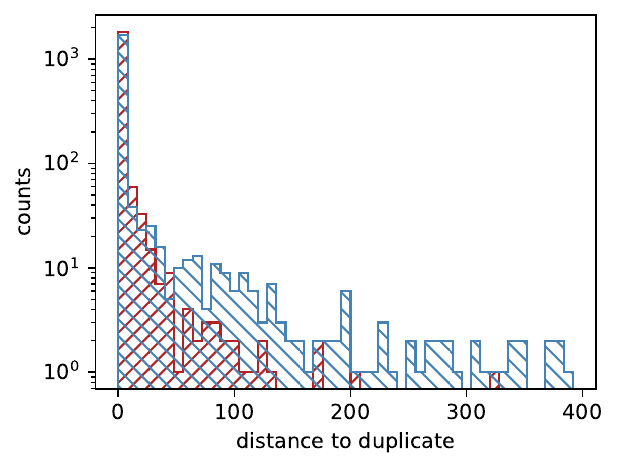}
\includegraphics[width=0.99\textwidth]{figures/distances/titlesD1D5.png}
\includegraphics[width=0.19\textwidth]{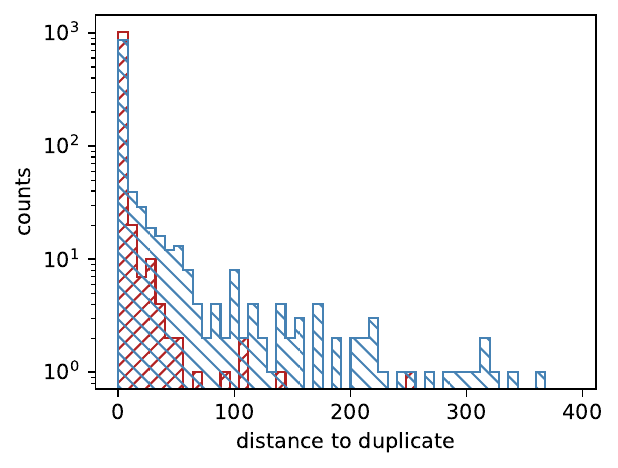}
\includegraphics[width=0.19\textwidth]{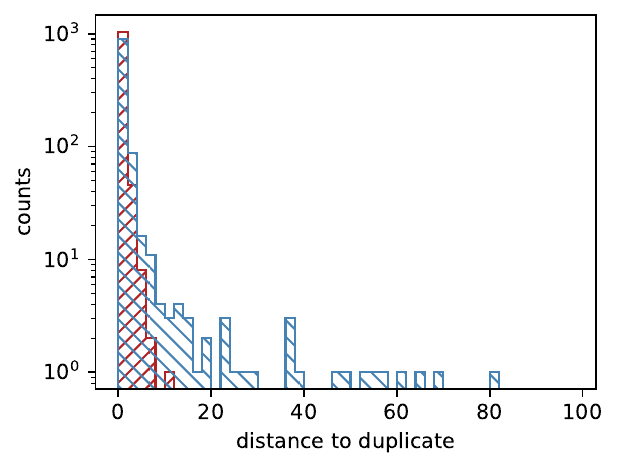}
\includegraphics[width=0.19\textwidth]{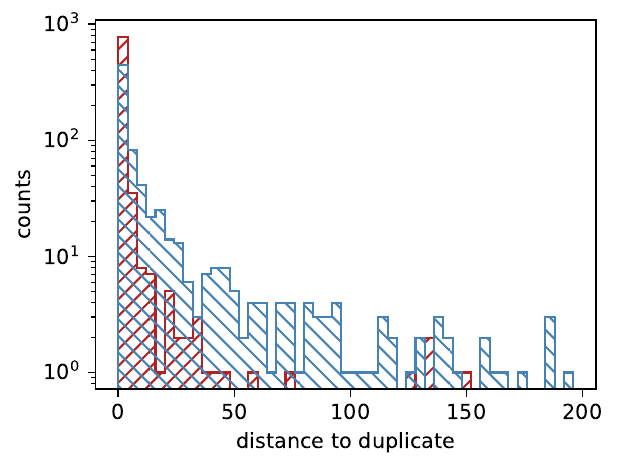}
\includegraphics[width=0.19\textwidth]{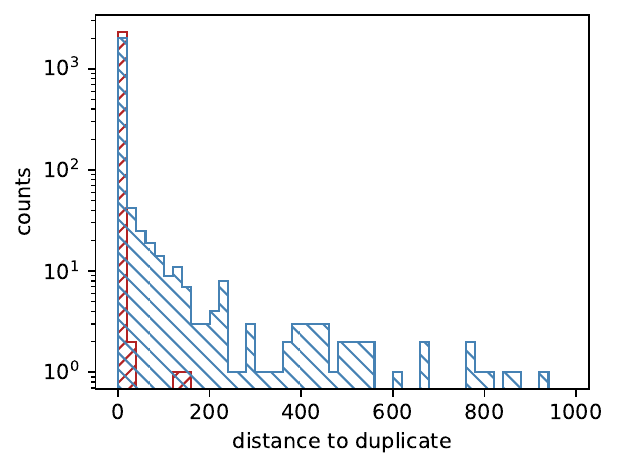}
\includegraphics[width=0.19\textwidth]{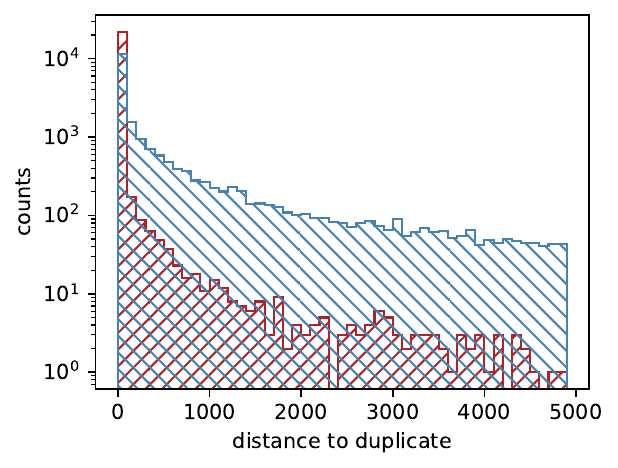}
\includegraphics[width=0.99\textwidth]{figures/distances/titlesD6D10.png}
\vspace{-10pt}
\caption{The distribution of distances between duplicate pairs across all datasets with schema-agnostic settings when reversing the datasets, i.e., when indexing $\mathcal{E}_2$ and querying with $\mathcal{E}_1$.}
\vspace{-10pt}
\label{fig:distanceDistributionRSA}
\end{figure*}

\begin{figure*}[t]
\centering
\includegraphics[width=0.4\textwidth]{figures/distances/legend.png}\\
\includegraphics[width=0.16\textwidth]{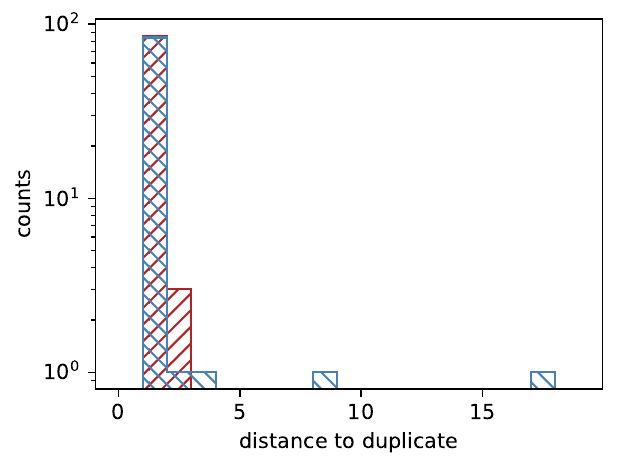}
\includegraphics[width=0.16\textwidth]{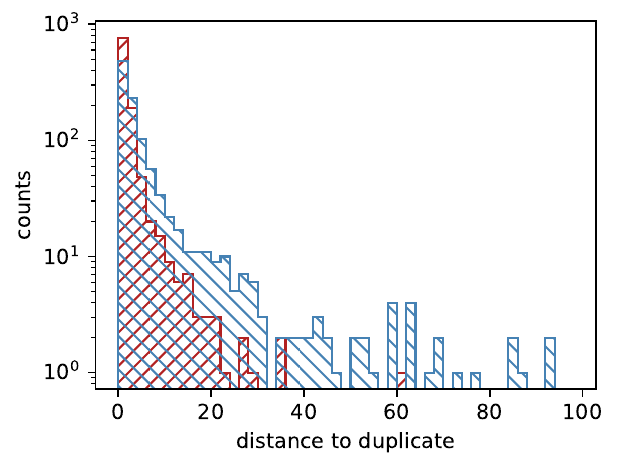}
\includegraphics[width=0.16\textwidth]{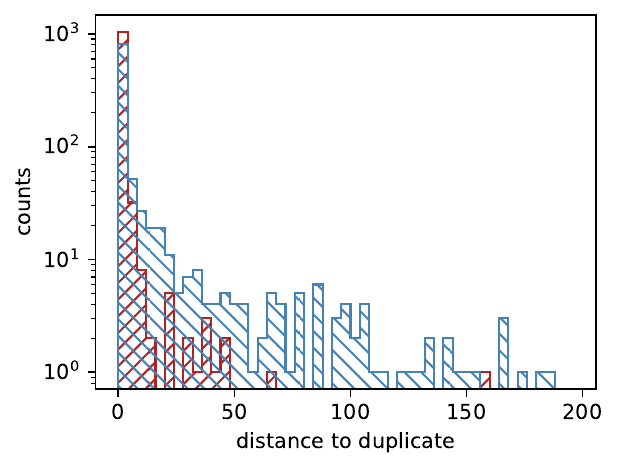}
\includegraphics[width=0.16\textwidth]{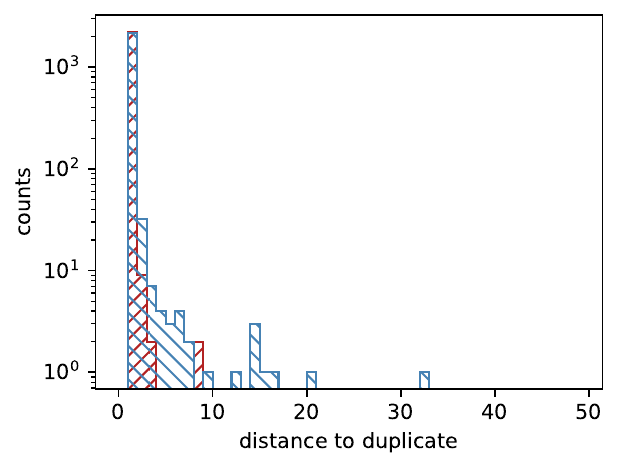}
\includegraphics[width=0.16\textwidth]{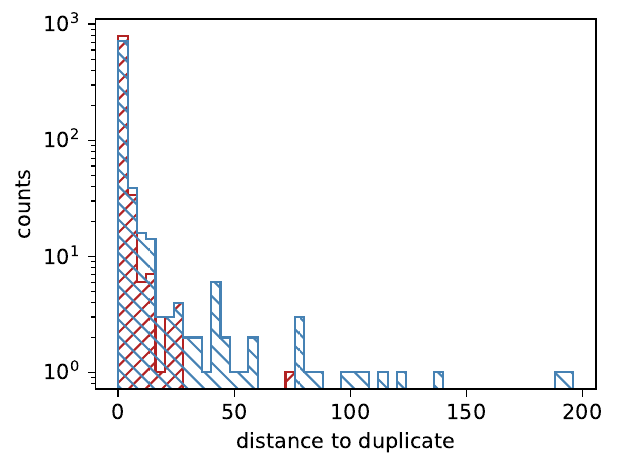}
\includegraphics[width=0.16\textwidth]{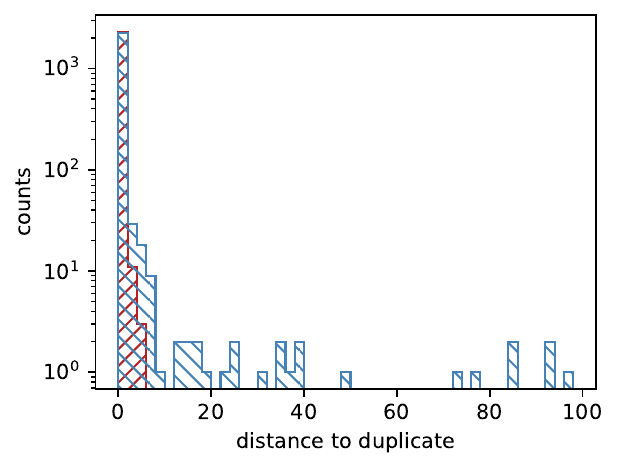}
\includegraphics[width=0.99\textwidth]{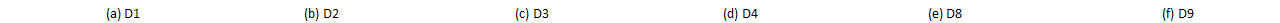}
\includegraphics[width=0.16\textwidth]{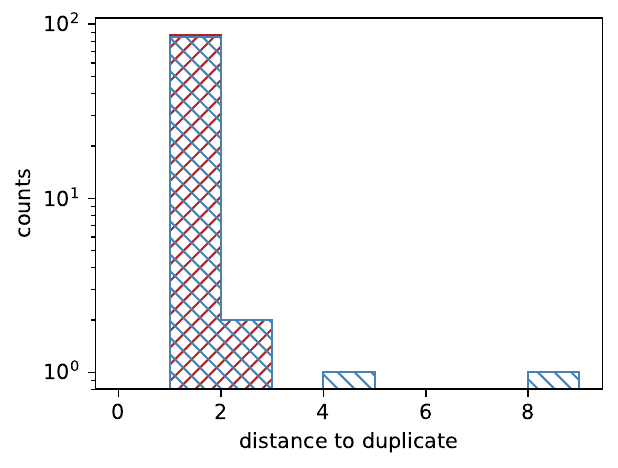}
\includegraphics[width=0.16\textwidth]{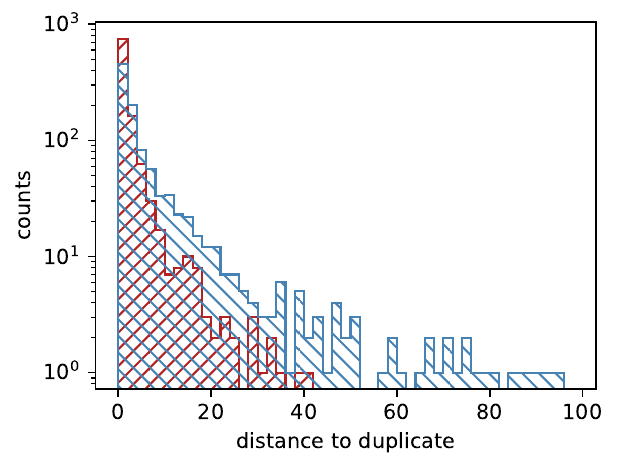}
\includegraphics[width=0.16\textwidth]{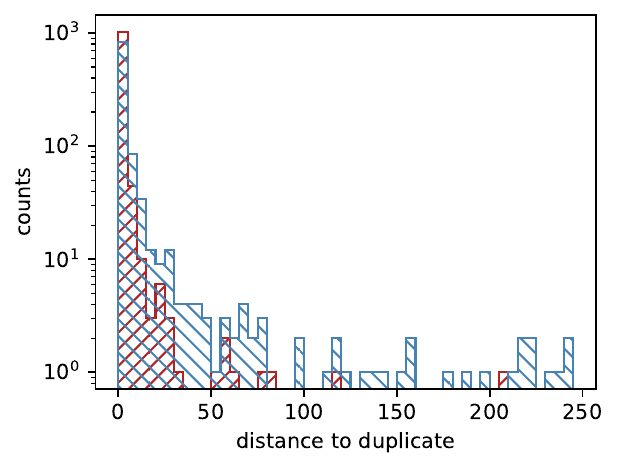}
\includegraphics[width=0.16\textwidth]{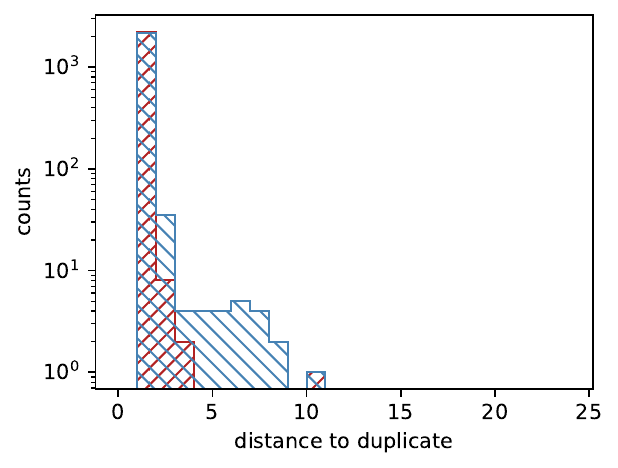}
\includegraphics[width=0.16\textwidth]{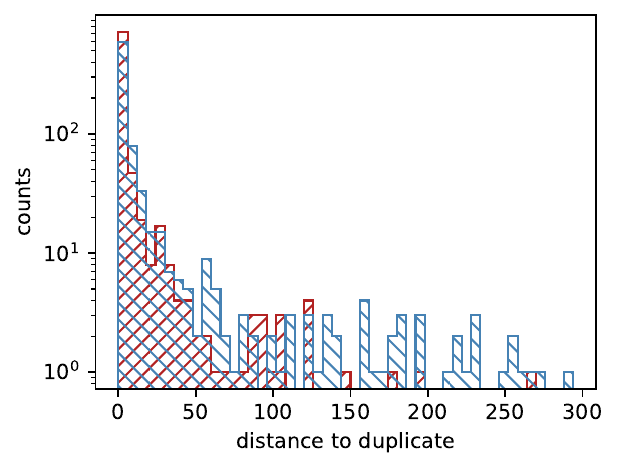}
\includegraphics[width=0.16\textwidth]{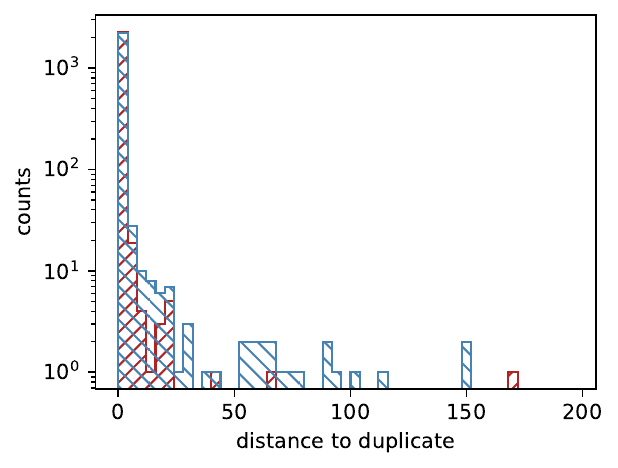}
\includegraphics[width=0.99\textwidth]{figures/distances/sbTitles.png}
\vspace{-10pt}
\caption{The distribution of distances between duplicate pairs across all datasets with schema-based settings when indexing $\mathcal{E}_1$ and querying with $\mathcal{E}_2$ (upper line) and when reversing the datasets, i.e., indexing $\mathcal{E}_2$ and querying with $\mathcal{E}_1$ (lower line).}
\label{fig:distanceDistributionSB}
\end{figure*}

The distributions of ranking position across all datasets with schema agnostic settings appear in Figure \ref{fig:distanceDistributionSA}. The horizontal axis corresponds to the ranking position of the matches, with $x=0$ denoting that the duplicate is placed at the top of the ranking list of candidates. The vertical axis corresponds to the number of duplicate pairs per ranking position. Figure \ref{fig:distanceDistributionRSA} depicts the same diagrams over all datasets with schema-agnostic settings when reversing the role of the input datasets (i.e., indexing $\mathcal{E}_2$ and querying with $\mathcal{E}_1$), while the schema-based settings appear in Figure \ref{fig:distanceDistributionSB}. 

The syntactic representation is represented by the configuration of \textsf{DkNN}, i.e., multiset of character five-grams (C5GM) in combination with 
cosine similarity. As explained in Section \ref{sec:quantitativeAnalysis}, these settings achieve the best performance, on average, for \textsf{kNNJ} and semantic ones. For the semantic representations, we use pre-trained 300-dimensional fastText embeddings in combination with Euclidean distance, as computed by the brute-force approach. These settings are representative of all relevant methods (\textsf{FAISS}, \textsf{SCANN} and \textsf{DeepBlocker}). 

We observe that in the vast majority of cases, the syntactic representations have a higher concentration of duplicates on the top ranking positions. This justifies the superiority of \textsf{kNNJ} over \textsf{FAISS}, \textsf{SCANN} and \textsf{DeepBlocker}. There are only a handful of exceptions to this pattern.

In the schema-agnostic settings, we observe three exceptions: (i) in $D_{a1}$, where \textsf{DeepBlocker} achieves a higher precision than \textsf{kNNJ} (0.247 vs 0.224), (ii) in $D_{a4}$, where all methods have practically identical precision, and (ii) in $D_{a7}$, where \textsf{DeepBlocker} and \textsf{kNNJ} have an equivalent precision. In all these cases, we observe that the concentration of duplicate pairs on $x=0$ is equally high for both syntactic and semantic representations. 

The same applies to the three exceptions in the schema-based settings. These are: (i) in $D_{b1}$, where \textsf{FAISS} and \textsf{SCANN} exceed the precision of \textsf{kNNJ}, (ii) $D_{b4}$, where all cardinality-based NN methods outperform \textsf{kNNJ}, and (iii) $D_{b9}$, where \textsf{kNNJ} underperforms \textsf{FAISS} and \textsf{SCANN}, as in $D_{b1}$.

These patterns verify that \textit{syntactic representations typically provide more accurate evidence for filtering than semantic ones.}

\section*{C. Run-time Analysis}

\begin{figure*}[t]
\centering
\includegraphics[width=0.26\textwidth]{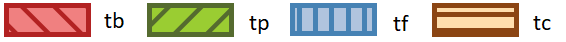}
\hspace{12pt}
\includegraphics[width=0.2\textwidth]{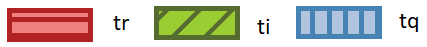}
\hspace{24pt}
 \includegraphics[width=0.2\textwidth]{figures/runtimes_color/nnLegend.png}\\
\includegraphics[width=0.26\textwidth]{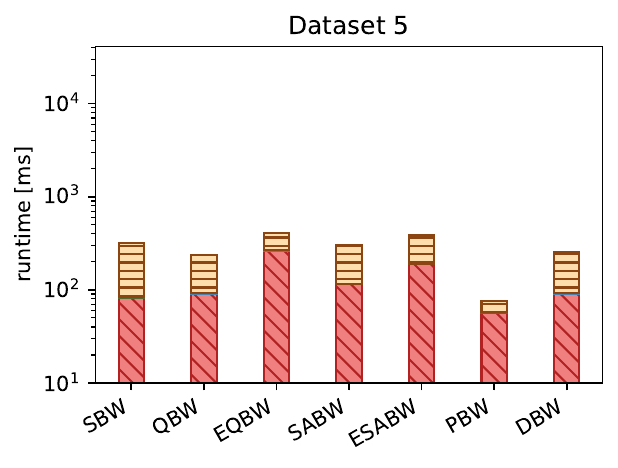}
\includegraphics[width=0.26\textwidth]{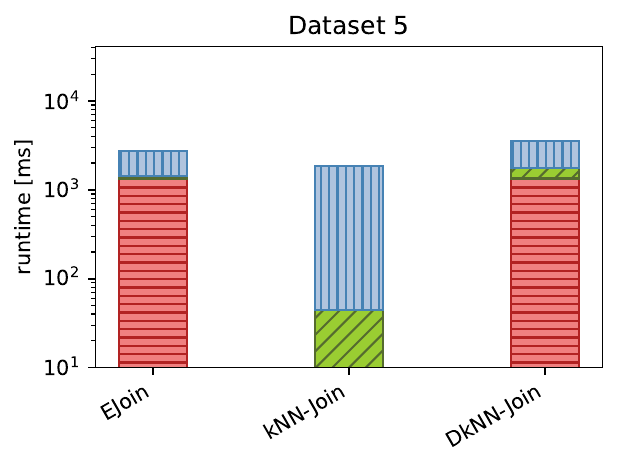}
\includegraphics[width=0.26\textwidth]{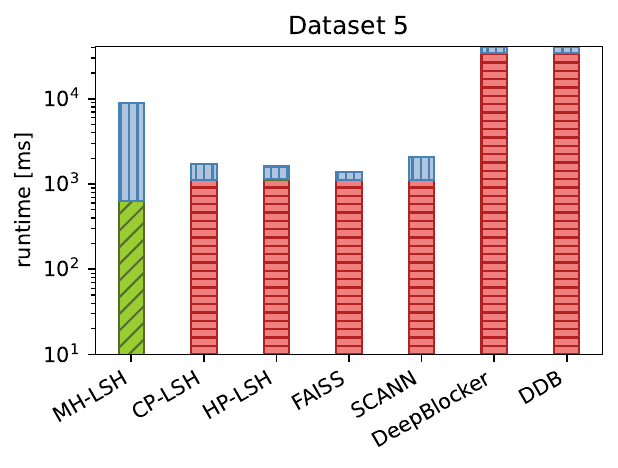}
\includegraphics[width=0.26\textwidth]{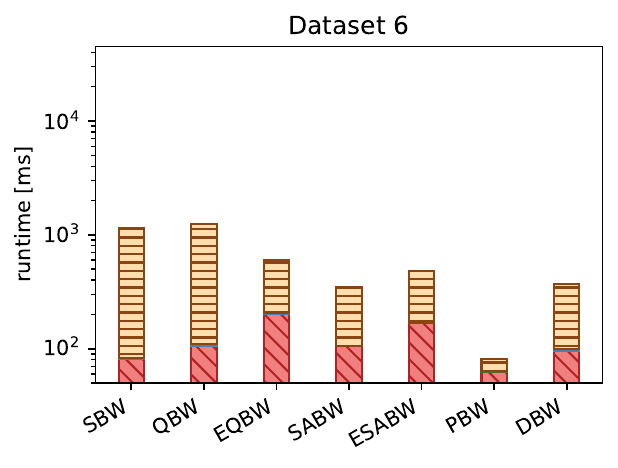}
\includegraphics[width=0.26\textwidth]{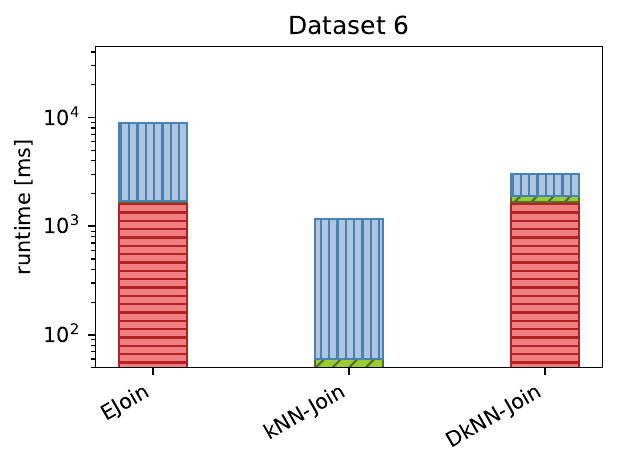}
\includegraphics[width=0.26\textwidth]{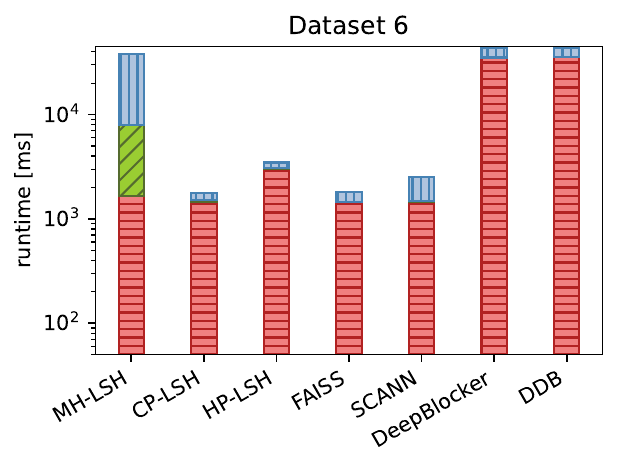}
\includegraphics[width=0.26\textwidth]{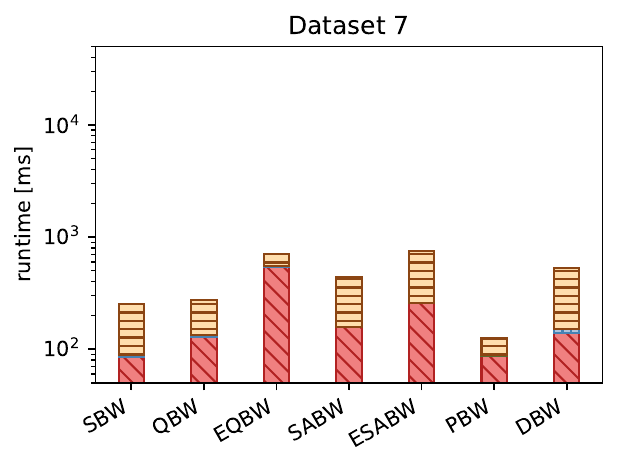}
\includegraphics[width=0.26\textwidth]{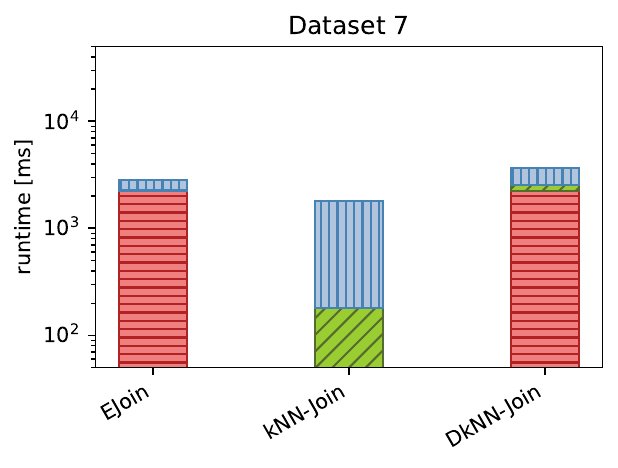}
\includegraphics[width=0.26\textwidth]{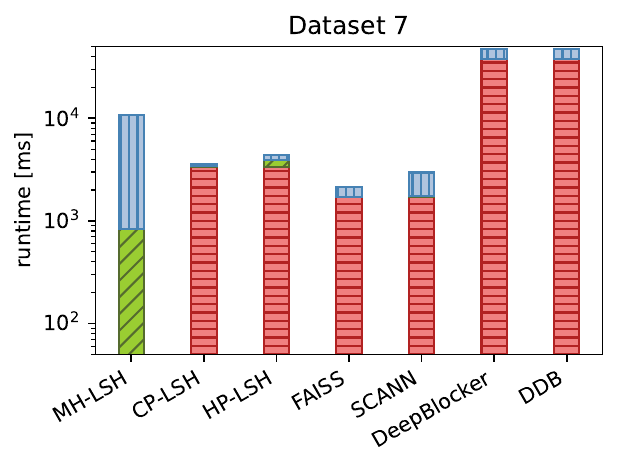}
\includegraphics[width=0.26\textwidth]{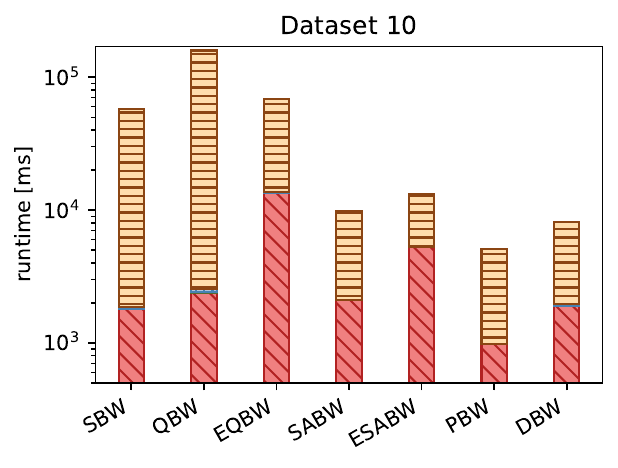}
\includegraphics[width=0.26\textwidth]{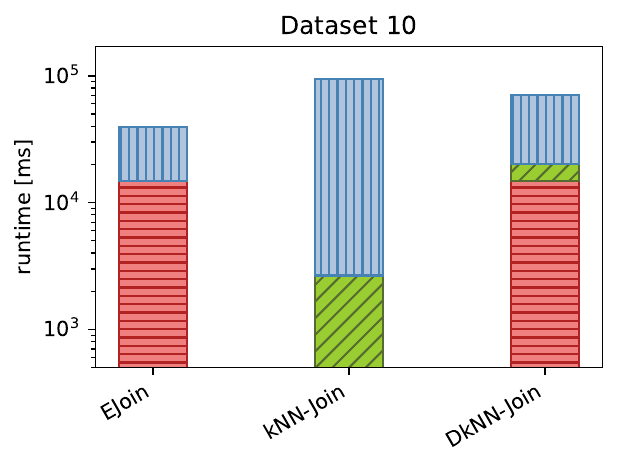}
\includegraphics[width=0.26\textwidth]{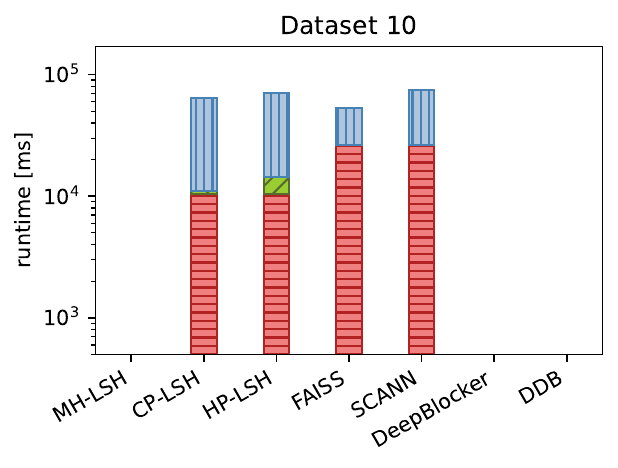}
\vspace{-10pt}
\caption{The break-down of the overall run-time of the blocking workflows (left column), the sparse NN methods (middle column) and the dense NN methods (right column) for the schema-agnostic settings of $D_5$-$D_7$ and $D_{10}$.}
\label{fig:breakDownA}
\end{figure*}

\begin{figure*}[t]
\centering
\includegraphics[width=0.26\textwidth]{figures/runtimes_color/brtLegend.png}
\hspace{12pt}
\includegraphics[width=0.2\textwidth]{figures/runtimes_color/nnLegend.png}
\hspace{24pt}
\includegraphics[width=0.2\textwidth]{figures/runtimes_color/nnLegend.png}\\
\includegraphics[width=0.26\textwidth]{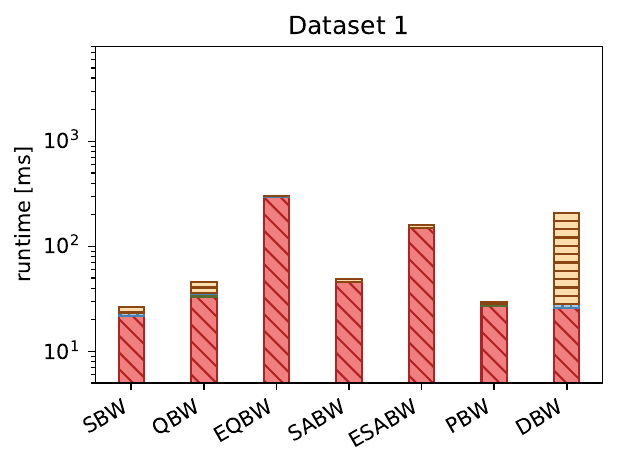}
\includegraphics[width=0.26\textwidth]{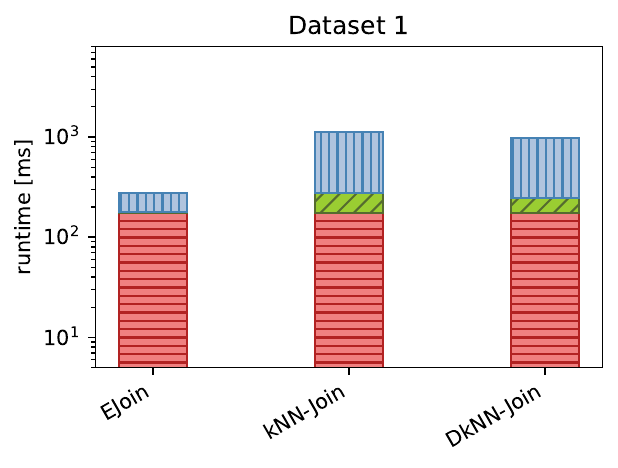}
\includegraphics[width=0.26\textwidth]{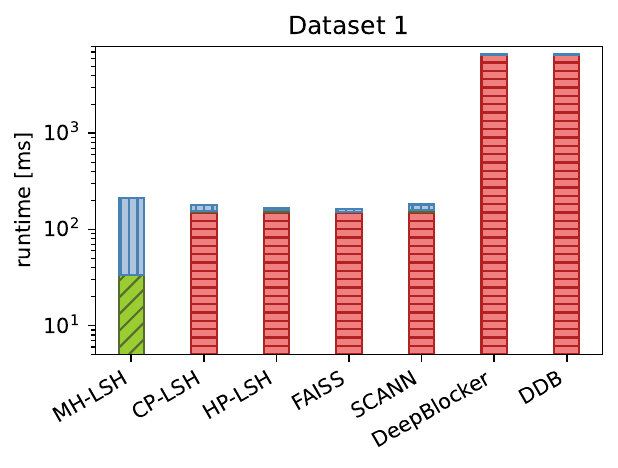}
\includegraphics[width=0.26\textwidth]{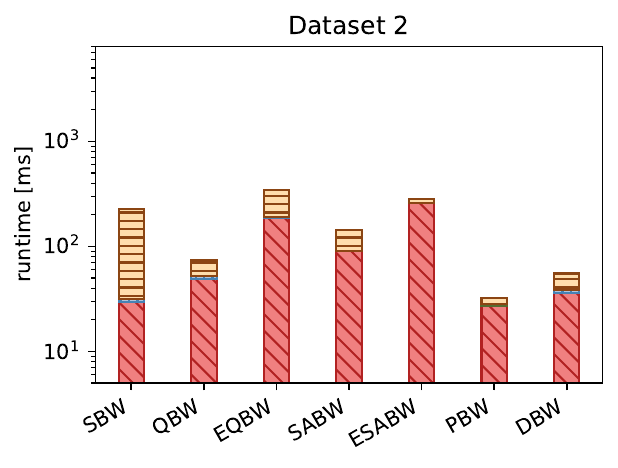}
\includegraphics[width=0.26\textwidth]{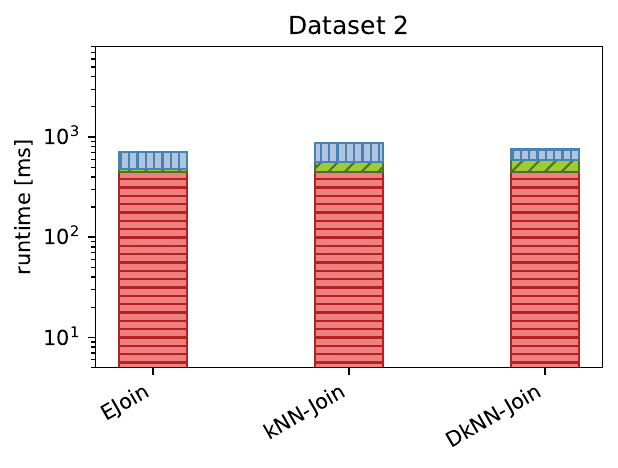}
\includegraphics[width=0.26\textwidth]{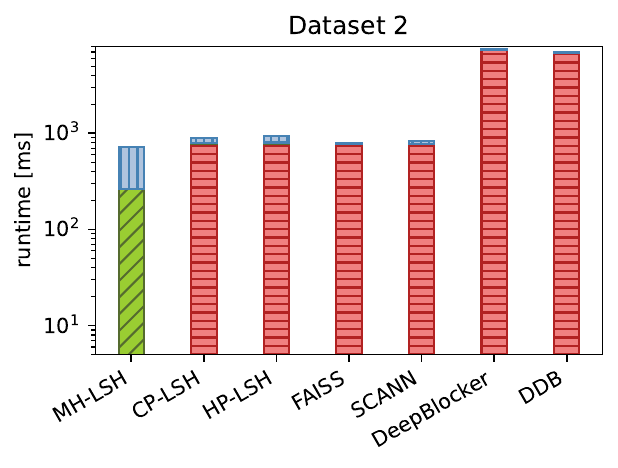}
\includegraphics[width=0.26\textwidth]{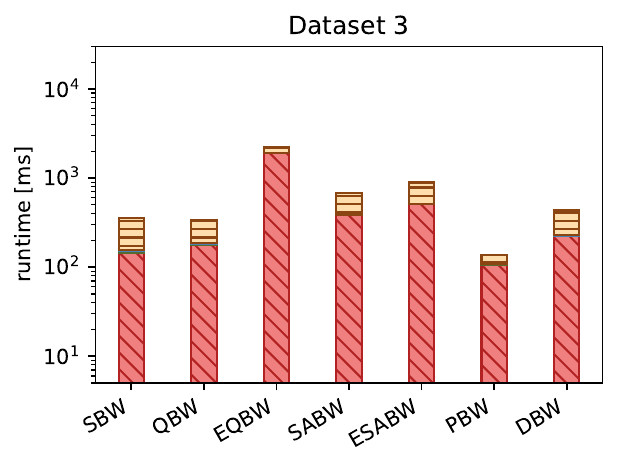}
\includegraphics[width=0.26\textwidth]{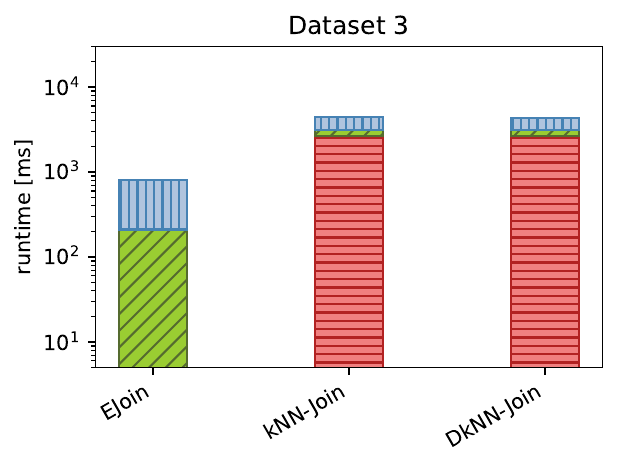}
\includegraphics[width=0.26\textwidth]{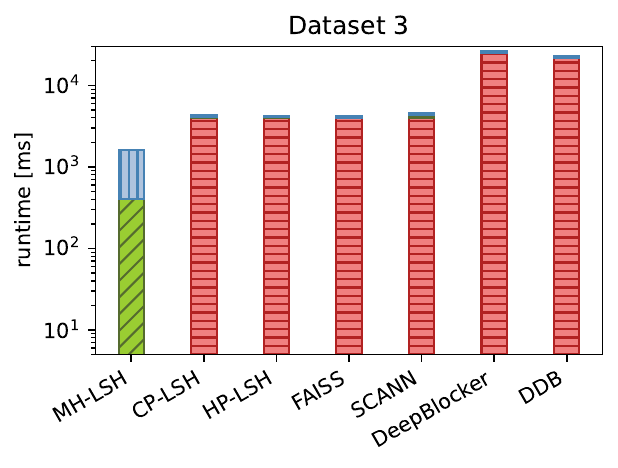}
\includegraphics[width=0.26\textwidth]{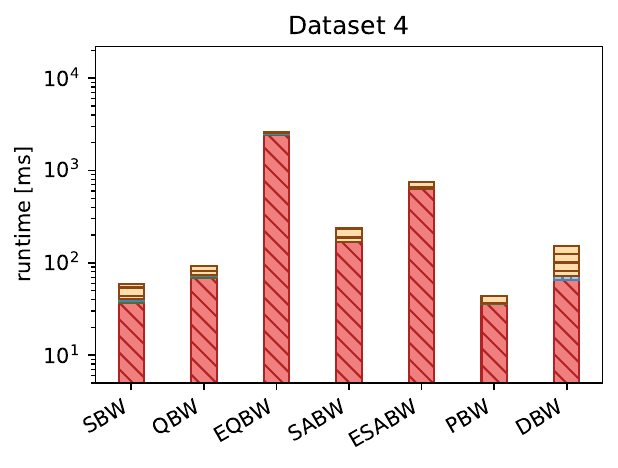}
\includegraphics[width=0.26\textwidth]{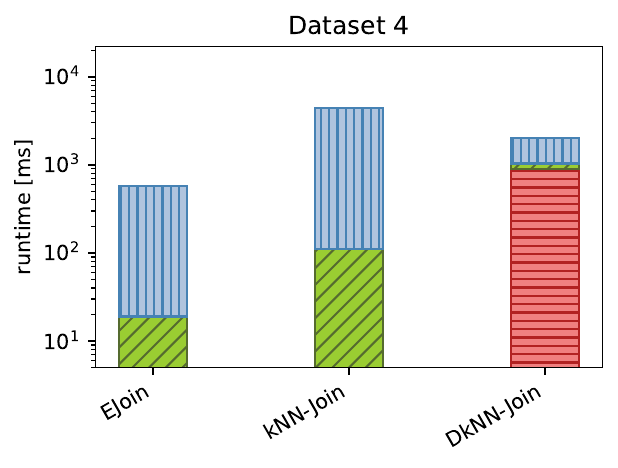}
\includegraphics[width=0.26\textwidth]{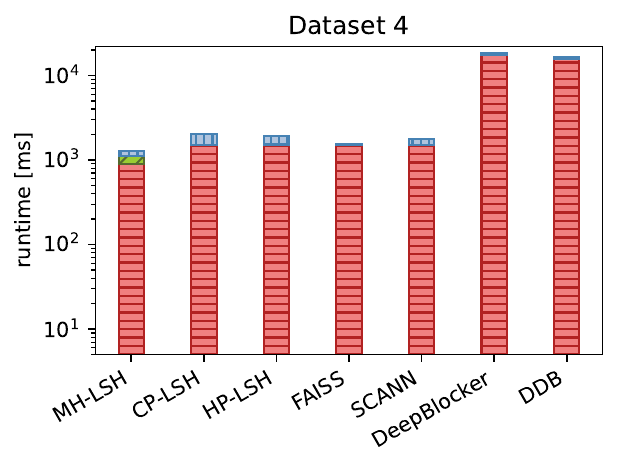}
\includegraphics[width=0.26\textwidth]{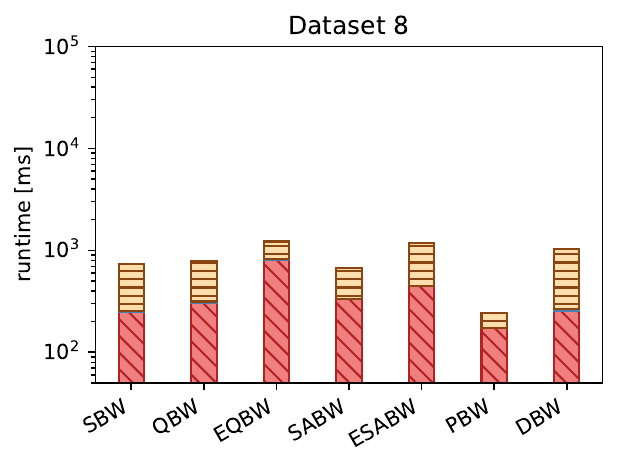}
\includegraphics[width=0.26\textwidth]{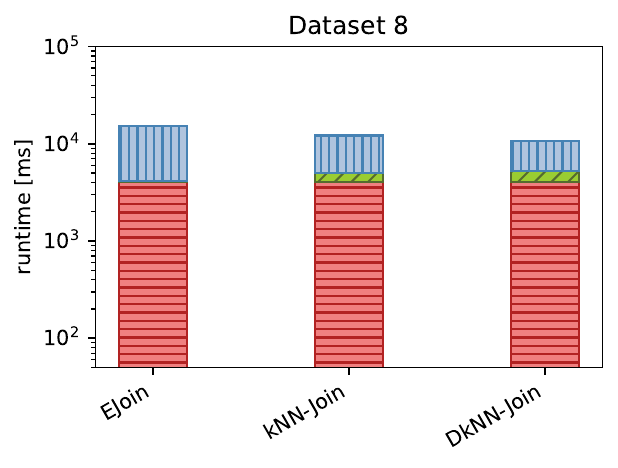}
\includegraphics[width=0.26\textwidth]{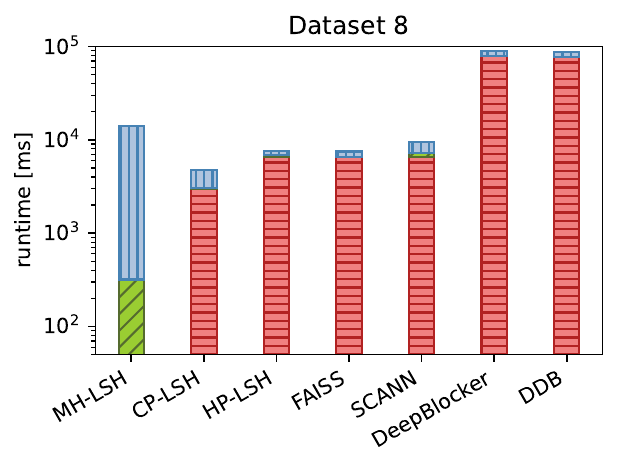}
\includegraphics[width=0.26\textwidth]{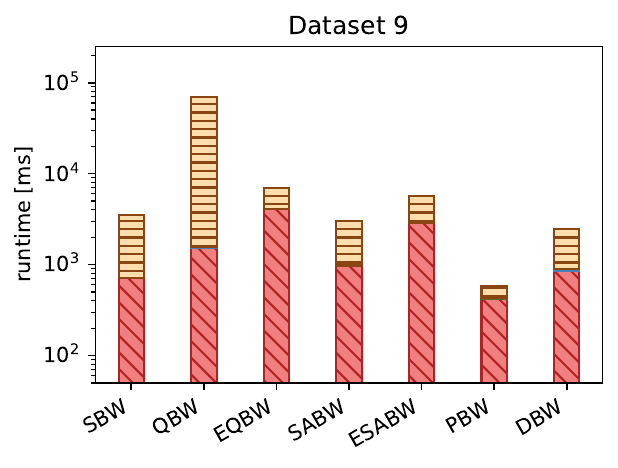}
\includegraphics[width=0.26\textwidth]{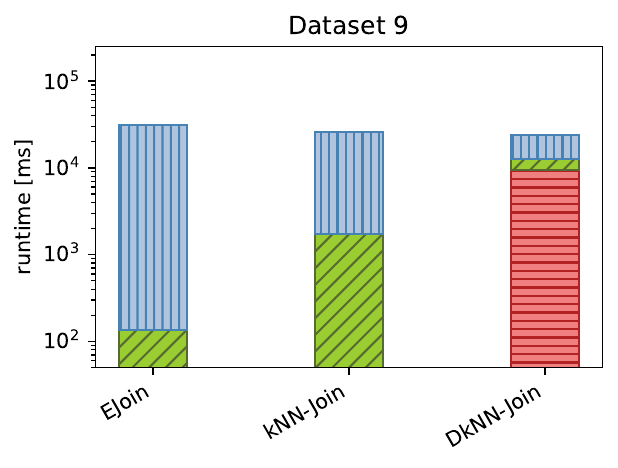}
\includegraphics[width=0.26\textwidth]{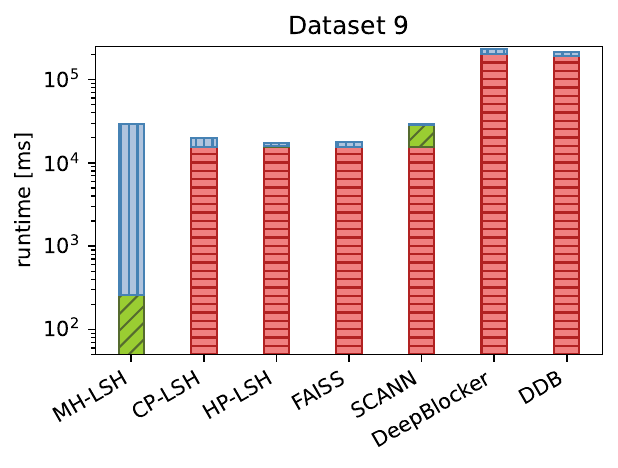}
\vspace{-10pt}
\caption{The break-down of the overall run-time of the blocking workflows (left column), the sparse NN methods (middle column) and the dense NN methods (right column) for the schema-agnostic settings of $D_1$-$D_4$ and $D_8$-$D_9$.}
\vspace{-10pt}
\label{fig:breakDownB}
\end{figure*}

\begin{figure*}[t]
\centering
\includegraphics[width=0.26\textwidth]{figures/runtimes_color/brtLegend.png}
\hspace{12pt}
\includegraphics[width=0.2\textwidth]{figures/runtimes_color/nnLegend.png}
\hspace{24pt}
\includegraphics[width=0.2\textwidth]{figures/runtimes_color/nnLegend.png}\\
\includegraphics[width=0.26\textwidth]{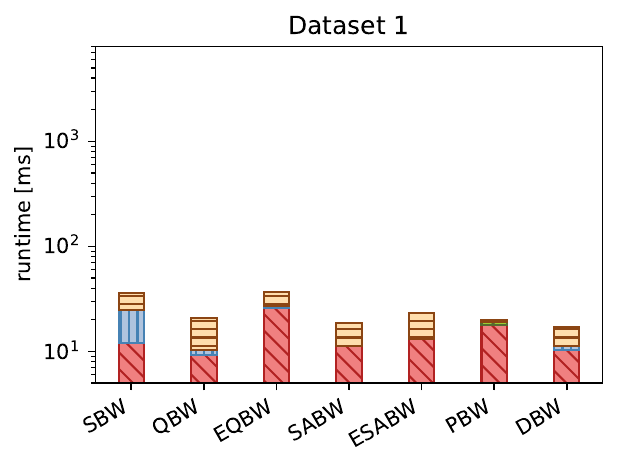}
\includegraphics[width=0.26\textwidth]{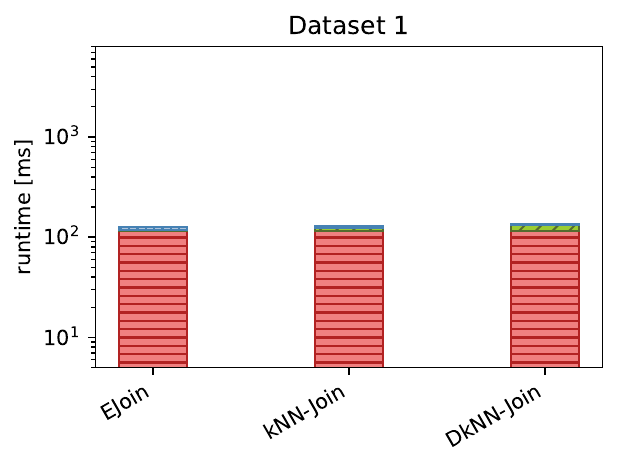}
\includegraphics[width=0.26\textwidth]{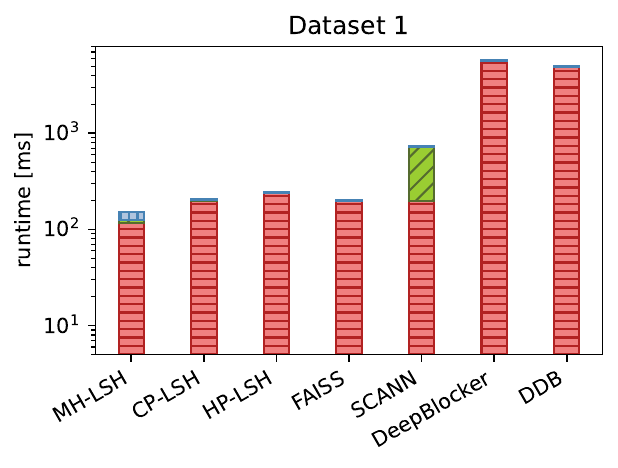}
\includegraphics[width=0.26\textwidth]{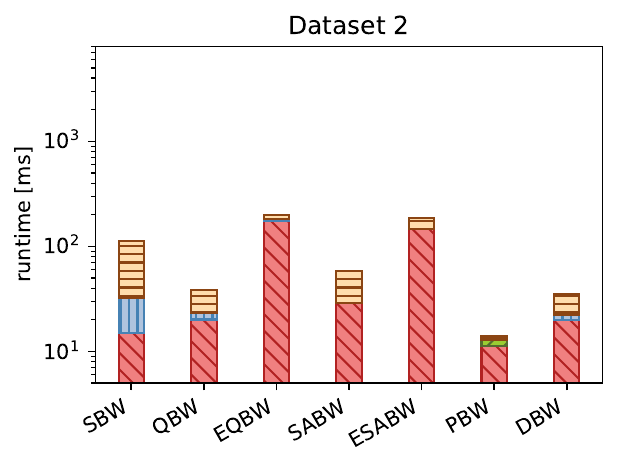}
\includegraphics[width=0.26\textwidth]{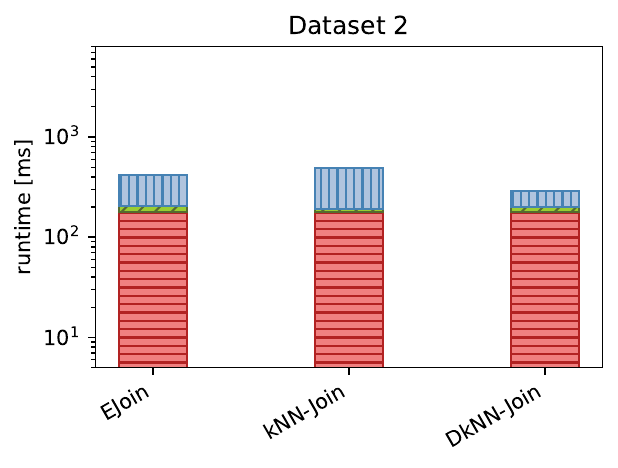}
\includegraphics[width=0.26\textwidth]{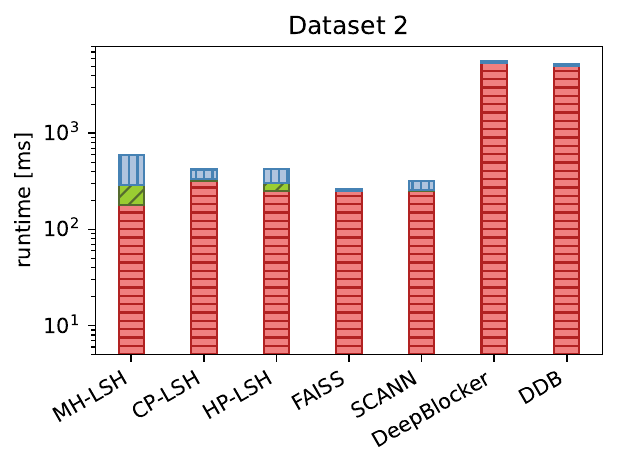}
\includegraphics[width=0.26\textwidth]{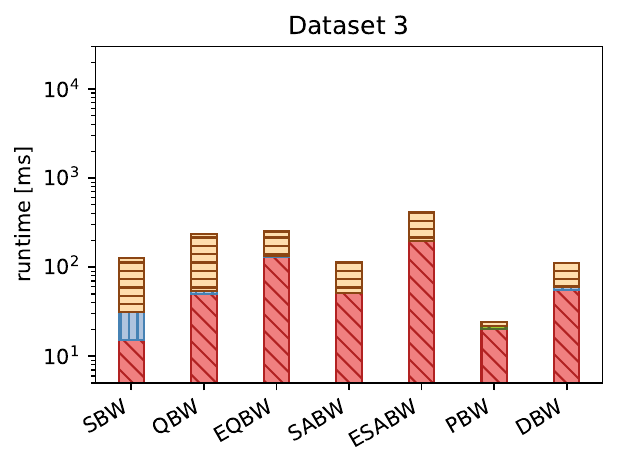}
\includegraphics[width=0.26\textwidth]{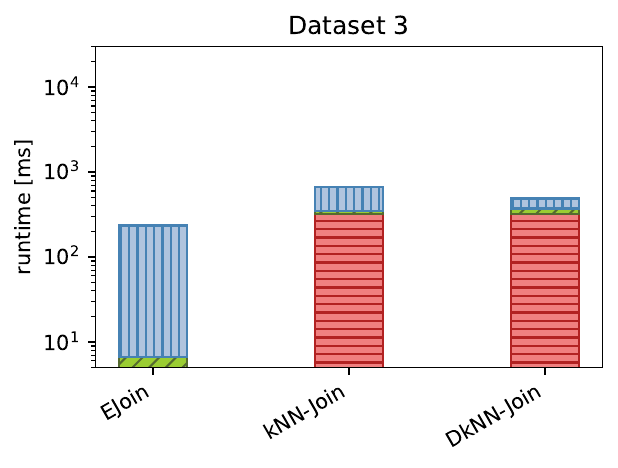}
\includegraphics[width=0.26\textwidth]{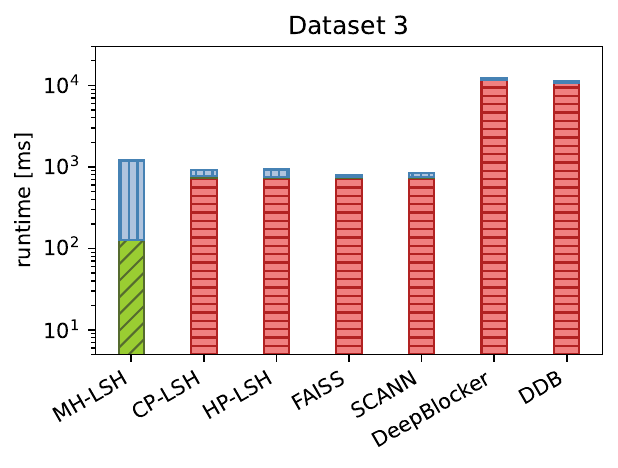}
\includegraphics[width=0.26\textwidth]{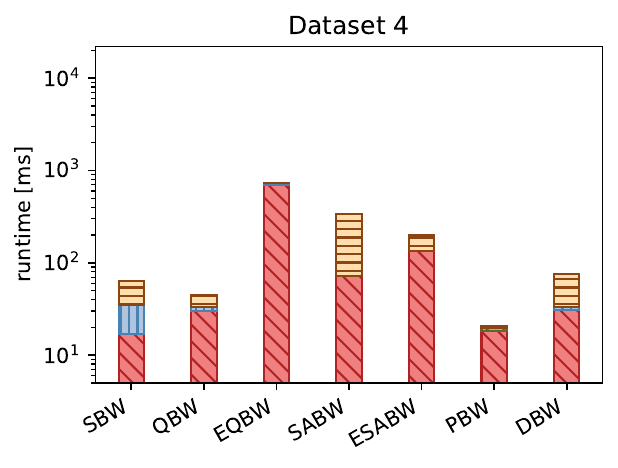}
\includegraphics[width=0.26\textwidth]{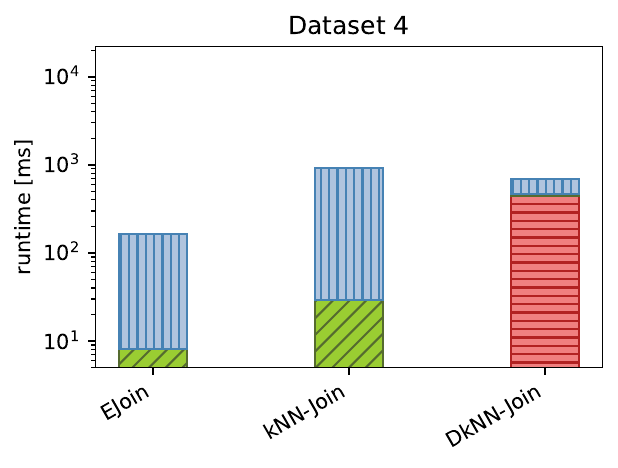}
\includegraphics[width=0.26\textwidth]{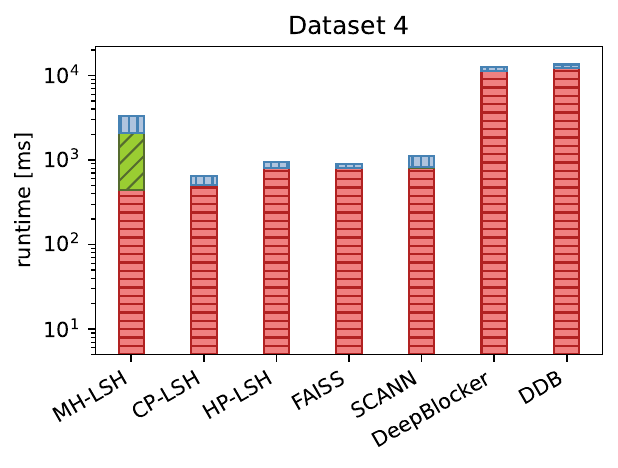}
\includegraphics[width=0.26\textwidth]{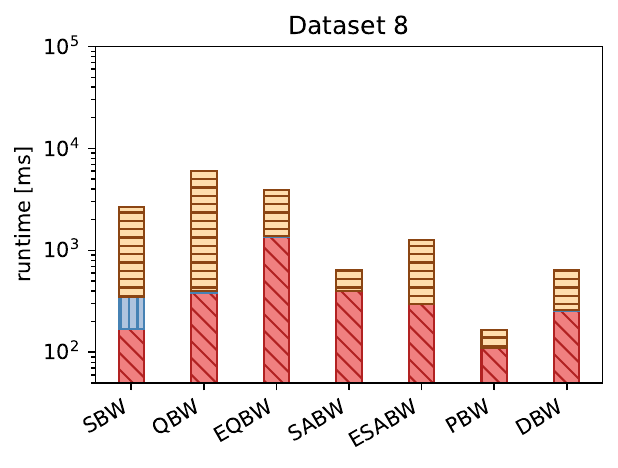}
\includegraphics[width=0.26\textwidth]{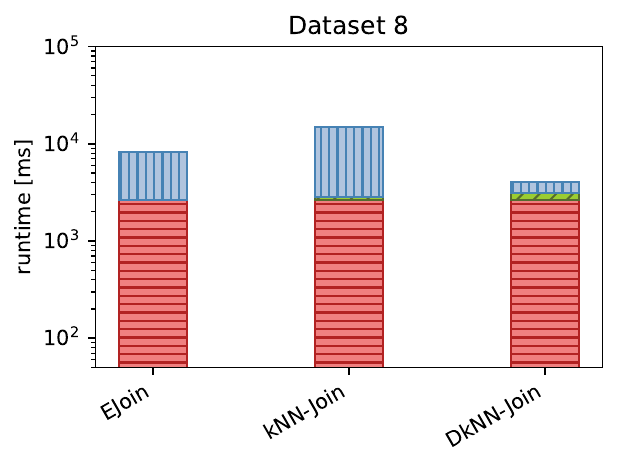}
\includegraphics[width=0.26\textwidth]{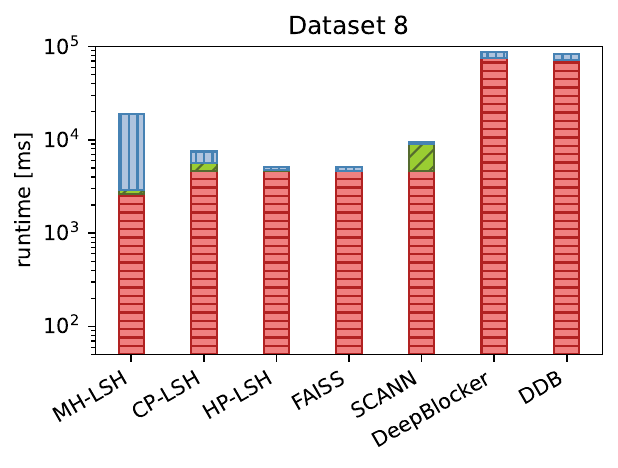}
\includegraphics[width=0.26\textwidth]{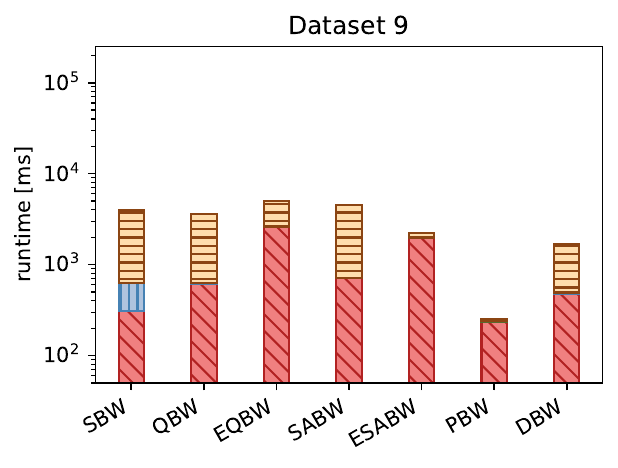}
\includegraphics[width=0.26\textwidth]{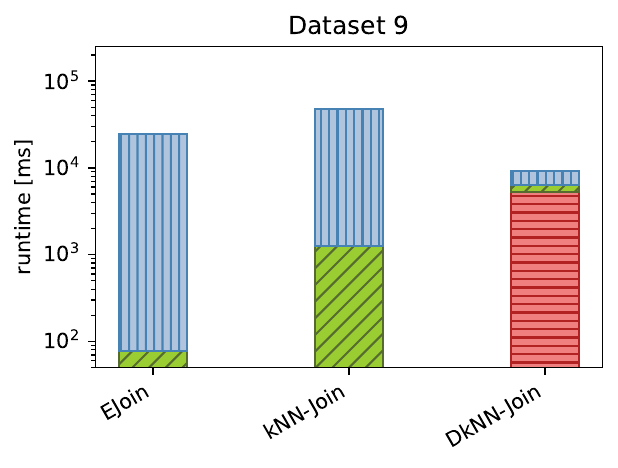}
\includegraphics[width=0.26\textwidth]{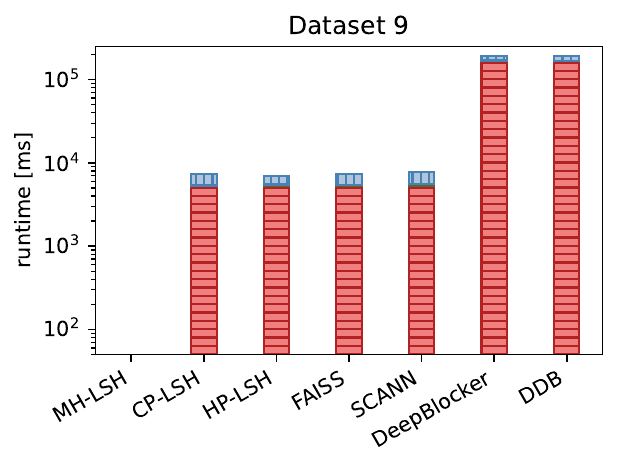}
\vspace{-10pt}
\caption{The break-down of the overall run-time of the blocking workflows (left column), the sparse NN methods (middle column) and the dense NN methods (right column) for the schema-based settings of $D_1$-$D_4$ and $D_8$-$D_9$.}
\vspace{-10pt}
\label{fig:breakDownC}
\end{figure*}

We now examine the breakdown of the run-time for every filtering method. 
The breakdown for all filtering methods, including the baseline ones, appears in Figure \ref{fig:breakDownA} for the schema-agnostic settings of the datasets $D_5$-$D_7$ and $D_{10}$, which lack the schema-based settings. For the remaining datasets, the breakdown of the schema-agnostic and the schema-based settings is presented in Figures \ref{fig:breakDownB} and  \ref{fig:breakDownC}, respectively. For the cardinality-based, the indexing and querying time corresponds to the best configuration (see $RVS$ in Table \ref{tb:joinConfiguration} for \textsf{kNNJ} and Table \ref{tb:nnConfiguration} for \textsf{FAISS}, \textsf{SCANN} and \textsf{DeepBlocker}). Note that every line corresponds to a different dataset, with a different diagram per type of filtering methods, but these diagrams share the same scale so that the comparison between all methods is straightforward. For this reason, the same scale per dataset is used by both schema-agnostic and schema-based settings in Figures \ref{fig:breakDownB} and  \ref{fig:breakDownC}.

Starting with the blocking workflows, their overall run-time consists of the run-times of the four steps in Figure \ref{fig:bworkflow}:
\begin{enumerate}
    \item the block building time ($t_b$),
    \item the block purging time ($t_p$),
    \item the block filtering time ($t_f$), and
    \item the comparison cleaning time ($t_c$).
\end{enumerate}

We observe that the block cleaning methods are quite fast, due to their coarse-grained functionality, corresponding to a tiny portion of the overall run-time. On average, the block purging time accounts for less than 0.9\% for all methods for both schema-agnostic and schema-based settings. The only exception is the fastest blocking workflow, \textsf{PBW}, where its contribution to $RT$ raises to 3.2\%, on average. Block Filtering involves a more fine-grained functionality that operates at the level of individual entities. As a result, it accounts for a larger portion of $RT$, which however remains lower than 2\% in all cases. The only exceptions are the schema-based settings of \textsf{SBW} and \textsf{DQBW}, where its cost raises to 17.9\% and 3.8\%, respectively, due to the absence of Block Purging (i.e., Block Filtering processes all the initial blocks). 

Regarding the relative cost of block building and comparison cleaning, we distinguish the blocking workflows in two main groups: one with the methods performing fast signature extraction (i.e., \textsf{SBW}, \textsf{QBW} and \textsf{DQBW}), and one with the methods incorporating an elaborate signature extraction (i.e., \textsf{EQBW} and \textsf{ESABW}). For the former, the cost of comparison cleaning outweighs the cost of block building: on average, it accounts for 68.8\% and 58.9\% of the overall $RT$ for \textsf{SBW} and \textsf{QBW}/\textsf{DQBW}, respectively. In contrast, the cost of block building is much higher than comparison cleaning for \textsf{EQBW} and \textsf{ESABW}, accounting for 65.2\% and 58.3\%, respectively, on average. This should be attributed not only to the more complex process of signature extraction, but also to the lower number of candidate pairs in the resulting set of blocks. To this category also belongs \textsf{SBW}, whose block building takes up 76.4\% of the overall run-time. However, this should be attributed to the very low cost of its comparison cleaning, which simply applies Comparison Propagation (i.e., it discards redundant candidate pairs, without assigning scores to remove superfluous ones, too). In the middle of these two categories lies \textsf{SABW}, where the cost of block building and comparison cleaning similar (46.6\% and 53.4\%, respectively, on average). The reason is that its block building phase is relatively complex, but produces a very small set of candidate pairs. As a result, it is often one of the fastest blocking workflows.

The overall run-time for the rest of the filtering methods is divided into: 
\begin{enumerate}
    \item the pre-processing time ($t_r$), which includes the cost of stop-word removal, stemming and the transformation of attribute values into embedding vectors (if applicable).
    \item the indexing time ($t_i$), and
    \item the querying time ($t_q$).
\end{enumerate}

As regards the sparse NN methods, we observe that the cost of indexing is consistently the lowest one among the three steps. On average, across all datasets and schema settings, $t_i$ accounts for 3.6\%, 5.6\% and 9.9\% of the overall run-time of $\varepsilon$-, \textsf{kNN}- and \textsf{DkNNJ}, respectively. The higher portion for \textsf{DkNN} stems from the complex representation it uses, i.e., the multiset of five-grams. The second most time-consuming step is pre-processing, which on average, accounts for 31.4\%, 22.7\% and 52.18\% of the total $RT$ for $\varepsilon$-, \textsf{kNN}- and \textsf{DkNN}, respectively. The discrepancy between the first two methods and the last one is caused by the default configurations of the baseline method, which always includes cleaning. In contrast, the fine-tuned methods apply cleaning in 2/3 and 1/2 of the cases, respectively (see Table \ref{tb:joinConfiguration}). The contribution of $t_r$ to $RT$ would be much higher for both methods, if they consistently applied cleaning. Note also that the breakdown of \textsf{DkNN} demonstrates that stop-word removal and stemming have a high computational cost. Finally, the rest of $RT$ corresponds to the querying phase, which occupies 65.0\%, 72.3\% and 37.9\% of the overall run-time, on average, for $\varepsilon$-, \textsf{kNN}- and \textsf{DkNN}, respectively. Note that the portion of $t_q$ is higher for \textsf{kNNJ} than for $\varepsilon$-\textsf{Join}, because its querying phase is more costly, due to the sorting of candidate pairs. This portion fluctuates considerably for both methods ($\sigma=$0.299 in both cases), as it depends on the threshold they use. The variance is significantly reduced for \textsf{DkNN} ($\sigma=$0.186), because it uses the same cardinality-threshold in all cases. The low value of this threshold is another reason for the dominance of the pre-processing time in the case of \textsf{DkNN}.

Finally, for the dense NN methods we observe that the pre-processing time dominates their run-time to a significant extent. The reason is that this step now includes the cost of creating the semantic representations of entities based on the pre-trained fastText embeddings. Its portion ranges from 66\% (\textsf{SCANN}) to 86.4\% (\textsf{FAISS}) and 91.0\% (\textsf{DeepBlocker} and \textsf{DDB}). The portion is lower for \textsf{SCANN}, due to the time-consuming partitioning it performs in both the indexing and the query phase. In contrast, it is higher for \textsf{FAISS}, due to the very low cost of the other two phases. For the same reason, $t_r$ accounts for a larger portion of $RT$ in the schema-based settings, where both indexing and querying are much faster. However, in the case of \textsf{DeepBlocker} and \textsf{DDB}, the high portion of $t_r$ should be attributed to the Autoencoder, which learns the tuple embedding module, i.e., it transforms values into fastText embeddings, it creates an artificial dataset and then learns a neural model. As expected, the indexing phase is the fastest step for all methods, accounting for $<$0.5\% (\textsf{FAISS}, \textsf{DeepBlocker} and \textsf{DDB}) to 11.9\% (\textsf{SCANN}) of the overall run-time. The rest of $RT$ corresponds to the querying phase, whose portion ranges from 9.0\% (\textsf{DeepBlocker} and \textsf{DDB}) to 22.7\% (\textsf{CP-LSH}). Note that the only exception to these patterns is \textsf{MH-LSH}, which does not involve the cost of semantic representations. As a result, its $t_r$ is reduced to 14.9\% of $RT$, with the querying phase dominating its run-time (69.7\%), due to the very large number of candidate pairs it generates.

\section*{D. Detailed Configuration}

In this section, we present the detailed configuration of every filtering method that corresponds to its performance in Table \ref{tb:results}.

Table \ref{tb:bwConfigurations} reports the configuration of blocking workflows across all datasets and schema settings. \textsf{$BP$} stands for the use of Block Purging or not (it is a parameter-free approach), \textsf{$BFr$} indicates the filtering ratio used by Block Filtering, \textsf{$PA$} denotes the pruning algorithm that is used by the Meta-blocking step and \textsf{$WS$} stands for the corresponding weighting scheme. Recall that the the domain of each parameter appears in Table \ref{tb:bwConfigurationDomain}.

The configuration of sparse NN methods is listed in Table \ref{tb:joinConfiguration}. $CL$ stands for the use of pre-processing for cleaning an attribute value, $SM$ for the similarity measure, $RM$ for the representation model that is used, $t$ for the similarity threshold and $K$ for the number of nearest neighbors, while $RVS$ indicates whether $\mathcal{E}_2$ is indexed and $\mathcal{E}_1$ is used as the query set, instead of the opposite. The domain of each parameter appears in Table \ref{tb:joinConfigurationDomain}.

Finally, the configuration of the dense NN methods is reported in Table \ref{tb:nnConfiguration}. Note that $CL$ denotes the use of pre-processing for cleaning an attribute value, $t$ the corresponding Jaccard similarity threshold, $k$ the size of the k-shingles that are used as representation model and $RVS$ whether the indexed and the query dataset should be reversed or not. The domain of each parameter appears in Table \ref{tb:nnConfigurationDomain}.

\section*{E. Configuration Analysis}

Based on the fine-tuning experiments that were performed over $D_{c1}$-$D_{c10}$, we can draw several useful conclusions regarding the configuration of the filtering techniques.

For \underline{the blocking workflows}, the following rules of thumb can be used when fine-tuning them:

$\bullet$ Block Purging should never be used in schema-based settings. Even in schema-agnostic settings, it should be avoided, due to its aggresive pruning. 

$\bullet$ Block Filtering offers a more suitable alternative for coarse-grained block cleaning. It should always be part of a blocking workflow, albeit in combination with a relative high ratio (i.e., $BFr \geq 0.5$ in most cases).

$\bullet$ Among the comparison cleaning techniques, RCNP and BLAST constitute the best choice in most cases. Typically, they are combined with the weighting schemes~$\chi^2$~and~ARCS.

$\bullet$ For Q-Grams Blocking, large values for q should be used regardless of the schema settings, i.e., $q=6$ in most cases. The same applies to Extended Q-Grams Blocking, except for the schema-based settings, where $q$ should be set to 3. At the same time, the threshold parameter should be set to $t=0.9$.

$\bullet$ Similarly, large values for $l_{min}$ should be typically used for Suffix Arrays Blocking and Extended Suffix Arrays Blocking. For both algorithms, the maximum block size $b_{max}$ should be set in proportion to the number of input entities.

For \underline{sparse NN methods}, the following guidelines should be used when fine-tuning them:

$\bullet$ The cosine similarity is the best measure in the vast majority of cases for both $\epsilon$- and kNN-Joins.

$\bullet$ Stop word removal and stemming should be typically applied, at least as a means of reducing the search space. 

$\bullet$ In the case of kNN-Join, the largest dataset should be indexed, and the smallest one should be used for querying. In this way, very small values for k (i.e., $k<5$) suffice for achieving high recall and precision.

$\bullet$ For $\epsilon$-Joins, a low similarity threshold should be used in the case of schema-agnostic settings (0.35 on average) and a higher one for schema-based settings (0.55).

$\bullet$ Both joins should be combined with character n-grams for tokenization. The size of n should be 2 or 3 for schema-based settings and 4 or 5 for schema-agnostic ones. In the latter case, multi-sets should be used instead of bags.

For \underline{dense NN methods}, the following advice could be used when fine-tuning them:

$\bullet$ Stop word removal and stemming should be typically applied to the input data.

$\bullet$ For MinHash LSH, the size of $k$-shingles should be set to 2, while the number of bands and rows should be set to 32 and 16, respectively.

$\bullet$ For Cross-polytope and Hyperplane LSH, the number of tables and hashes should be set in proportion to the size of the input data. The number of probes can be automatically configured in order to achieve the desired level of recall.

$\bullet$ For FAISS, SCANN and DeepBlocker, the number of candidates can be reduced by indexing the largest dataset and querying it with the smallest one. The number of candidates per query entity, k, should be proportional to the number of input entities, with the schema-based settings calling for smaller values than the schema-agnostic ones. 

$\bullet$ FAISS should be combined with a Flat index, while its embedding vectors should always be normalized and combined with the Euclidean distance.

$\bullet$ Similarly, SCANN should be combined with a brute-force index over Euclidean distances.

$\bullet$ DeepBlocker should be coupled with AutoEncoder, which lowers the memory requirements and accelerates its run-time.

\section*{F. Scalability Analysis}

In this section, we examine how the effectiveness of the filtering techniques evolves as the size of the input data increases from 10$^4$ ($D_{10K}$) to 2$\cdot$10$^6$ ($D_{2M}$) entities. Figures \ref{fig:scalabilityRecall} and \ref{fig:scalabilityPrecision} report the recall ($PC$) and precision ($PQ$), respectively, of all methods over each dataset. 

Starting with the blocking workflows on the left diagrams, we observe that they can be distinguished into three groups. The first one includes \textsf{SABW} and \textsf{ESABW}, whose recall drops steadily, from 0.91 over $D_{10K}$ to around 0.55 over $D_{2M}$. This should be attributed to their proactive functionality, which sets an upper limit on the sizes of their blocks, $b_{max}$. This limit is actually fixed and independent of the number of input entities, ensuring their high time efficiency at the cost of more missed duplicate pairs, as the input size increases. Their precision, though, remains relative stable, fluctuating between 0.43 and 0.33 in most cases.

The second group of blocking workflows includes \textsf{QBW}, \textsf{DBW} and \textsf{PBW}. They all perform a very conservative pruning of candidate pairs -- especially \textsf{PBW}, which consequently scales only up to $D_{300K}$. As a result, they maintain high recall across all datasets, consistently exceeding the corresponding threshold (0.9). On the downside, their precision drops significantly as the size of the input increases. For instance, \textsf{QBW}'s $PQ$ drops from 0.44 over $D_{10K}$ to $\sim0.001$ over $D_{2M}$. \textsf{EQBW} follows the methods of this group in close distance, but trades slightly lower recall for significantly higher precision ($PC$=0.86 and $PQ$=0.03 over $D_{2M}$).

Finally, \textsf{SBW} forms a group on its own. It emphasizes precision, maintaining it between 0.55 and 0.60 across all datasets. Due to its careful pruning of candidates, though, its recall drops gradually from 0.91 over $D_{10K}$ to 0.81 over $D_{2M}$. As a result, \textsf{SBW} offers the best balance between recall and precision among the blocking workflows.

Regarding the sparse NN methods, we observe that they all achieve very high recall across all datasets. \textsf{DkNN-Join} and $\epsilon$-Join consistently exceed the recall threshold, while the $PC$ of \textsf{kNN-Join} drops gradually, reaching 0.81 over $D_{2M}$. However, $\epsilon$-Join and \textsf{kNN-Join} exhibit perfect precision across all datasets, which indicates that they perform the most accurate pruning among all filtering techniques. Even for \textsf{DkNN-Join}, $PQ$ is stable and relatively high (0.16) across all datasets. This high effectiveness comes at the cost of low time efficiency, as shown in Figure \ref{fig:scalability}.

Finally, we observe the following patterns in the case of the dense NN methods: (i) The LSH variants are the only techniques of this category that satisfy the recall threshold across all datasets they can process. This stems from the excessively large number of candidate pairs they generate, which prevents them from scaling to the largest datasets. (ii) The remaining techniques combine very low precision with a steadily decreasing recall. The most effective approach is \textsf{SCANN}, whose recall drops to 0.70 over $D_{2M}$, while its precision amounts to 0.02. It is followed in close distance by \textsf{FAISS}, which trades the fastest functionality for slightly lower recall and precision (0.67 and 0.018 over $D_{2M}$, respectively).

Overall, the highly efficient functionality of \textsf{FAISS} and \textsf{SCANN} comes at the cost of an effectiveness that is inversely proportional to size of the input data. In contrast, the sparse NN methods exhibit the highest robustness and effectiveness at the cost of the highest run-time. In the middle of these two extremes lie the blocking workflows, with \textsf{SBW} offering the best balance between $PC$ and $PQ$ for a reasonable run-time.

%% file: tables/results.tex
\begin{table*}[t]\centering
\scriptsize
  \setlength{\tabcolsep}{2.2pt}
    \caption{The performance of the blocking workflows, the sparse and the dense NN methods.}
    \vspace{-8pt}
    \begin{tabular}{ | l | r  r  r  r  r  r  r  r  r  r || r  r  r  r  r  r  |}
		\cline{2-17}
		\multicolumn{1}{c|}{} &
		\multicolumn{10}{c||}{\textbf{Schema-agnostic Settings}}&
		\multicolumn{6}{c|}{\textbf{Schema-based Settings}}\\
		\multicolumn{1}{c|}{} &
		\multicolumn{1}{c}{$\mathbf{D_{a1}}$} &
		\multicolumn{1}{c}{$\mathbf{D_{a2}}$} &
		\multicolumn{1}{c}{$\mathbf{D_{a3}}$} &
		\multicolumn{1}{c}{$\mathbf{D_{a4}}$} &
		\multicolumn{1}{c}{$\mathbf{D_{a5}}$} &
        \multicolumn{1}{c}{$\mathbf{D_{a6}}$} &
        \multicolumn{1}{c}{$\mathbf{D_{a7}}$} &
        \multicolumn{1}{c}{$\mathbf{D_{a8}}$} &
        \multicolumn{1}{c}{$\mathbf{D_{a9}}$} &
        \multicolumn{1}{c||}{$\mathbf{D_{a10}}$} &
        \multicolumn{1}{c}{$\mathbf{D_{b1}}$} &
		\multicolumn{1}{c}{$\mathbf{D_{b2}}$} &
		\multicolumn{1}{c}{$\mathbf{D_{b3}}$} &
		\multicolumn{1}{c}{$\mathbf{D_{b4}}$} &
        \multicolumn{1}{c}{$\mathbf{D_{b8}}$} &
        \multicolumn{1}{c|}{$\mathbf{D_{b9}}$} \\
		\hline
        SBW & 1.000 & 0.902 & 0.901 & 0.903 & 0.922 & 0.924 & 0.909 & 0.904 & 0.915 & 0.904 & 0.933 & 0.901 & 0.922 & 0.976 & 0.904 & 0.906 \\
        QBW & 0.978 & 0.912 & 0.902 & 0.915 & 0.910 & 0.910 & 0.903 & 0.904 & 0.993 & 0.900 & 0.933 & 0.914 & 0.915 & 0.952 & 0.903 & 0.928 \\
        EQBW & 0.910 & 0.901 & 0.905 & 0.918 & 0.905 & 0.904 & 0.905 & 0.900 & 0.913 & 0.900 & 0.944 & 0.902 & 0.905 & 0.913 & 0.903 & 0.910 \\
        SABW & 1.000 & 0.901 & 0.900 & 0.969 & 0.900 & 0.903 & 0.901 & 0.912 & 0.902 & 0.900 & 0.910 & 0.903 & 0.909 & 0.997 & 0.900 & 0.900 \\
        ESABW & 0.921 & 0.901 & 0.902 & 0.902 & 0.900 & 0.900 & 0.901 & 0.900 & 0.900 & 0.901 & 0.910 & 0.901 & 0.901 & 0.902 & 0.900 & 0.900 \\
        PBW & 1.000 & 0.981 & 0.971 & 0.999 & 0.998 & 0.995 & 0.991 & 1.000 & 1.000 & 0.980 & 
        0.989 & {\color{red}0.746} & 0.936 & {\color{red}0.838} & 0.996 & 0.935 \\
        DBW & 1.000 & 0.930 & 0.936 & 1.000 & 0.943 & 0.894 & 0.984 & 0.996 & 0.997 & 0.922 & {\color{red}0.607} & {\color{red}0.888} & 0.894 & 0.999 & 0.931 & 0.995 \\
        \hline
        $\epsilon$-Join & 0.921 & 0.903 & 0.911 & 0.901 & 0.911 & 0.901 & 0.900 & 0.902 & 0.901 & 0.902 & 0.900	& 0.901 & 0.900 & 0.913 & 0.907 & 0.901 \\
        kNNJ & 1.000 & 0.924 & 0.900 & 0.996 & 0.961 & 0.910 & 0.972 & 0.910 & 0.957 & 0.903 & 0.978 & 0.925 & 0.914 & 0.994 & 0.900 & 0.970  \\
        DkNN & 1.000 & 0.915 & {\color{red}0.804} & 0.998 & {\color{red}0.882} & 0.921 & 0.984 & 0.939 & 0.995 & {\color{red}0.887} & 1.000 & 0.916 & 0.934 & 0.998 & {\color{red}0.831} & 0.989 \\
        \hline
        MH-LSH & 0.910 & 0.987 & 0.903 & 0.903 & 0.903 & 0.985 & 0.936 & 0.948 & 0.923 & - & 0.900 & 0.942 & 0.936 & 0.973 & 0.955 & - \\
        CP-LSH & 0.910 & 0.908 & 0.900 & 0.900 & 0.901 & 0.905 & 0.900 & 0.900 & 0.901 & 0.901 & 0.910 & 0.911 & 0.902 & 0.901 & 0.917 & 0.908 \\
        HP-LSH & 0.910 & 0.901 & 0.900 & 0.900 & 0.900 & 0.900 & 0.900 & 0.900 & 0.900 & 0.900 & 0.910 & 0.900 & 0.900 & 0.917 & 0.900 & 0.900 \\
        FAISS & 0.933 & 0.902 & 0.900 & 0.961 & 0.901 & 0.900 & 0.908 & 0.900 & 0.900 & 0.900 & 0.955 & 0.906 & 0.900 & 0.971 & 0.900 & 0.912 \\
        SCANN & 0.933 & 0.902 & 0.900 & 0.961 & 0.901 & 0.900 & 0.908 & 0.900 & 0.900 & 0.900 & 0.966 & 0.906 & 0.900 & 0.970 & 0.900 & 0.912 \\
        DeepBlocker & 0.943 & 0.903 & 0.900 & 0.983 & 0.901 & 0.900 & 0.926 & 0.900 & 0.908 & - & 0.976 & 0.900 & 0.900 & 0.964 & 0.900 & 0.919 \\ 
        DDB & 0.970 & {\color{red}0.732} & {\color{red}0.652} & 0.996 & {\color{red}0.822} & {\color{red}0.745} & 0.949 & 0.938 & 0.966 & -  & 0.992 & {\color{red}0.800} & {\color{red}0.839} & 0.990 & 0.900 & 0.972\\
        \hline
        \multicolumn{17}{c}{\textbf{(a) recall ($\mathbf{PC}$)} -- values in red correspond to $PC\ll0.9$} \\
        \hline
        SBW & 0.533 & 0.216 & \underline{0.017} & \textbf{\underline{0.957}} & \textbf{\underline{0.382}} & \textbf{\underline{0.189}} & \underline{0.154} & \underline{0.117} & \underline{0.470} & \textbf{\underline{0.475}} & \textbf{\underline{0.769}} & 0.259 & 0.211 & 0.822 & 0.028 & \underline{0.524} \\
        QBW & 0.465 & 0.740 & 0.013 & 0.897 & 0.210 & 0.078 & 0.112 & 0.116 & 0.254	 & 0.347 & 0.755 & 0.750 & \textbf{\underline{0.240}} & 0.783 & \underline{0.030} & 0.232 \\
        EQBW & 0.757 & 0.204 & 0.012 & 0.926 & 0.220 & 0.078 & 0.124 & 0.087 & 0.149 & 0.390 & 0.764 & 0.261 & 0.188 & \underline{0.854} & 0.021 & 0.182 \\
        SABW &  \textbf{\underline{0.767}} & 0.384 & 0.015 & 0.804 & 0.217 & 0.065 & 0.146 & 0.096 & 0.322 & 	0.020 & 0.757 & 0.390 & 0.226 & 0.695 & 0.010 & 0.014 \\
        ESABW & 0.469 & \textbf{\underline{0.759}} & 0.010 & 0.751 & 0.201 & 0.059 & 0.136 & 0.088 & 0.130	 & 0.014 & 0.743 & \textbf{\underline{0.780}} & 0.131 & 0.545 & 0.009 & 0.010 \\
        PBW & 0.307 & 0.015 & 0.002 & 0.020 & 0.006 & 0.004 & 0.003 & 4.5$\cdot$10$^{-4}$ & 0.001 & 3.3$\cdot$10$^{-4}$ & 0.162 & {\color{red}0.175} & 0.047 & {\color{red}0.230} & 5.8$\cdot$10$^{-4}$ & 0.005 \\
        DBW & 2.7$\cdot$10$^{-4}$ & 0.065 & 0.005 & 0.042 & 0.036 & 0.008 & 0.008 & 0.002 & 0.003 & 0.009 & {\color{red}0.199} & {\color{red}0.163} & 0.069 & 0.063 & 0.005 & 0.003 \\
        \hline
        $\epsilon$-Join & \underline{0.732} & 0.095 & 0.010 & 0.945 & 0.018 & 0.001 & \textbf{\underline{0.192}} & 0.068 & 0.765 & 0.033 & \underline{0.381} & 0.147 & 0.144 & \underline{0.886} & 0.020 & \underline{0.669} \\
        kNNJ & 0.224 & \underline{0.229} & \textbf{\underline{0.028}} & \underline{0.954} & \underline{0.305} & \underline{0.122} & 0.130 & \textbf{\underline{0.150}} & \textbf{\underline{0.877}} & \underline{0.149} & 0.309 & \underline{0.295} & \textbf{\underline{0.240}} & 0.836 & \textbf{\underline{0.049}} & 0.647 \\
        DkNN & 0.047 & 0.181 & {\color{red}0.130} & 0.190 & {\color{red}0.053} & 0.024 & 0.026 & 0.062 & 0.182 & {\color{red}0.147} & 0.100 & 0.173 & 0.149 & 0.187 & {\color{red}0.054} & 0.166\\
        \hline
        MH-LSH & 2.6$\cdot$10$^{-4}$ & 0.001 & 2.7$\cdot$10$^{-4}$ & 0.005 & 6.6$\cdot$10$^{-5}$ & 2.7$\cdot$10$^{-5}$ & 3.4$\cdot$10$^{-5}$ & 1.6$\cdot$10$^{-5}$ & 2.1$\cdot$10$^{-5}$ & - & 0.007 & 0.001 & 2.9$\cdot$10$^{-4}$ & 0.036 & 1.7$\cdot$10$^{-5}$ & -\\
        CP-LSH & 0.003 & 0.006 & 0.001 & 0.079 & 0.001 & 2.1$\cdot$10$^{-4}$ & 0.002 & 4.0$\cdot$10$^{-4}$ & 2.2$\cdot$10$^{-4}$ & 7.8$\cdot$10$^{-5}$ & 0.130 & 0.008 & 0.003 & 0.876 & 0.001 & 0.002 \\
        HP-LSH & 0.002 & 0.004 & 0.001 & 0.059 & 4.4$\cdot$10$^{-4}$ & 2.1$\cdot$10$^{-4}$ & 0.001 & 2.6$\cdot$10$^{-4}$ & 1.5$\cdot$10$^{-4}$ & 7.3$\cdot$10$^{-5}$ & 0.061 & 0.007 & 0.002 & 0.859 & 4.0$\cdot$10$^{-4}$ & 0.024 \\
        FAISS & 0.082 & \underline{0.032} & 0.001 & 0.932 & \underline{0.012} & \underline{0.005} & 0.041 & 0.001 & 0.012 & \underline{1.5$\cdot$10$^{-4}$} & 0.376 & \underline{0.050} & 0.024 & \textbf{\underline{0.942}} & 0.004 & \textbf{\underline{0.836}} \\
        SCANN & 0.082 & \underline{0.032} & 0.001 & 0.932 & \underline{0.012} & \underline{0.005} & 0.041 & 0.002 & 0.013 & \underline{1.5$\cdot$10$^{-4}$} & \underline{0.381} & \underline{0.050} & 0.024 & 0.941 & 0.005 & \textbf{\underline{0.836}} \\
        DeepBlocker & \underline{0.247} & 0.026 & \underline{0.002} & \underline{0.953} & 0.011 & 0.003 & \underline{0.130} & \underline{0.018} & \underline{0.167} & - & 0.256 & 0.029 & \underline{0.073} & 0.935 & \underline{0.012} & 0.211 \\
        DDB & 0.008 & {\color{red}0.146} & {\color{red}0.047} & 0.169 & {\color{red}0.053} & {\color{red}0.020} & 0.027 & 0.007 & 0.007 & - & 0.008 & {\color{red}0.160} & {\color{red}0.061} & 0.168 & 0.007 & 0.007 \\
        \hline
        \multicolumn{17}{c}{\textbf{(b) $\mathbf{PQ}$} -- values in red correspond to $PC\ll0.9$} \\
        \hline
        SBW & 27 ms & 225 ms & 359 ms & 58 ms & 316 ms & 1.1 s & 252 ms & 741 ms & 3.5 s & 57.7 s & 36 ms & 113 ms & 127 ms & 64 ms & 2.7 s & 4.0 s \\
        QBW & 46 ms & 75 ms & 340 ms & 92 ms & 234 ms & 1.2 s & 273 ms & 787 ms & 70.0 s & 158.6 s & 21 ms & 38 ms & 237 ms & 45 ms & 6.0 s & 3.6 s \\
        EQBW & 303 ms & 347 ms & 2.2 s & 2.6 s & 411 ms & 603 ms & 704 ms & 1.2 s & 6.9 s & 68.1 s & 37 ms	 & 198 ms & 251 ms & 738 ms & 3.9 s & 5.0 s \\
        SABW & 49 ms & 142 ms & 676 ms & 238 ms & 305 ms & 347 ms & 436 ms & 674 ms & 3.0 s & 9.8 s & 19 ms & 58 ms & 114 ms & 337 ms & 645 ms & 4.5 s \\
        ESABW & 161 ms & 281 ms & 891 ms & 745 ms & 388 ms & 481 ms & 750 ms & 1.2 s & 5.6 s & 13.3 s & 23 ms & 187 ms & 418 ms & 201 ms & 1.3 s & 2.2 s \\
        PBW &  29 ms & 32 ms & 135 ms & 44 ms & 76 ms & 81 ms & 125 ms & 243 ms & 579 ms & 5.1 s & 20 ms & {\color{red}14 ms} & 24 ms & {\color{red}21 ms} & 164 ms & 255 ms \\
        DBW & 209 ms & 55 ms & 435 ms & 153 ms & 252 ms & 368 ms & 526 ms & 1.0 s & 2.5 s & 8.2 s & {\color{red}17 ms} & {\color{red}35 ms} & 111 ms & 76 ms & 637 ms & 1.7 s \\
        \hline
        $\epsilon$-Join & 278 ms & 703 ms & 811 ms & 575 ms & 2.8 s & 8.8 s & 2.8 s & 15.1 s & 30.9 s & 39.4 s & 128 ms & 418 ms & 235 ms & 163 ms & 8.2 s & 24.6 s \\
        kNNJ & 1.1 s & 874 ms & 4.5 s & 4.4 s & 1.8 s & 1.2 s & 1.8 s & 12.2 s & 26.2 s & 93.9 s & 130 ms & 490 ms & 660 ms & 921 ms & 15.0 s & 47.6 s \\
        DkNN & 969 ms & 750 ms & {\color{red}4.4 s} & 2.0 s & {\color{red}3.6 s} & 3.0 s & 3.6 s & 10.7 s & 24.1 s & {\color{red}70.2 s} & 136	ms & 291 ms & 488 ms & 698 ms & {\color{red}4.1 s} & 9.1 s \\
        \hline
        MH-LSH & 212 ms & 717 ms & 1.6 s & 1.3 s & 9.0 s & 38.4 s & 10.7 s & 14.2 s & 29.4 s & - & 153 ms & 598 ms & 1.2 s & 3.3 s & 18.8 s & -\\
        CP-LSH & 181 ms & 902 ms & 4.3 s & 2.0 s & 1.7 s & 1.8 s & 3.6 s & 4.8 s & 19.9 s & 63.9 s & 208 ms & 420 ms & 927 ms & 646 ms & 7.5 s & 7.4 s \\
        HP-LSH & 168 ms & 940 ms & 4.2 s & 1.9 s & 1.6 s & 3.5 s & 4.4 s & 7.5 s & 17.5 s & 71.1 s & 246 ms & 430 ms & 930 ms & 943 ms & 5.2 s & 7.0 s \\
        FAISS & 164 ms & 790 ms & 4.2 s & 1.5 s & 1.4 s & 1.8 s & 2.1 s & 7.5 s & 17.7 s & 53.3 s & 204 ms & 264 ms & 787 ms & 901 ms & 5.1 s & 7.4 s \\
        SCANN & 182 ms & 828 ms & 4.6 s & 1.8 s & 2.1 s & 2.5 s & 3.0 s & 9.5 s & 29.5 s & 74.4 s & 741 ms & 319	ms & 841 ms & 1.1 s & 9.4 s & 7.8 s \\
        DeepBlocker & 6.7	s & 7.5	s & 26.1 s & 18.6 s & 40.9 s & 43.4	s & 47.0 s & 89.0 s & 230.9	s & - &  5.8 s & 5.6 s & 12.2 s & 12.5 s & 86.5 s & 194.0 s \\
        DDB & 6.7 s & {\color{red}7.1 s} & {\color{red}22.7 s} & 16.7 s & {\color{red}40.7 s} & {\color{red}43.8 s} & 47.1 s & 87.2 s & 216.1	s & - &  5.0 s & {\color{red}5.2 s} & {\color{red}11.4 s} & 13.5 s & 82.5	s & 191.2 s \\
        \hline
        \multicolumn{17}{c}{\textbf{(c)  the overall run-time ($RT$) in milliseconds ($ms$) or seconds ($s$)} -- values in red correspond to $PC\ll0.9$} \\
        \hline
        SBW & 167 & 4,493 & \underline{59,685} & \textbf{\underline{2,100}} & \textbf{\underline{4,750}} & \textbf{\underline{5,238}} & \underline{6,453} & \underline{6,596} & \underline{4,492} & \textbf{\underline{43,444}} & \textbf{\underline{108}} & 3,739 & 4,832 & 2,639 & 27,707 & \underline{3,992} \\
        QBW & 187 & 1,326 & 78,072 & 2,268 & 8,518 & 12,569 & 8,868 & 6,660 & 9,038 & 59,275 & 110 & 1,312 & \textbf{\underline{4,217}} & 2,704 & \underline{25,819} & 9,231 \\
        EQBW & 107 & 4,746 & 81,496 & 2,205 & 8,084 & 12,459 & 7,976 & 8,869 & 14,164 & 52,739 & 110 & 3,719 & 5,307 & \underline{2,377} & 36,076 & 11,530 \\
        SABW & \textbf{\underline{116}} & 2,529 & 68,021 & 2,682 & 8,164 & 15,001 & 6,751 & 8,141 & 6,474 & 	1.0$\cdot$10$^6$ & 107 & 2,491 & 4,437 & 3,192 & 79,778 & 1.5$\cdot$10$^5$ \\
        ESABW & 175 & \textbf{\underline{1,277}} & 96,064 & 2,674 & 8,819 & 16,473 & 7,261 & 8,705 & 16,016	& 1.5$\cdot$10$^6$ & 109 & \textbf{\underline{1,243}} & 7,567 & 3,683 & 88,236 & 2.0$\cdot$10$^5$ \\
        PBW & 290 & 69,340 & 4.6$\cdot$10$^5$ & 1.1$\cdot$10$^5$ & 3.2$\cdot$10$^5$ & 2.6$\cdot$10$^5$ & 4.2$\cdot$10$^5$ & 1.9$\cdot$10$^6$ & 2.3$\cdot$10$^6$ & 6.9$\cdot$10$^7$ & 210 & {\color{red}6,571} & 6,885 & {\color{red}2,290} & 38,630 & 3,111 \\
        DBW & 3.3$\cdot$10$^5$ & 15,350 & 2.0$\cdot$10$^5$ & 53,203 & 51,573 & 1.2$\cdot$10$^5$ & 1.3$\cdot$10$^5$ & 4.1$\cdot$10$^5$ & 8.4$\cdot$10$^5$ & 2.4$\cdot$10$^6$ & {\color{red}272} & {\color{red}5,859} & 14,302 & 35,114 & 1.6$\cdot$10$^5$ & 7.0$\cdot$10$^5$ \\
        \hline
        $\epsilon$-Join & \underline{112} & 10,194 & 1.02$\cdot$10$^5$ & 2,120 & 98,786 & 6.8$\cdot$10$^5$ & \textbf{\underline{5,132}} & 11,332 & 2,719 & 6.2$\cdot$10$^5$ & \underline{210} & 6,571 & 6,885 & \underline{2,290} & 38,630 & \underline{3,111} \\
        kNNj & 398 & \underline{4,345} & \textbf{\underline{35,333}} & \underline{2,323} & \underline{6,195} & \underline{8,013} & 8,194 & \textbf{\underline{5,163}} & \textbf{\underline{2,520}} & \underline{1.4$\cdot$10$^5$} & 282 & \underline{3,375} & \textbf{\underline{4,210}} & 2,642 & \textbf{\underline{15,581}} & 3,462 \\
        DkNN & 1,909 & 5,451 & {\color{red}6,844} & 11,649 & {\color{red}32,920} & 41,450 & 41,034 & 12,884 & 12,638 & {\color{red}1.3$\cdot$10$^5$} & 870 & 5,686 & 6,909 & 11,929 & {\color{red}13,277} & 13,593 \\
        \hline
        MH-LSH & 3.8$\cdot$10$^5$ & 1.1$\cdot$10$^6$ & 3.7$\cdot$10$^6$ & 3.9$\cdot$10$^5$ & 2.7$\cdot$10$^7$ & 3.9$\cdot$10$^7$ & 3.1$\cdot$10$^7$ & 4.9$\cdot$10$^7$ & 1.0$\cdot$10$^8$ & - & 11,411 & 1.0$\cdot$10$^6$ & 3.6$\cdot$10$^6$ & 64,398 & 4.9$\cdot$10$^6$ & - \\
        CP-LSH & 23,590 & 1.6$\cdot$10$^5$ & 1.2$\cdot$10$^6$ & 25,344 & 3.2$\cdot$10$^6$ & 4.7$\cdot$10$^6$ & 5.7$\cdot$10$^5$ & 1.9$\cdot$10$^6$ & 9.6$\cdot$10$^6$ & 2.6$\cdot$10$^8$ & 622 & 1.3$\cdot$10$^5$ & 3.5$\cdot$10$^5$ & 2,286	 & 6.9$\cdot$10$^5$ & 1.3$\cdot$10$^6$ \\
        HP-LSH & 34,118 & 2.4$\cdot$10$^5$ & 1.8$\cdot$10$^6$ & 34,148 & 4.1$\cdot$10$^6$ & 4.6$\cdot$10$^6$ & 6.7$\cdot$10$^5$ & 3.0$\cdot$10$^6$ & 1.4$\cdot$10$^7$ & 2.8$\cdot$10$^8$ & 1,328 & 1.3$\cdot$10$^5$ & 6.4$\cdot$10$^5$ & 2,375 & 1.9$\cdot$10$^6$ & 87,651 \\
        FAISS & 1,017 & \underline{30,128} & 1.7$\cdot$10$^6$ & 2,294 & \underline{1.5$\cdot$10$^5$} & \underline{2.0$\cdot$10$^5$} & 24,224 & 5.3$\cdot$10$^5$ & 1.8$\cdot$10$^5$ & \underline{1.3$\cdot$10$^8$} & 226 & \underline{19,368} & 41,974 & \textbf{\underline{2,293}} & 2.0$\cdot$10$^5$ & \textbf{\underline{2,516}} \\
        SCANN & 1,017 & \underline{30,128} & 1.4$\cdot$10$^6$ & 2,294 & \underline{1.5$\cdot$10$^5$} & \underline{2.0$\cdot$10$^5$} & 24,224 & 4.2$\cdot$10$^5$ & 1.6$\cdot$10$^6$ & \underline{1.3$\cdot$10$^8$} & \underline{226} & \underline{19,368} & 41,974 & 2,293 & 1.5$\cdot$10$^5$ & \textbf{\underline{2,516}} \\
        DeepBlocker & \underline{339} & 37,660 & \underline{5.5$\cdot$10$^5$} & \underline{2,294} & 1.6$\cdot$10$^5$ & 3.2$\cdot$10$^5$ & \underline{7,810} & \underline{43,418} & \underline{12,580} & - & 339 & 33,356 & \underline{13,540} & 2,294 & \underline{63,850} & 10,064 \\
        DDB & 11,280 & {\color{red}5,380} & {\color{red}15,195} & 13,080 & {\color{red}30,280} & {\color{red}39,050} & 39,050 & 1.1$\cdot$10$^5$ & 3.1$\cdot$10$^6$ & - & 11,280 & {\color{red}5,380} & {\color{red}15,195} & 13,080 & 1.1$\cdot$10$^5$ & 3.1$\cdot$10$^6$ \\
         \hline
        \multicolumn{17}{c}{\textbf{(d) The actual number of candidate pairs per method, dataset and schema settings.}}
	\end{tabular}
	\label{tb:results}
	\vspace{-16pt}
\end{table*}

%% file: tables/blockProcessing.tex
\begin{table*}[t]\centering
\scriptsize
\setlength{\tabcolsep}{3pt}
    \caption{The best configuration of each blocking workflow across all datasets and schema settings. }
    \vspace{-8pt}
	\begin{tabular}{ | c | l | r | r | r | r | r | r | r | r | r | r | r || r | r | r | r | r | r |}
		\cline{3-19}
		\multicolumn{2}{c|}{} &
		\multicolumn{11}{c||}{\textbf{Schema-agnostic}}&
		\multicolumn{6}{c|}{\textbf{Schema-based}}\\
		\multicolumn{2}{c|}{} &
		\multicolumn{1}{c|}{$\mathbf{D_{a1}}$} &
		\multicolumn{1}{c|}{$\mathbf{D_{a2}}$} &
		\multicolumn{1}{c|}{$\mathbf{D_{a3}}$} &
		\multicolumn{1}{c|}{$\mathbf{D_{a4}}$} &
		\multicolumn{1}{c|}{$\mathbf{D_{a5}}$} &
        \multicolumn{1}{c|}{$\mathbf{D_{a6}}$} &
        \multicolumn{1}{c|}{$\mathbf{D_{a7}}$} &
        \multicolumn{1}{c|}{$\mathbf{D_{a8}}$} &
        \multicolumn{1}{c|}{$\mathbf{D_{a9}}$} &
        \multicolumn{1}{c|}{$\mathbf{D_{a10}}$} &
        \multicolumn{1}{c||}{$\mathbf{D_{10K}}$} &
        \multicolumn{1}{c|}{$\mathbf{D_{b1}}$} &
		\multicolumn{1}{c|}{$\mathbf{D_{b2}}$} &
		\multicolumn{1}{c|}{$\mathbf{D_{b3}}$} &
		\multicolumn{1}{c|}{$\mathbf{D_{b4}}$} &
        \multicolumn{1}{c|}{$\mathbf{D_{b8}}$} &
        \multicolumn{1}{c|}{$\mathbf{D_{b9}}$} \\
		\hline
        \hline
        & $BP$ & - & - & \checkmark & \checkmark & \checkmark & - & - & - & \checkmark & - & \checkmark & - & - & - & - & - & - \\
        Standard & $BFr$ & 0.050 & 0.875 & 0.925 & 0.225 & 1.000 & 0.975 & 0.350 & 0.225 & 0.625 & 0.800 & - & 0.900 & 0.875 & 0.800 & 0.250 & 0.650 & 0.525 \\
        Blocking & $PA$ & WEP & BLAST & RCNP & RCNP & RCNP & RCNP & RCNP & RCNP & RCNP & BLAST & RCNP & RCNP & BLAST & BLAST & RWNP & BLAST & RCNP \\
        & $WS$ & ARCS & $\chi^2$ & $\chi^2$ & EJS & CBS & CBS & CBS & ARCS & JS & $\chi^2$ & CBS & CBS & $\chi^2$ & $\chi^2$ & ECBS & $\chi^2$ & CBS \\
        \hline
        \multirow{5}{*}{\shortstack[c]{Q-Grams\\ Blocking}} & $q$ & 4  & 4  & 6 & 6 & 6 & 6 & 6 & 5 & 3 & 4 & 5 & 6 & 4 & 3 & 6 & 3 & 6 \\
        & $BP$ & \checkmark & - & \checkmark & \checkmark & - & - & - & \checkmark & - & \checkmark & - & - & - & - & - & \checkmark & - \\
        & $BFr$ & 0.325 & 0.325 & 0.875 & 0.100 & 0.425 & 0.875 & 0.225 & 0.275 & 0.525 & 0.950 & 0.475 & 0.950 & 0.475 & 0.450 & 0.100 & 0.675 & 0.325\\
        & $PA$ & RCNP & BLAST & CEP & WEP & RCNP & RCNP & RCNP & RCNP & BLAST & BLAST & WEP & 
        RCNP & BLAST & BLAST & WEP & RCNP & RCNP \\
        & $WS$ & CBS & ARCS & $\chi^2$ & EJS & ARCS & ECBS & ECBS & ARCS & $\chi^2$ & $\chi^2$ & $\chi^2$ & CBS & ARCS & $\chi^2$ & EJS & $\chi^2$ & ECBS \\
        \hline
        \multirow{6}{*}{\shortstack[c]{Extended\\ Q-Grams\\ Blocking}} & $q$ & 4 & 4 & 3 & 2 & 6 & 6 & 6 & 6 & 6 & 3 & 5 & 3 & 3 & 4 & 3 & 3 & 6 \\
        & $t$ & 0.80 & 0.90 & 0.90 & 0.85 & 0.90 & 0.90 & 0.80 & 0.80 & 0.90 & 0.90 & 0.90 & 0.90 & 0.80 & 0.90 & 0.85 & 0.90 & 0.90 \\
        & $BP$ & - & - & - & \checkmark & \checkmark & - & - & - & \checkmark & - & - &- & - & - & - & - & -\\
        & $BFr$ & 0.025 & 0.900 & 0.500 & 0.025 & 0.775 & 0.675 & 0.175 & 0.175 & 0.475 & 0.750 & 0.625 & 0.975 & 0.325 & 0.750 & 0.025 & 0.750 & 0.400 \\
        & $PA$ & WEP & BLAST & WNP & WEP & RCNP & RCNP & RCNP & RCNP & RCNP & BLAST & WEP & 
        RCNP & WNP & BLAST & WEP & BLAST & RCNP \\
        & $WS$ & ECBS & $\chi^2$ & ARCS & EJS & ARCS & ARCS & EJS & ARCS & ECBS & $\chi^2$ & ARCS & CBS & ARCS & $\chi^2$ & EJS & $\chi^2$ & ECBS \\
        \hline
        \multirow{4}{*}{\shortstack[c]{Suffix\\ Arrays\\ Blocking}} & $l_{min}$ & 4 & 2 & 3 & 2 & 6 & 6 & 6 & 5 & 6 & 6 & 4 & 6 & 2 & 2 & 2 & 3 & 6 \\
        & $b_{max}$ & 2 & 46 & 65 & 16 & 35 & 79 & 20 & 52 & 100 & 96 & 8 & 10 & 35 & 46 & 97 & 92 & 100 \\
        & $PA$ & WEP & BLAST & RCNP & BLAST & RWNP & RWNP & BLAST & RCNP & RCNP & WNP & WEP &RCNP & BLAST & BLAST & BLAST & WEP & WNP \\
        & $WS$ & ECBS & $\chi^2$ & $\chi^2$ & $\chi^2$ & ARCS & ARCS & ARCS & ARCS & $\chi^2$ & $\chi^2$ & EJS & CBS & $\chi^2$ & $\chi^2$ & $\chi^2$ & $\chi^2$ & CBS \\
        \hline
        Extended & $l_{min}$ & 2 & 3 & 6 & 2 & 6 & 6 & 6 & 6 & 6 & 5 & 5 & 6 & 2 & 2 & 6 & 6 & 6 \\
        Suffix & $b_{max}$ & 3 & 10 & 84 & 8 & 39 & 86 & 23 & 24 & 84 & 100 & 8 & 100 & 18 & 98 & 20 & 91 & 100 \\
        Arrays & $PA$ & RWNP & BLAST & WNP & BLAST & RWNP & RWNP & BLAST & RWNP & RCNP & CNP & WEP & RCNP & BLAST & BLAST & BLAST & WNP & CEP \\
        Blocking & $WS$ & ARCS & ARCS & ARCS & $\chi^2$ & ARCS & ARCS & ARCS & ARCS & $\chi^2$ & $\chi^2$ & $\chi^2$ & CBS & ARCS & JS & ARCS & ARCS & ECBS \\
		\hline
	\end{tabular}
	\vspace{-10pt}
	\label{tb:bwConfigurations}
\end{table*}

%% file: tables/simJoins.tex
\begin{table*}[t]\centering
\scriptsize
\setlength{\tabcolsep}{3.5pt}
    \caption{The best configuration per sparse NN method across all datasets and schema settings. }
    \vspace{-8pt}
	\begin{tabular}{ | c | l | r | r | r | r | r | r | r | r | r | r | r || r | r | r | r | r | r | }
		\cline{3-19}
		\multicolumn{2}{c|}{} &
		\multicolumn{11}{c||}{\textbf{Schema-agnostic}}&
		\multicolumn{6}{c|}{\textbf{Schema-based}}\\
		\multicolumn{2}{c|}{} &
		\multicolumn{1}{c|}{$\mathbf{D_{a1}}$} &
		\multicolumn{1}{c|}{$\mathbf{D_{a2}}$} &
		\multicolumn{1}{c|}{$\mathbf{D_{a3}}$} &
		\multicolumn{1}{c|}{$\mathbf{D_{a4}}$} &
		\multicolumn{1}{c|}{$\mathbf{D_{a5}}$} &
        \multicolumn{1}{c|}{$\mathbf{D_{a6}}$} &
        \multicolumn{1}{c|}{$\mathbf{D_{a7}}$} &
        \multicolumn{1}{c|}{$\mathbf{D_{a8}}$} &
        \multicolumn{1}{c|}{$\mathbf{D_{a9}}$} &
        \multicolumn{1}{c|}{$\mathbf{D_{a10}}$} &
        \multicolumn{1}{c||}{$\mathbf{D_{10K}}$} &
        \multicolumn{1}{c|}{$\mathbf{D_{b1}}$} &
		\multicolumn{1}{c|}{$\mathbf{D_{b2}}$} &
		\multicolumn{1}{c|}{$\mathbf{D_{b3}}$} &
		\multicolumn{1}{c|}{$\mathbf{D_{b4}}$} &
        \multicolumn{1}{c|}{$\mathbf{D_{b8}}$} &
        \multicolumn{1}{c|}{$\mathbf{D_{b9}}$} \\
		\hline
        \hline
        \multirow{4}{*}{$\epsilon$-Join} & $CL$ & \checkmark & \checkmark & \checkmark & - & \checkmark & - & - & \checkmark & \checkmark & \checkmark & - & \checkmark & \checkmark & - & - & \checkmark  & - \\
        & $RM$ & T1G & C3G & C5G & T1G & C5GM & C2G & T1GM & C3GM & C3GM & T1G & C2G & C4G & C3GM & C3G & T1G & C3G & C3GM \\
        & $SM$ & Cosine &  Cosine &  Cosine & Jaccard &  Cosine &  Cosine &  Cosine & Jaccard & Jaccard &  Cosine & Jaccard & 
        Cosine & Cosine & Cosine & Cosine & Cosine & Cosine \\
        & $t$ & 0.82 & 0.26 & 0.08 & 0.58 & 0.16 & 0.34 & 0.49 & 0.28 & 0.35 & 0.15 & 0.44 &  0.63 & 0.38 & 0.39 & 1.00 & 0.41 & 0.81 \\
        \hline
        \multirow{5}{*}{kNN-Join} & 
        $CL$ & \checkmark & \checkmark & \checkmark & - & - & - & - & \checkmark & - & - & - & \checkmark & \checkmark & \checkmark & - & \checkmark & - \\
        & $RVS$ & \checkmark & - & \checkmark & - & - & - & - & \checkmark & \checkmark & \checkmark & - & \checkmark & - & \checkmark & - & \checkmark & \checkmark\\
        & $RM$ & C4GM & C3GM & G5GM & C2GM & C5G & C5G & C5G & C4GM & C4G & C4G & C2GM & 
        C5G & C2G & C3G & C3G & C2G & C2GM \\
        & $SM$ & Dice & Cosine & Cosine & Cosine & Cosine & Cosine & Cosine & Cosine & Cosine & Cosine & Jaccard & Cosine & Cosine & Cosine & Cosine & Cosine & Cosine \\
        & $K$ & 1 & 4 & 26 & 1 & 1 & 1 & 1 & 2 & 1 & 5 & 6 & 1 & 3 & 3 & 1 & 6 & 1 \\
		\hline
	\end{tabular}
	\vspace{-10pt}
	\label{tb:joinConfiguration}
\end{table*}

%% file: tables/nnConfiguration.tex
\begin{table*}[t]\centering
\scriptsize
\setlength{\tabcolsep}{4pt}
    \caption{The best configuration per dense NN method across all datasets and schema settings. }
    \vspace{-8pt}
	\begin{tabular}{ | c | l | r | r | r | r | r | r | r | r | r | r | r || r | r | r | r | r | r | }
		\cline{3-19}
		\multicolumn{2}{c|}{} &
		\multicolumn{11}{c||}{\textbf{Schema-agnostic}}&
		\multicolumn{6}{c|}{\textbf{Schema-based}}\\
		\multicolumn{2}{c|}{} &
		\multicolumn{1}{c|}{$\mathbf{D_{a1}}$} &
		\multicolumn{1}{c|}{$\mathbf{D_{a2}}$} &
		\multicolumn{1}{c|}{$\mathbf{D_{a3}}$} &
		\multicolumn{1}{c|}{$\mathbf{D_{a4}}$} &
		\multicolumn{1}{c|}{$\mathbf{D_{a5}}$} &
        \multicolumn{1}{c|}{$\mathbf{D_{a6}}$} &
        \multicolumn{1}{c|}{$\mathbf{D_{a7}}$} &
        \multicolumn{1}{c|}{$\mathbf{D_{a8}}$} &
        \multicolumn{1}{c|}{$\mathbf{D_{a9}}$} &
        \multicolumn{1}{c|}{$\mathbf{D_{a10}}$} &
        \multicolumn{1}{c||}{$\mathbf{D_{10K}}$} &
        \multicolumn{1}{c|}{$\mathbf{D_{b1}}$} &
		\multicolumn{1}{c|}{$\mathbf{D_{b2}}$} &
		\multicolumn{1}{c|}{$\mathbf{D_{b3}}$} &
		\multicolumn{1}{c|}{$\mathbf{D_{b4}}$} &
        \multicolumn{1}{c|}{$\mathbf{D_{b8}}$} &
        \multicolumn{1}{c|}{$\mathbf{D_{b9}}$} \\
		\hline
        \hline
        & $CL$ & - & - & - & \checkmark & - & \checkmark & - & - & - & - & \checkmark & \checkmark & \checkmark & - & \checkmark & \checkmark & -\\
        \textsf{MinHash} & $\#bands$ & 4 & 32 & 16 & 4 & 32 & 32 & 16 & 32 & 16	& - & 32 & 4 & 32 & 32 & 2	 & 32  & -\\
        \textsf{LSH} & $\#rows$ & 64 & 8 & 8 & 128 & 16 & 8 & 16 & 16 & 16 & - & 16 & 128 & 16 & 16 & 256 & 16 & - \\
         & $k$ & 2 & 2 & 2 & 2 & 2 & 5 & 2 & 2 & 2 & - & 2 & 2 & 2 & 2 & 5 & 2 & -\\
        \hline
         & $CL$ & - & \checkmark & \checkmark & \checkmark & - & - & \checkmark & - & \checkmark & - & \checkmark & \checkmark & \checkmark & \checkmark & - & \checkmark & - \\
       & $\#tables$ & 16 & 60 & 20 & 5 & 10 & 50 & 8 & 50 & 50 & 100 & 128 & 24 & 43 & 65 & 1 & 500 & 5 \\
        Cross-Polytope  & $\#hashes$ & 2 & 1 & 1 & 2 & 1 & 1 & 2 & 2 & 1 & 1 & 2 & 2 & 1 & 1 & 3 & 2 & 1 \\
        LSH & cp dimension & 32 & 256 & 256 & 256 & 512 & 256 & 128 & 128 & 512 & 128 & 256 & 128 & 256 & 512 & 16 & 128 & 512 \\
        & $\#probes$ & 180 & 60 & 159 & 400 & 87 & 50 & 339 & 1548 & 51 & 114 & 2,497 & 24 & 43 & 65 & 1 & 500 & 6 \\
		\hline
    	 & $CL$ & - & \checkmark & \checkmark & \checkmark & - & \checkmark & \checkmark & \checkmark & \checkmark & - & \checkmark & - & - & \checkmark & \checkmark & \checkmark & - \\
         Hyperplane & $\#tables$ & 13 & 50 & 19 & 28 & 17 & 200 & 200 & 150 & 100 & 200 & 512 & 13 & 100 & 19 & 2 & 150 & 100 \\
         LSH & $\#hashes$ & 15 & 11 & 9 & 18 & 12 & 14 & 16 & 15 & 13 & 12 & 19 & 14 & 11 & 12 & 20 & 14 & 19 \\
         & $\#probes$ & 297 & 385 & 275 & 602 & 461 & 1,450 & 226 & 878 & 643 & 1,961 & 9,056 & 33 & 181 & 577 & 2 & 360 & 111 \\
        \hline
        & $CL$ & - & \checkmark & \checkmark & \checkmark & - & - & - & \checkmark & \checkmark & \checkmark & \checkmark & \checkmark & - & \checkmark & \checkmark & \checkmark & - \\
        FAISS & $RVS$ & \checkmark & - & - & - & \checkmark & \checkmark & \checkmark & - & \checkmark & - & - & \checkmark & - & \checkmark & - & \checkmark & \checkmark \\
        & $K$ & 3 & 28 & 545 & 1 & 30 & 40 & 4 & 24 & 71 & 4,860 & 104 & 1 & 18 & 31 & 1 & 78 & 1 \\
        \hline
        & $CL$ & - & \checkmark & \checkmark & \checkmark & - & - & - & \checkmark & \checkmark & \checkmark & \checkmark & \checkmark & - & \checkmark & \checkmark & \checkmark & - \\
        & $RVS$ & \checkmark & - & - & - & \checkmark & \checkmark & \checkmark & - & \checkmark  & \checkmark - & & \checkmark & - & \checkmark & - & \checkmark & \checkmark \\
        SCANN & $K$ & 3 & 28 & 475 & 1 & 30 & 40 & 4 & 19 & 63 & 4,860 & 37 & 1 & 18 & 31 & 1 & 60 & 1 \\
        & index & BF & BF & AH & BF & BF & BF & BF & AH & AH & BF & AH & AH & BF & BF & BF & AH & BF \\
        & similarity & DP & L2$^2$ & L2$^2$ & L2$^2$ & L2$^2$ & L2$^2$ & DP & L2$^2$ & L2$^2$ & DP & L2$^2$ &  L2$^2$ & L2$^2$ & L2$^2$ & DP & L2$^2$ & DP \\
        \hline
        & $CL$ & - & \checkmark & \checkmark & \checkmark & \checkmark & - & \checkmark & \checkmark & \checkmark & - & \checkmark & \checkmark & - & - & \checkmark & - & \checkmark \\
        DeepBlocker & $RVS$ & - & \checkmark & \checkmark & \checkmark & - & - & \checkmark	&  - & - & - & - & - & \checkmark & - & \checkmark & - & - \\
        & $K$ & 1 & 35 & 180 & 1  & 31  & 63  & 1  & 17  & 5  & - & 65 & 1  & 31  & 10  & 1  & 25 & 4 \\
        \hline
	\end{tabular}
	\vspace{-10pt}
	\label{tb:nnConfiguration}
\end{table*}